\documentclass[reprint, aps, pre, superscriptaddress]{revtex4-2}
\usepackage[utf8]{inputenc}
\usepackage[english]{babel}
\usepackage{hyperref}

\usepackage{mathtools}
\usepackage{amssymb}
\usepackage{physics}
\usepackage{array}

\usepackage{graphicx}
\graphicspath{{./images/}}
\usepackage[dvipsnames]{xcolor}
\usepackage{tikz}
    \usetikzlibrary{calc} 
    \usetikzlibrary{arrows.meta}
    \usetikzlibrary{decorations.pathmorphing}     \usetikzlibrary{decorations.markings}

\definecolor{MyWhite}{rgb}{1.0, 1.0, 1.0}
\definecolor{MyBlack}{rgb}{0.0, 0.0, 0.0}
\definecolor{MyRed}{rgb}{0.8, 0.2, 0.2}
\definecolor{MyBlue}{rgb}{0.0, 0.3, 0.7}
\definecolor{MyGreen}{rgb}{0.2, 0.7, 0.2}
\definecolor{MyYellow}{rgb}{1.0, 0.9, 0.2}
\definecolor{MyPurple}{rgb}{0.6, 0.0, 0.6}
\definecolor{MyOrange}{rgb}{1.0, 0.6, 0.0}

\renewcommand{\bm}[1]{\mathbf{#1}} 
\let\defaultv = \v

\renewcommand{\d}{\text{d}} 
\renewcommand{\Tr}{\text{Tr}} 

\newcommand{\n}{\underline{n}} 
\newcommand{\p}{\bm{p}} 
\newcommand{\pp}{\bm{p}^{\prime}} 
\newcommand{\q}{N} 

\newcommand{\X}{X} 
\newcommand{\Y}{Y} 

\newcommand{\M}[1]{\bm{M}_{\text{#1}}} 
\newcommand{\U}{\bm{U}} 
\renewcommand{\L}{\bm{L}_{1}} 
\newcommand{\R}{\bm{R}_{2N}} 

\renewcommand{\v}{\bm{v}} 
\newcommand{\w}{\bm{w}} 
\newcommand{\V}{\bm{V}} 
\newcommand{\W}{\bm{W}} 
\renewcommand{\S}{\bm{S}} 
\newcommand{\T}{\bm{T}} 
\renewcommand{\r}{\ket*{r}} 
\renewcommand{\l}{\bra*{l}} 
\newcommand{\rp}{\ket*{r^{\prime}}} 
\newcommand{\lp}{\bra*{l^{\prime}}} 

\newcommand{\tM}[1]{\tilde{\bm{M}}_{\text{#1}}} 
\newcommand{\tU}{\tilde{\bm{U}}} 
\newcommand{\tL}{\tilde{\bm{L}}_{1}} 
\newcommand{\tR}{\tilde{\bm{R}}_{2N}} 
\newcommand{\A}{\bm{A}} 
\newcommand{\B}{\bm{B}} 
\newcommand{\tp}{\tilde{\bm{p}}} 
\newcommand{\tpp}{\tilde{\bm{p}}^{\prime}} 
\newcommand{\tv}{\tilde{\bm{v}}} 
\newcommand{\tw}{\tilde{\bm{w}}} 
\newcommand{\tV}{\tilde{\bm{V}}} 
\newcommand{\tW}{\tilde{\bm{W}}} 
\renewcommand{\tr}{\ket*{\tilde{R}}} 
\newcommand{\tl}{\bra*{\tilde{L}}} 
\newcommand{\trp}{\ket*{\tilde{R}^{\prime}}} 
\newcommand{\tlp}{\bra*{\tilde{L}^{\prime}}} 
\newcommand{\tZ}{\tilde{\bm{Z}}} 
\newcommand{\tz}{\tilde{\bm{z}}} 
\newcommand{\tq}{\tilde{\bm{q}}} 

\def\int{\tikz[scale=0.3, baseline=1.8]{
    \draw[draw=MyBlack!25] (-0.2, -0.2) grid (1.2, 1.2);
    \draw[draw=MyBlack!75, fill=MyYellow] (0.0, 0.0) rectangle (1.0, 1.0);
}}

\def\cemp{\tikz[scale=0.25, baseline=1.2]{
    \draw[draw=MyBlack!25] (-0.2, -0.2) grid (1.2, 1.2);
    \draw[draw=MyBlack!75, fill=MyWhite] (0.0, 0.0) rectangle (1.0, 1.0);
}}
\def\cocc{\tikz[scale=0.25, baseline=1.2]{
    \draw[draw=MyBlack!25] (-0.2, -0.2) grid (1.2, 1.2);
    \draw[draw=MyBlack!75, fill=MyBlack] (0.0, 0.0) rectangle (1.0, 1.0);
}}
\def\cempocc{\tikz[scale=0.15, baseline=1.8]{
    \draw[draw=MyBlack!25] (-0.2, -0.2) grid (1.2, 2.2);
    \draw[draw=MyBlack!75, fill=MyWhite] (0.0, 0.0) rectangle (1.0, 1.0);
    \draw[draw=MyBlack!75, fill=MyBlack] (0.0, 1.0) rectangle (1.0, 2.0);
}}
\def\coccemp{\tikz[scale=0.15, baseline=1.8]{
    \draw[draw=MyBlack!25] (-0.2, -0.2) grid (1.2, 2.2);
    \draw[draw=MyBlack!75, fill=MyBlack] (0.0, 0.0) rectangle (1.0, 1.0);
    \draw[draw=MyBlack!75, fill=MyWhite] (0.0, 1.0) rectangle (1.0, 2.0);
}}

\def\empemp{\tikz[scale=0.4, baseline=8.8]{
    \draw[draw=MyBlack!25] (-0.2, -0.2) grid (1.2, 2.2);
    \draw[draw=MyBlack!75, fill=MyWhite] (0.0, 0.0) rectangle (1.0, 1.0);
    \draw[draw=MyBlack!75, fill=MyWhite] (0.0, 1.0) rectangle (1.0, 2.0);
}}
\def\empoccalt{\tikz[scale=0.4, baseline=8.8]{
    \draw[draw=MyBlack!25] (-0.2, -0.2) grid (1.2, 2.2);
    \draw[draw=MyBlack!75, fill=MyWhite] (0.0, 0.0) rectangle (1.0, 1.0);
    \draw[draw=MyBlack!75, fill=MyBlack] (0.0, 1.0) rectangle (1.0, 2.0);
}}
\def\occempalt{\tikz[scale=0.4, baseline=8.8]{
    \draw[draw=MyBlack!25] (-0.2, -0.2) grid (1.2, 2.2);
    \draw[draw=MyBlack!75, fill=MyBlack] (0.0, 0.0) rectangle (1.0, 1.0);
    \draw[draw=MyBlack!75, fill=MyWhite] (0.0, 1.0) rectangle (1.0, 2.0);
}}
\def\occocc{\tikz[scale=0.4, baseline=8.8]{
    \draw[draw=MyBlack!25] (-0.2, -0.2) grid (1.2, 2.2);
    \draw[draw=MyBlack!75, fill=MyBlack] (0.0, 0.0) rectangle (1.0, 1.0);
    \draw[draw=MyBlack!75, fill=MyBlack] (0.0, 1.0) rectangle (1.0, 2.0);
}}

\def\empoccempalt{\tikz[scale=0.4, baseline=14.6]{
    \draw[draw=MyBlack!25] (-0.2, -0.2) grid (1.2, 3.2);
    \draw[draw=MyBlack!75, fill=MyWhite] (0.0, 0.0) rectangle (1.0, 1.0);
    \draw[draw=MyBlack!75, fill=MyBlack] (0.0, 1.0) rectangle (1.0, 2.0);
    \draw[draw=MyBlack!75, fill=MyWhite] (0.0, 2.0) rectangle (1.0, 3.0);
}}
\def\occempoccalt{\tikz[scale=0.4, baseline=14.6]{
    \draw[draw=MyBlack!25] (-0.2, -0.2) grid (1.2, 3.2);
    \draw[draw=MyBlack!75, fill=MyBlack] (0.0, 0.0) rectangle (1.0, 1.0);
    \draw[draw=MyBlack!75, fill=MyWhite] (0.0, 1.0) rectangle (1.0, 2.0);
    \draw[draw=MyBlack!75, fill=MyBlack] (0.0, 2.0) rectangle (1.0, 3.0);
}}

\begin{document}

\preprint{APS/123-QED}

\title{Exact solution of the Rule 150 reversible cellular automaton}

\author{Joseph W. P. Wilkinson}
\affiliation{School of Physics and Astronomy, University of Nottingham, Nottingham, NG7 2RD, United Kingdom}
\affiliation{Centre for the Mathematics and Theoretical Physics of Quantum Non-equilibrium Systems, University of Nottingham, Nottingham, NG7 2RD, United Kingdom}

\author{Toma\defaultv{z} Prosen}
\affiliation{Department of Physics, Faculty of Mathematics and Physics, University of Ljubljana, SI-1000 Ljubljana, Slovenia}

\author{Juan P. Garrahan}
\affiliation{School of Physics and Astronomy, University of Nottingham, Nottingham, NG7 2RD, United Kingdom}
\affiliation{Centre for the Mathematics and Theoretical Physics of Quantum Non-equilibrium Systems, University of Nottingham, Nottingham, NG7 2RD, United Kingdom}

\date{\today}

\begin{abstract}
    We study the dynamics and statistics of the Rule 150 reversible cellular automaton (RCA). This is a one-dimensional lattice system of binary variables with synchronous (Floquet) dynamics, that corresponds to a bulk deterministic and reversible discretized version of the kinetically constrained ``exclusive one-spin facilitated" (XOR) Fredrickson-Andersen (FA) model, where the local dynamics is restricted: a site flips if and only if its adjacent sites are in different states from each other. Similar to other RCA that have been recently studied, such as Rule~54 and Rule~201, the Rule~150 RCA is integrable, however, in contrast is noninteracting: the emergent quasiparticles, which are identified by the domain walls, behave as free fermions. This property allows us to solve the model by means of matrix product ans\"atze. In particular, we find the exact equilibrium and nonequilibrium stationary states for systems with closed (periodic) and open (stochastic) boundaries, respectively, resolve the full spectrum of the time evolution operator and, therefore, gain access to the relaxation dynamics, and obtain the exact large deviation statistics of dynamical observables in the long time limit.
\end{abstract}

\maketitle

\section{Introduction}\label{sec:introduction} 

In this paper we study the Rule 150 reversible cellular automaton (RCA) any solve many of its dynamical properties exactly. The model is defined on a one-dimensional lattice of sites of binary variables with deterministic and reversible discrete classical ``circuit'' dynamics. The naming of this RCA is due to the classification introduced in Ref.~\cite{Bobenko1993}, according to the specific dynamical rule.

The Rule 150 RCA is similar in many respects to other recently studied RCA, specifically, Rule 54~\cite{Prosen2016, Inoue2018, Prosen2017, Buca2019, Friedman2019, Gopalakrishnan2018, Gopalakrishnan2018b, Klobas2019, Klobas2020b, Alba2019, Alba2020, Klobas2020} (for a review see Ref.~\cite{Buca2021}) and Rule 201~\cite{Iadecola2020, Wilkinson2020}: (i) its dynamics is defined in terms of local space and time reversible gates applied periodically (in this sense it can be thought of as a driven Floquet system); (ii) the local dynamical rules impose kinetic constraints similar to those of known stochastic kinetically constrained models (KCM~\cite{Ritort2003, Garrahan2011, Garrahan2018}), particularly, variations of the Fredrickson-Andersen (FA) model: the ``exclusive one-spin facilitated" FA (XOR-FA) model~\cite{Causer2020} in the case of Rule 150, and the ``one-spin facilitated" FA (FA or OR-FA)~\cite{Fredrickson1984} and ``two-spin facilitated" FA (PXP or, simply, AND-FA)~\cite{Fendley2004} models, respectively, for Rules 54~\cite{Buca2019} and 201~\cite{Wilkinson2020}; and (iii) the Rule 150 RCA is integrable~\cite{Gombor2021}, but in contrast to Rules 54 and 201, its quasiparticles are noninteracting~\cite{Gopalakrishnan2018b}.

Properties (i) and (ii) mean that the Rule 150 RCA can alternatively be called the ``Floquet-XOR-FA'' model, as Rules 54 and 201 can, respectively, be called the Floquet-FA~\cite{Gopalakrishnan2018} and Floquet-PXP~\cite{Wilkinson2020}. Property (iii) implies that we can readily solve the Rule 150 RCA exactly, whereby the noninteracting nature of the emergent quasiparticles makes the solutions simpler than those for Rules 54 and 201. This is precisely what we do here using matrix product ans\"{a}tze. We consider the cases for periodic boundary conditions, for which the overall dynamics is completely deterministic, and open boundary conditions, where the dynamics becomes stochastic at the boundaries. We find the exact stationary states, for systems both in and out of equilibrium, obtain closed expressions for the complete spectrum of the Markov operator generating time evolution and, subsequently, resolve the relaxation dynamics, and compute the exact large deviation statistics for long time dynamical observables.

The study of RCA models like Rules 150, 54, and 201 relates to several other areas of interest. The first of these is slow dynamics due to physical constraints. Stochastic kinetically constrained models (KCM)~\cite{Fredrickson1984, Palmer1984, Jackle1991, Cancrini2008} (for a detailed review, see Refs.~\cite{Ritort2003, Garrahan2011, Garrahan2018}) are simple models for the kind of slow dynamically heterogeneous relaxation of classical glasses. Given that these RCA can be considered to be discrete, deterministic, and reversible counterparts to KCM, a natural question is to what extent they share features with those constrained models, for example, with the existence of phase transitions in their dynamical large deviations. This helps us to understand which properties are determined by kinetic constraints compared to those governed by the nature of the dynamics (e.g., stochastic vs. deterministic and integrable vs. ergodic). The second related area are ``circuit" systems of the brick-wall type, where dynamics is defined in terms of local gates applied synchronously throughout the system. Recently, this has become a much studied problem in the fields of quantum many-body physics, where the gates correspond to either unitary or dissipative transformations, as the CA can be used as tractable systems to address questions regarding, for example, entanglement growth, localization, operator spreading, chaos, and integrability~\cite{Nahum1, Nahum2, DeLuca, Bertini2019, Pollmann, Pollman2, sunderhauf, Khemani, Pretko}. In particular, circuit models exhibiting space-time duality are specially amenable to analytic solutions~\cite{Bertini2019, Krajnik2020, Klobas2020, Klobas2021}. The third related area is that of quantum KCM for the exploration of issues associated to quantum relaxation, nonergodicity, and nonthermal eigenstates~\cite{Horssen2015,Lan2018,Turner2018,Pancotti2020}. 

The main objective of this paper is to provide a clear, comprehensive, and self-contained study of the dynamics of the Rule 150 RCA. The simplicity of the model allows us to present numerous exact results (e.g., the stationary states, dynamical spectrum, and large deviations) which, for the more complex Rules 54 and 201, required several separate articles. In that sense, this current paper serves as an entry point for studying integrable RCA. The paper is organized as so. In Section~\ref{sec:model}, we introduce the model and define the discrete dynamics. In Sections~\ref{sec:equilibrium-stationary-states} and~\ref{sec:non-equilibrium-stationary-state}, we find the exact solution for the stationary states under closed periodic and open stochastic boundary conditions. In Section~\ref{sec:decay-modes}, we obtain exact analytic expressions for the entire spectrum of the stochastic time evolution operator and study the relaxation dynamics of the system in both the thermodynamic and long time limits. Section~\ref{sec:large-deviations} then presents the exact dynamical large deviation statistics of space and time extensive observables, whilst Section~\ref{sec:conclusions} provides our conclusions and several appendices contain miscellaneous other directly related results.

\bigskip

\noindent\textit{Note added:} As this paper was being completed, Ref.~\cite{Gombor2021} appeared proving that the Rule 150 RCA is Yang-Baxter integrable. 

\bigskip

\section{Model}\label{sec:model} 

In this section we introduce and define the model that we study throughout this paper.

\subsection{Dynamics}\label{sec:dynamics}

We consider a system, defined on a $(1 + 1)$-dimensional discrete square space-time lattice of even size $2N$ of sites, $x = 1, \ldots, 2N$, of binary variables, $n_{x} = 0, 1$. At discrete time $t$, the configuration $\n^{t}$ of the system is represented by a binary string,
\begin{equation}\label{eq:binary-representation}
    \n^{t} \equiv (n_{1}^{t}, n_{2}^{t}, \ldots, n_{2N}^{t}) \in \{0, 1\}^{2N},
\end{equation}
where the site $x$ at time $t$ is referred to as being \textit{empty} (or unexcited) if $n_{x}^{t} = 0$ and \textit{occupied} (or excited) if $n_{x}^{t} = 1$. We assume the system is initially closed and has periodic boundary conditions (PBC), imposed by identifying sites $n_{x + 2N}^{t} \equiv n_{x}^{t}$.

The time evolution of the system is defined in discrete time and consists of two distinct time steps. In the first, $\n^{2t} \to \n^{2t + 1}$, referred to as the even time step, only sites with \textit{even} index are updated, that is, sites with odd index are left unaltered, whereas in the second, $\n^{2t + 1} \to \n^{2t + 2}$, the odd time step, only sites with \textit{odd} index are updated. A full step of time evolution, $\n^{2t} \to \n^{2t + 2}$, is then defined by the composition of an even and odd time step, respectively. This discrete staggered dynamics is generated by the local space-time (or ``parity"~\cite{Gopalakrishnan2018d}) mapping,
\begin{equation}\label{eq:site-map}
    n_{x}^{t + 1} =
    \begin{cases}
        f_{x}^{t}, & x + t = 0 \pmod{2}, \\
        n_{x}^{t}, & x + t = 1 \pmod{2}, \\
    \end{cases}
\end{equation} 
where we have introduced the shorthand notation,
\begin{equation}\label{eq:bulk-update}
    f_{x}^{t} \equiv f(n_{x - 1}^{t}, n_{x}^{t}, n_{x + 1}^{t}),
\end{equation}
to denote a three-site function acting on site $x$ at time $t$. The dynamics is given by the discrete, deterministic, and reversible \textit{Rule 150 reversible cellular automaton} (RCA), identified by the local update rule,
\begin{equation}\label{eq:rule-150}
    f_{x}^{t} = n_{x - 1}^{t} + n_{x}^{t} + n_{x + 1}^{t} \pmod{2}.
\end{equation}

\begin{figure}[t]
    \centering
    \includegraphics{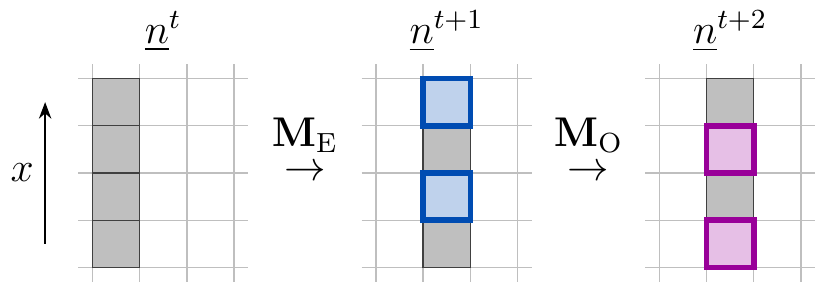}
    \caption{\textbf{Time evolution.} Schematic representation of the discrete time evolution of $2N = 4$ sites of the lattice under one full step of time evolution (i.e., two successive time steps). In the first, the \textit{even time step}, only sites with even indices are updated, while during the second, the \textit{odd time step}, only odd indexed sites are updated. Blue and purple borders indicate which sites have been updated by the local three-site function $f_{x}^{t}$ in the even and odd time steps, respectively.}
    \label{fig:evolution}
\end{figure}

\begin{figure}[t]
    \centering
    \includegraphics{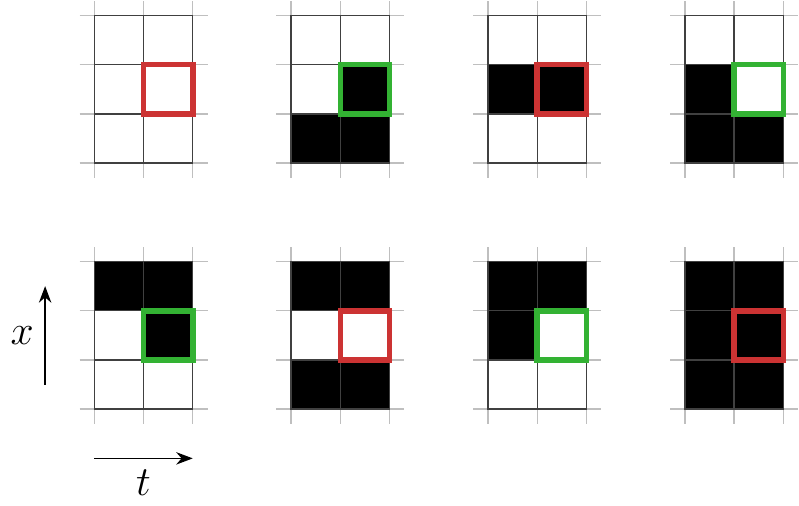}
    \caption{\textbf{Rule 150.} Illustration of the Rule 150 cellular automaton, as defined in Eq.~\eqref{eq:rule-150}, where white and black squares represent empty and occupied sites, respectively. In each diagram, only the central site is updated; green and red borders indicate whether the site has changed or not under the action of the deterministic local function $f_{x}^{t}$. Note also the discrete local symmetries of the model: spatial-inversion (``up-down"), time-reversal (``left-right"), and particle-hole (``black-white").}
    \label{fig:rule-150}
\end{figure}

It is convenient to represent the time evolution of the lattice geometrically, as shown schematically in Figure~\ref{fig:evolution}. It then follows that the local update rule in Eq.~\eqref{eq:rule-150} can be expressed diagrammatically, as illustrated in Figure~\ref{fig:rule-150}, by representing the empty and occupied sites with white and black squares, respectively, where the squares on the left of each diagram correspond to the local subconfigurations of sites at time $t$, i.e., $(n_{x - 1}^{t}, n_{x}^{t}, n_{x + 1}^{t})$, while the squares on the right are the same subset of sites at $t + 1$, that is, after the local update rule~\eqref{eq:rule-150} acts on the triplet of sites, i.e., $(n_{x - 1}^{t + 1}, n_{x}^{t + 1}, n_{x + 1}^{t + 1}) \equiv (n_{x - 1}^{t}, f_{x}^{t}, n_{x + 1}^{t})$. In addition to efficiently representing the discrete dynamics of Rule 150, Figure~\ref{fig:rule-150} also illustrates the local symmetries exhibited by the model. Explicitly, a \textit{spatial-inversion} symmetry,
\begin{equation}\label{eq:model-spatial-inversion-symmetry}
    f(n_{x - 1}^{t}, n_{x}^{t}, n_{x - 1}^{t}) = f(n_{x + 1}^{t}, n_{x}^{t}, n_{x - 1}^{t}),
\end{equation}
a \textit{time-reversal} symmetry,
\begin{equation}\label{eq:model-time-reversal-symmetry}
     n_{x}^{t} = f(n_{x - 1}^{t}, f(n_{x - 1}^{t}, n_{x}^{t}, n_{x + 1}^{t}), n_{x + 1}^{t}),
\end{equation}
and a \textit{particle-hole} symmetry,
\begin{equation}\label{eq:model-particle-hole-symmetry}
\begin{aligned}
    f(n_{x - 1}^{t}, n_{x}^{t}, n_{x - 1}^{t}) & \\ & \hspace{-18pt} = 1 - f(1 - n_{x - 1}^{t}, 1 - n_{x}^{t}, 1 - n_{x + 1}^{t}),
\end{aligned}
\end{equation}
which respectively manifest through the invariance of the local dynamics under the exchange of sites $x - 1 \leftrightarrow x + 1$, times $t - 1 \leftrightarrow t + 1$, and variables $0 \leftrightarrow 1$.

From a dynamical perspective, the local update~\eqref{eq:rule-150} can be understood as a \textit{kinetic constraint} whereby a site flips if and only if one of the sites adjacent to it is empty, with the other occupied. We can, therefore, interpret Rule 150 as a discrete, deterministic, and reversible version of the ``exclusive one-spin facilitated'' Fredrickson-Andersen (or XOR-FA) model~\cite{Fredrickson1984, Gopalakrishnan2018b, Causer2020}. This contrasts the ``one-spin facilitated" Fredrickson-Andersen (FA) model associated to the extensively studied Rule 54 RCA~\cite{Prosen2016, Prosen2017, Inoue2018, Gopalakrishnan2018, Gopalakrishnan2018b, Alba2019, Buca2019, Klobas2019, Klobas2020a, Klobas2020b}: a site can flip if either of its nearest neighbouring sites are occupied. As the map~\eqref{eq:site-map} is applied periodically, we refer to the dynamics as \textit{Floquet}, hence, the Floquet-XOR-FA model.

\subsection{Statistical states}\label{sec:states}

The \textit{statistical states} of the system are defined as probability distributions over the set of configurations $\n$, and are represented by vectors in $(\mathbb{R}^{2})^{\otimes{2N}}$,
\begin{equation}
    \p^{t} = \sum_{n} p_{n}^{t} \bm{e}_{n}, \qquad \bm{e}_{n} = \bigotimes_{x = 1}^{2N} \bm{e}_{n_{x}},
\end{equation}
where $\bm{e}_0$ and $\bm{e}_1$ are basis vectors in $\mathbb{R}^2$, and the nonnegative and normalized components,
\begin{equation}\label{eq:probability-conditions}
    p_{n}^{t} \geq 0, \qquad \sum_{n} p_{n}^{t} = 1,
\end{equation}
correspond to the probabilities of the configurations $\n$ at time $t$, given by the binary representation of the integer, $\smash{n = \sum_{x = 1}^{2N} 2^{2N - x} n_{x}}$. The probabilities over the configurations can then be written equivalently as
\begin{equation}
    p_{n}^{t} \equiv p^t_{\n} \equiv p^t_{n_{1}, n_{2}, \ldots, n_{2N}}.
\end{equation}

The time evolution of the statistical states is defined locally in terms of an $8 \times 8$ permutation matrix $\U$ acting on the vector space $(\mathbb{R}^{2})^{\otimes 3}$ (i.e., three sites of the lattice) that encodes the local update rule in Eq.~\eqref{eq:rule-150},
\begin{equation}
\begin{aligned}
    [\U]_{m_{x - 1} m_{x} m_{x + 1}, n_{x - 1} n_{x} n_{x + 1}} & \\ 
    & \hspace{-36pt} = \delta_{m_{x - 1}, n_{x - 1}} \delta_{m_{x}, f_{x}} \delta_{m_{x + 1}, n_{x + 1}}.
\end{aligned}
\end{equation}
Explicitly, the local time evolution operator is given by
\begin{equation}\label{eq:local-time-evolution-operator}
    \U =
    \begin{bmatrix}
        1 & & & & & & & \\
        & & & 1 & & & & \\
        & & 1 & & & & & \\
        & 1 & & & & & & \\
        & & & & & & 1 & \\
        & & & & & 1 & & \\
        & & & & 1 & & & \\
        & & & & & & & 1 \\
    \end{bmatrix},
\end{equation}
which we remark is symmetric and involutory and, therefore, orthogonal,
\begin{equation}
    \U = \U^{\mathrm{T}} = \U^{-1}, \qquad \U^{2} = \bm{I}^{\otimes 3},
\end{equation}
where $\bm{I}$ is the $2 \times 2$ identity matrix acting on the elementary space $\mathbb{R}^{2}$ (i.e., a single site of the lattice).

The full time evolution of the state $\p^{t}$ is then given by the discrete \textit{Floquet master equation},
\begin{equation}\label{eq:model-floquet-master-equation}
    \p^{t + 1} = 
    \begin{cases}
        \M{E} \p^{t}, & t = 0 \pmod{2}, \\
        \M{O} \p^{t}, & t = 1 \pmod{2}, \\
    \end{cases}
\end{equation}
where $\M{E}$ and $\M{O}$ are products of local operators acting on even and odd sites on the even and odd time steps, respectively [cf. Eq.~\eqref{eq:site-map}],
\begin{equation}\label{eq:periodic-time-step-operators}
    \M{E} = \prod_{x = 1}^{N} \U_{2x}, \qquad
    \M{O} = \prod_{x = 1}^{N} \U_{2x - 1},
\end{equation}
with the shorthand notation $\U_{x}$ introduced to denote the local operator $\U$ acting nontrivially on the site $x$,
\begin{equation}
    \U_{x} = \bm{I}^{\otimes (x - 2)} \otimes \U \otimes \bm{I}^{\otimes (2N - x - 1)}.
\end{equation}

Notice that whilst $\U_{x}$ acts on just three adjacent sites of the lattice $(x - 1, x, x + 1)$ it only affects site $x$ and so satisfies the following compatibility conditions,
\begin{equation}\label{eq:bulk-compatibility-condition}
    [\U_{2x}, \U_{2x + 2j}] = 0, \qquad [\U_{2x - 1}, \U_{2x + 2j - 1}] = 0,
\end{equation}
which implies that the order in which the $\U_{x}$ are applied in the even and odd time steps is irrelevant. Additionally, the discrete local symmetries of the update~\eqref{eq:rule-150} guarantee that the time evolution operator $\U$ satisfies the following symmetry conditions,
\begin{equation}\label{eq:model-local-symmetry-conditions}
    [\U, \bm{J}_{\mathrm{S}}] = 0, \qquad [\U, \bm{J}_{\mathrm{T}}] = 0, \qquad [\U, \bm{J}_{\mathrm{P}}] = 0,
\end{equation}
where the $8 \times 8$ matrices $\bm{J}_{\mathrm{S}}$, $\bm{J}_{\mathrm{T}}$, and $\bm{J}_{\mathrm{P}}$ are, respectively, the generators of the spatial-inversion, time-reversal, and particle-hole symmetries [cf. Eqs.~\eqref{eq:model-spatial-inversion-symmetry},~\eqref{eq:model-time-reversal-symmetry}, and~\eqref{eq:model-particle-hole-symmetry}], which are given explicitly in Appendix~\ref{app:symmetries}. Theses subsequently manifest in the dynamics of the \textit{Floquet operator},
\begin{equation}\label{eq:model-floquet-master-operator}
    \M{} \equiv \M{O} \M{E},
\end{equation}
in terms of a combination of a spatial-inversion and time-reversal symmetry, reminiscent of the associated symmetries of the local time evolution operator $\U$, and a similar particle-hole symmetry. Explicitly,
\begin{equation}\label{eq:model-global-symmetry-conditions}
    [\M{}, \mathcal{J}_{\mathrm{ST}}] = 0, \qquad [\M{}, \mathcal{J}_{\mathrm{P}}] = 0,
\end{equation}
where $\mathcal{J}_{\mathrm{ST}} \equiv \mathcal{J}_{\mathrm{S}} \mathcal{J}_{\mathrm{T}}$ and $\mathcal{J}_{\mathrm{P}}$ are the respective generators of the symmetries. Moreover, the dynamics of the model exhibits a further \textit{number-parity} symmetry,
\begin{equation}\label{eq:model-global-number-parity-symmetry}
    [\M{}, \mathcal{J}_{\mathrm{N}}] = 0,
\end{equation}
which conserves the parity of the number of excited sites. For more details on the symmetries, see Appendix~\ref{app:symmetries}.

\subsection{Quasiparticles}\label{sec:quasiparticles}

\begin{figure}[t]
    \centering
    \includegraphics{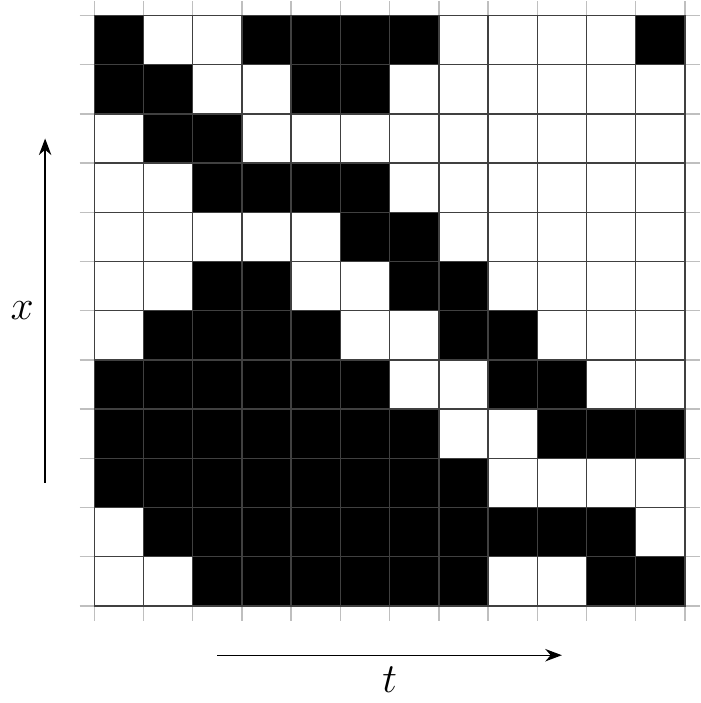}
    \caption{\textbf{Dynamics.} An example of the discrete time evolution of a random initial configuration with periodic boundary conditions. We intuitively interpret the pairs of adjacent sites located at the interfaces between sets of empty and occupied sites (i.e., $\protect\cempocc$, $\protect\coccemp$) as up- and down-moving \textit{quasiparticles} that are noninteracting and propagate ballistically with velocities of $v^{\pm} = \pm 1$, respectively. Notice that after colliding, the sites of the quasiparticles are exchanged (i.e., $\protect\cempocc \leftrightarrow \protect\coccemp$).}
    \label{fig:trajectory}
\end{figure}

The graphical representation for the model introduced in Figure~\ref{fig:evolution} immediately offers an intuitive interpretation of the discrete dynamics in terms of up- and down-moving \textit{quasiparticles} (see, e.g., Figure~\ref{fig:trajectory}), which propagate ballistically with constant velocities of $v^{\pm} = \pm 1$ and interact trivially without scattering. We can, therefore, interpret the model as a discretized \textit{Fermi gas} (i.e., an ensemble of noninteracting fermions in discrete space and time). The quasiparticles, or \textit{solitons}, are identified as pairs of adjacent sites located at the interfaces between sets of empty and occupied sites (i.e., the \textit{domain walls}), as highlighted in Figure~\ref{fig:trajectory}. Specifically,
\begin{equation}
    (0, 1) \equiv \empoccalt, \qquad (1, 0) \equiv \occempalt.
\end{equation}
Whether a quasiparticle is \textit{positive} (i.e., an up-mover) or \textit{negative} (i.e., down-mover) depends explicitly on the parity of the sum of the space and time indices, as succinctly detailed by the following expression,
\begin{equation}\label{eq:quasiparticle-detection}
    (n_{x}^{t}, 1 - n_{x}^{t}) \equiv
    \begin{cases}
        \text{negative}, & x + t = 0 \pmod{2}, \\
        \text{positive}, & x + t = 1 \pmod{2}. \\
    \end{cases}
\end{equation}
It then follows that quasiparticles only collide if they have opposite velocities. Specifically, the interactions between quasiparticles are necessarily \textit{two-body}, involving exactly one up-mover and one down-mover, and are given by the partial overlap of the subconfigurations representing the positive and negative quasiparticles. Explicitly,
\begin{equation}
    (0, 1, 0) \equiv \empoccempalt, \qquad (1, 0, 1) \equiv \occempoccalt.
\end{equation}
The remaining sites between quasiparticles, namely, the subsets of empty and occupied sites,
\begin{equation}
    (\ldots, 0, 0, \ldots) \equiv
    \begin{matrix}
        \vdots \\
        \empemp \\
        \vdots \\
    \end{matrix},
    \qquad (\ldots, 1, 1, \ldots) \equiv
    \begin{matrix}
        \vdots \\
        \occocc \\
        \vdots \\
    \end{matrix},
\end{equation}
are then collectively referred to as \textit{vacua}.

Due to the even system size $2N$ and PBC the numbers of positive and negative quasiparticles in a configuration are \textit{constrained} and must satisfy the following identity,
\begin{equation}\label{eq:quasiparticle-constraint}
    \q^{+}_{n} - \q^{-}_{n} = 0 \pmod{2},
\end{equation}
where $\q^{+}_{n}$ and $\q^{-}_{n}$ count the total number of positive and negative quasiparticles, respectively, in the configuration $\n$. To prove this, we introduce a graph representation for the lattice and demonstrate that all closed walks, which correspond to the configurations, are composed of cycles that necessarily satisfy the physical constraint~\eqref{eq:quasiparticle-constraint}. The details of this proof are presented in Appendix~\ref{app:quasiparticle-constraint}.

\section{Equilibrium stationary states for periodic boundary conditions}\label{sec:equilibrium-stationary-states}

A particularly interesting family of macroscopic states are those invariant under time evolution. In this section, we consider the \textit{equilibrium stationary states} (ESS). The simplest class of ESS, as we will show, can be constructed by introducing a pair of chemical potentials associated to the quasiparticles of each species which are conjugate to the numbers of positive and negative quasiparticles that are conserved by the deterministic dynamics and periodic boundary conditions (PBC). We demonstrate that these stationary states correspond to \textit{generalized Gibbs states}, which we show can be expressed in two equivalent forms. Namely, using a \textit{patch state ansatz} (PSA) and as a \textit{matrix product state} (MPS), as was done for Rule 54 in Refs.~\cite{Prosen2016} and~\cite{Prosen2017}, respectively. The principal benefit of the PSA is in its intuitive construction, which only requires that the states be stationary and exhibit short-range correlations. Moreover, it facilities a rigorous derivation for an efficient MPS representation of the state, which manifests a highly versatile algebraic structure that explicitly demonstrates the stationarity of the states without relying on the prior equivalence to the PSA.

\subsection{Patch state ansatz}\label{sec:patch-state-ansatz}

Given the staggering of the discrete time evolution, we require the stationary states to map into themselves after a full step of time evolution (i.e., a consecutive even and odd time step). Therefore, each ESS is associated to two vectors, $\p$ and $\pp$, which correspond to the even and odd time steps, respectively,
\begin{equation}\label{eq:stationarity-condition}
    \M{E} \p = \pp, \qquad \M{O} \pp = \p.
\end{equation}
For systems with PBC, the dynamics is reversible and so the conditions for time invariance~\eqref{eq:stationarity-condition} can be recast as
\begin{equation}\label{eq:periodic-stationarity-condition}
    \M{E} \p = \M{O} \p.
\end{equation}
We now propose the following patch state ansatz, similar to those introduced for Rule 54~\cite{Prosen2016} and Rule 201~\cite{Wilkinson2020}, for the components $p_{n}$ of the stationary state $\p$, that can be straightforwardly demonstrated to be the simplest ansatz of this form. Namely, the staggered product of $2N$ rank $2$ tensors exhibiting short-range correlations,
\begin{equation}\label{eq:psa}
    p_{n} = \frac{1}{Z} \big( \X_{n_{1}, n_{2}} \Y_{n_{2}, n_{3}} \cdots \X_{n_{2N - 1}, n_{2N}} \Y_{n_{2N}, n_{1}} \big), 
\end{equation}
where $\X_{n_{x}, n_{x + 1}}$ and $\Y_{n_{x}, n_{x + 1}}$ are the rank $2$ tensors to be determined, and $Z$ is the partition function given by the normalization.

In order to ensure that the stationarity condition~\eqref{eq:periodic-stationarity-condition} is satisfied, the following equality must hold for each and every configuration $\n$,
\begin{equation}\label{eq:psa-equations}
\begin{aligned}
    \X_{n_{1}, f_{2}} & \Y_{f_{2}, n_{3}} \cdots \X_{n_{2N - 1}, f_{2N}} \Y_{f_{2N}, n_{1}} \\
    & \qquad = \X_{f_{1}, n_{2}} \Y_{n_{2}, f_{3}} \cdots \X_{f_{2N - 1}, n_{2N}} \Y_{n_{2N}, f_{1}}.
\end{aligned}
\end{equation}
For $N > 1$, this set of equations is highly degenerate and overdetermined, and simplifies to the following conditions for the scalar components,
\begin{equation}\label{eq:psa-solutions}
    \X_{00} \Y_{00} = \X_{11} \Y_{11}, \qquad \X_{01} \Y_{10} = \X_{10} \Y_{01}.
\end{equation}
We recall that the probabilities $p_{n}$ are normalized by the partition function $Z$ and so we are free to set $\X_{00} \Y_{00} = 1$ which, together with Eq.~\eqref{eq:psa-solutions}, implies
\begin{equation}\label{eq:psa-normalization}
    \X_{00} \Y_{00} = \X_{11} \Y_{11} = 1.
\end{equation}          
Furthermore, we note that the scalar components are determined up to the following gauge transformation,
\begin{equation}\label{eq:psa-gauge-transformation}
\begin{aligned}
    \X_{n_{x}, n_{x + 1}} & \mapsto g_{n_{x}} \X_{n_{x}, n_{x + 1}} h^{-1}_{n_{x + 1}}, \\
    \Y_{n_{x}, n_{x + 1}} & \mapsto h_{n_{x}} \Y_{n_{x}, n_{x + 1}} g^{-1}_{n_{x + 1}}, 
\end{aligned}
\end{equation}
which, together with the normalization in Eq.~\eqref{eq:psa-normalization}, allows us to choose the following gauge,
\begin{equation}\label{eq:psa-gauge}
    \X_{00} = \Y_{00} = \X_{01} = \Y_{01} = 1.
\end{equation}
Combining the solutions to the system of equations~\eqref{eq:psa-solutions} with the chosen normalization~\eqref{eq:psa-normalization} and gauge~\eqref{eq:psa-gauge} yields the following two-parameter family of solutions,
\begin{equation}\label{eq:psa-arbitrary}
\begin{aligned}
    \X_{00} & = 1, \\
    \X_{01} & = 1, \\
    \X_{10} & = \xi \omega, \\
    \X_{11} & = \frac{\omega}{\xi},
\end{aligned} \qquad
\begin{aligned}
    \Y_{00} & = 1, \\
    \Y_{01} & = 1, \\
    \Y_{10} & = \xi \omega, \\
    \Y_{11} & = \frac{\xi}{\omega},
\end{aligned}
\end{equation}
where $\xi$ and $\omega$ are \textit{spectral parameters} which, due to the nonnegativity and normalizability of the probabilities $p_{n}$, are strictly positive (i.e., $\xi, \omega \in \mathbb{R}^{+}$). 

The conditions for stationarity~\eqref{eq:stationarity-condition}, together with the solutions~\eqref{eq:psa-arbitrary} imply that $\pp$, that is, the stationary state associated with the odd time step, takes on a form similar to $\p$, but with the patch tensors exchanged. Explicitly,
\begin{equation}
    p^{\prime}_{n} = \frac{1}{Z} \big(\Y_{n_{1}, n_{2}} \X_{n_{2}, n_{3}} \cdots \Y_{n_{2N - 1}, n_{2N}} \X_{n_{2N}, n_{1}} \big).
\end{equation}
We remark that interchanging the roles of the patch state tensors $\smash{\X_{n_{x}, n_{x + 1}} \leftrightarrow \Y_{n_{x}, n_{x + 1}}}$ is equivalent to exchanging the spectral parameters $\xi \leftrightarrow \omega$, and, therefore the states $\p \leftrightarrow \pp$. Hence, the PSA preserves the symmetry of the model, specifically, shifting the state one site in space is equivalent to evolving the state one step in time.

\subsection{\label{sec:conserved-charges} Conserved charges}

The parametrization chosen for the tensors in Eq.~\eqref{eq:psa-arbitrary} is arbitrary. However, these solutions exhibit a physically intuitive form, whereby the spectral parameters $\xi$ and $\omega$ can be expressed in terms of thermodynamic quantities,
\begin{equation}\label{eq:chemical-potentials}
\begin{aligned}
    \xi = \exp(-\mu^{+}), \qquad
    \omega = \exp(-\mu^{-}),
\end{aligned}
\end{equation}
with $\mu^{\pm}$ the chemical potentials associated to the positive and negative quasiparticles, respectively. To demonstrate this, we utilise the gauge freedom to transform the patch state tensor solutions into an equivalent form. Explicitly, we choose the gauge transformation
\begin{equation}
\begin{aligned}
    g_{0} & = 1, \\
    g_{1} & = \frac{1}{\omega},
\end{aligned} \qquad
\begin{aligned}
    h_{0} & = 1, \\
    h_{1} & = \frac{1}{\xi},
\end{aligned}
\end{equation}
which, by~\eqref{eq:psa-arbitrary}, yields
\begin{equation}\label{eq:psa-scalars}
\begin{aligned}
    \X_{00} & \mapsto 1, \\
    \X_{01} & \mapsto \xi, \\
    \X_{10} & \mapsto \xi, \\
    \X_{11} & \mapsto 1,
\end{aligned} \qquad
\begin{aligned}
    \Y_{00} & \mapsto 1, \\
    \Y_{01} & \mapsto \omega, \\
    \Y_{10} & \mapsto \omega, \\
    \Y_{11} & \mapsto 1.
\end{aligned}
\end{equation}
It follows from Eq.~\eqref{eq:quasiparticle-detection} that the number of each species of quasiparticle within a configuration $\n$ can be determined by the counts of the two site subconfigurations $(0, 1)$ and $(1, 0)$. Therefore, the newly parametrized solutions imply that the components $p_{n}$ of the stationary states $\p$ can be distributed according to a \textit{grand canonical ensemble},
\begin{equation}\label{eq:grand-canonical-ensemble}
    p_{n} = \frac{1}{Z} \xi^{\q^{+}_{n}} \omega^{\q^{-}_{n}},
\end{equation}
where the numbers of positive and negative quasiparticles $\q^{\pm}_{n}$ in the configurations $\n$ can be calculated directly by taking the logarithmic derivatives of the (unnormalized) probability components $p_{n}$ of the PSA. Explicitly,
\begin{equation}
\begin{aligned}
    \q^{+}_{n} & = \frac{\mathrm{d}}{\mathrm{d} \ln \xi} \ln \prod_{x = 1}^{N} \X_{n_{2x - 1}, n_{2x}}, \\
    \q^{-}_{n} & = \frac{\mathrm{d}}{\mathrm{d} \ln \omega} \ln \prod_{x = 1}^{N} \Y_{n_{2x}, n_{2x + 1}},
\end{aligned}
\end{equation}
which can equivalently be written as extensive sums over the locally conserved charges as
\begin{equation}
\begin{aligned}
    \q^{+}_{n} & = \sum_{x = 1}^{N} \frac{\mathrm{d}}{\mathrm{d} \xi} \X_{n_{2x - 1}, n_{2x}}, \\
    \q^{-}_{n} & = \sum_{x = 1}^{N} \frac{\mathrm{d}}{\mathrm{d} \omega} \Y_{n_{2x}, n_{2x + 1}}.
\end{aligned}
\end{equation}
It follows straightforwardly that the positive and negative quasiparticles are exactly the elementary local charges of the Floquet-XOR-FA model. Still, they do not represent a \textit{complete} set of local charges. Indeed, it can be readily shown that PSA tensors with ranks $r > 2$ yield similarly conserved charges that correspond to localized groups of noninteracting quasiparticles of the same species. Solving an equivalent set of equations to Eq.~\eqref{eq:psa-equations}, explicitly,
\begin{equation}\label{eq:psa-equations-general}
    p_{n} = \frac{1}{Z} \big( \X_{n_{1}, \ldots, n_{r}} \Y_{n_{2}, \ldots, n_{r + 1}} \cdots \Y_{n_{2N}, \ldots, n_{r - 1}} \big),
\end{equation}
we observe that the Floquet-XOR-FA model possesses an exponential number of locally conserved charges, as can be seen in Table~\ref{tab:conserved-charges}. We can then immediately deduce that the number of groups of noninteracting quasiparticles of the same species with support $r$, denoted by $\#_{r}$, reads
\begin{equation}
    \#_{r} = 2 \sum_{k = 1}^{\lfloor \frac{r}{2} \rfloor} \binom{\lfloor \frac{r}{2} \rfloor - 1}{k - 1} = 2 \big(2^{\lfloor \frac{r}{2} \rfloor - 1}\big) = 2^{\lfloor \frac{r}{2} \rfloor},
\end{equation}
where $k$ counts the number of quasiparticles of the same species in the localized group with support $r$. Intuitively, this can be understood simply as following directly from the physical properties of the quasiparticles. Specifically, the expression for $\#_{r}$ counts the total number of ways of arranging $k$ quasiparticles of the same species of size $2$ on $r$ sites for $k = 1, \ldots, \lfloor \frac{r}{2} \rfloor$ for each species of quasiparticle.

\begin{table}
\begin{ruledtabular}
\begin{tabular}{ccccccccc}
    $r$ & $2$ & $3$ & $4$ & $5$ & $6$ & $7$ & $8$ & $9$ \\
    \hline
    $\#_{r}$ & $2$ & $2$ & $4$ & $4$ & $8$ & $8$ & $16$ & $16$ \\
\end{tabular}
\end{ruledtabular}
\caption{\textbf{Locally conserved charges.} The number of locally conserved charges $\#_{r}$ with support $r$, obtained numerically by solving the sets of equations in Eq.~\eqref{eq:psa-equations-general} with rank $r$ tensors.}
\label{tab:conserved-charges}
\end{table}

\subsection{Matrix product ansatz}\label{sec:matrix-product-state}

As with Rules 54~\cite{Prosen2017} and 201~\cite{Wilkinson2020}, the stationary states can equivalently be expressed in terms of \textit{matrix product states},
\begin{equation}\label{eq:mps}
    p_{n} = \frac{1}{Z} \text{Tr} \big( \V_{n_{1}} \W_{n_{2}} \cdots \V_{n_{2N - 1}} \W_{n_{2N}} \big),
\end{equation} 
where $\V_{n_{x}}$ and $\W_{n_{x}}$ are matrices to be determined,
and $Z$ is the partition function. In order to efficiently derive the exact MPS construction and present the versatile algebraic cancellation scheme that explicitly demonstrates the stationarity of the states, it will prove convenient to introduce the following vectors of matrices, which correspond to the physical sites of the lattice,
\begin{equation}\label{eq:mps-matrices-tensor-product-notation}
    \v_{x} =
    \begin{bmatrix}
        \V_{0} \\
        \V_{1} \\
    \end{bmatrix}, \qquad
    \w_{x} =
    \begin{bmatrix}
        \W_{0} \\
        \W_{1} \\
    \end{bmatrix}.
\end{equation}
Using these vectors of matrices, we can compactly rewrite the stationary state $\p$ using \textit{tensor product notation},
\begin{equation}\label{eq:mps-tensor}
    \p = \frac{1}{Z} \text{Tr} \big[ \v_{1} \w_{2} \cdots \v_{2N - 1} \w_{2N} \big],
\end{equation}
where the subscripts denote which elementary space $\mathbb{R}^{2}$, i.e., which site $x$ of the lattice, of the tensor product the vector is an element of. Formally, Eq.~\eqref{eq:mps-tensor} reads,
\begin{equation}
    \p = \frac{1}{Z} \text{Tr} \big( \v \otimes \w \otimes \cdots \otimes \v \otimes \w \big),
\end{equation}
however, we choose to use explicit notation with the site subscripts for clarity.

In order to exactly construct the MPS from the PSA, we introduce a two-dimensional \textit{auxiliary space} which allows us to define $\V_{n_{x}}$ and $\W_{n_{x}}$ as $2 \times 2$ matrices, whose nonzero components are given by the PSA tensors,
\begin{equation}\label{eq:psa-to-mps}
\begin{aligned}
    [\V_{n_{x}}]_{n_{x}, n_{x + 1}} & \equiv \X_{n_{x}, n_{x + 1}}, \\
    [\W_{n_{x}}]_{n_{x}, n_{x + 1}} & \equiv \Y_{n_{x}, n_{x + 1}},
\end{aligned}
\end{equation}
which gives the following general class of $2 \times 2$ matrices,
\begin{equation}\label{eq:mps-matrix-components}
\begin{aligned}
    \V_{0} & =
    \begin{bmatrix}
        \X_{00} & \X_{01} \\
        0 & 0 \\
    \end{bmatrix}, \\
    \V_{1} & =
    \begin{bmatrix}
        0 & 0 \\
        \X_{10} & \X_{11} \\
    \end{bmatrix}, \qquad
\end{aligned}
\begin{aligned}
    \W_{0} & =
    \begin{bmatrix}
        \Y_{00} & \Y_{01} \\
        0 & 0 \\
    \end{bmatrix}, \\
    \W_{1} & =
    \begin{bmatrix}
        0 & 0 \\
        \Y_{10} & \Y_{11} \\
    \end{bmatrix}.
\end{aligned}
\end{equation}
Note that, by construction, Eq.~\eqref{eq:psa-to-mps} ensures equivalence between the MPS and PSA representations of the ESS,
\begin{equation}\label{eq:mps-psa-equivalence}
    \text{Tr} \big( \V_{n_{1}} \cdots \W_{n_{2N}} \big) \equiv \X_{n_{1}, n_{2}} \cdots \Y_{n_{2N}, n_{1}}.
\end{equation}
Explicitly, the matrices $\V_{n_{x}}$ and $\W_{n_{x}}$ read
\begin{equation}\label{eq:mps-matrices}
\begin{aligned}
    \V_{0} & =
    \begin{bmatrix}
        1 & \xi \\
        0 & 0 \\
    \end{bmatrix}, \\
    \V_{1} & =
    \begin{bmatrix}
        0 & 0 \\
        \xi & 1 \\
    \end{bmatrix},
\end{aligned} \qquad
\begin{aligned}
    \W_{0} & =
    \begin{bmatrix}
        1 & \omega \\
        0 & 0 \\
    \end{bmatrix}, \\
    \W_{1} & =
    \begin{bmatrix}
        0 & 0 \\
        \omega & 1 \\
    \end{bmatrix}.
\end{aligned}
\end{equation}

While the stationarity of the state $\p$ is directly implied by the equivalence between the two representations, the MPS is unique in that it exhibits an algebraic structure that allows us to explicitly demonstrate the stationarity. Namely, the matrices satisfy a \textit{cubic algebraic relation},
\begin{equation}\label{eq:mps-algebraic-relation-tensor}
    \U_{x} \big[ \v_{x - 1} \w_{x} \v_{x + 1} \S \big] = \v_{x - 1} \S \v_{x} \w_{x + 1},
\end{equation}
which compactly encodes the matrix product identities,
\begin{equation}\label{eq:mps-algebraic-relation}
    \V_{n_{x - 1}} \W_{f_{x}} \V_{n_{x + 1}} \S = \V_{n_{x - 1}} \S \V_{n_{x}} \W_{n_{x + 1}},
\end{equation}
obtained by explicitly writing out the physical space vectors in terms of their auxiliary space matrices. Here, we have introduced the \textit{delimiter matrix},
\begin{equation}\label{eq:delimiter-matrix}
    \S = \frac{1}{(1 - s^{-})(1 + s^{+})}
    \begin{bmatrix}
        1 - \xi \omega & \omega - \xi \\
        \omega - \xi & 1 - \xi \omega \\
    \end{bmatrix},
\end{equation}
which is defined by the bulk algebraic relations~\eqref{eq:mps-algebraic-relation}, with the parameters $s^{\pm}$ equal to either of the spectral parameters (i.e., $s^{+} = \xi$ or $\omega$ and $s^{-} = \xi$ or $\omega$). We can easily demonstrate that the inverse of the delimiter matrix $\S$ is given by exchanging the spectral parameters,
\begin{equation}\label{eq:delimiter-matrix-inverse}
    \S^{-1}(\xi, \omega) \equiv \S(\omega, \xi).
\end{equation}
Noticing that the MPS bulk matrices $\V_{n_{x}}$ and $\W_{n_{x}}$ are similarly given by an exchange of parameters,
\begin{equation}\label{eq:mps-parameter-exchange}
    \W_{n_{x}}(\xi, \omega) = \V_{n_{x}}(\omega, \xi),
\end{equation}
immediately implies a \textit{dual-relation},
\begin{equation}\label{eq:mps-algebraic-relation-tensor-primed}
    \U_{x} \big[ \w_{x - 1} \S^{-1} \w_{x} \v_{x + 1} \big] = \w_{x - 1} \v_{x} \w_{x + 1} \S^{-1},
\end{equation}
which explicitly encodes the following identities
\begin{equation}\label{eq:mps-algebraic-relation-primed}
    \W_{n_{x - 1}} \S^{-1} \W_{f_{x}} \V_{n_{x + 1}} = \W_{n_{x - 1}} \V_{n_{x}} \W_{n_{x + 1}} \S^{-1}.
\end{equation}
Before setting $s^{\pm}$, we must consider the cases $\xi \to 1$ and $\omega \to 1$ where the delimiter matrix and its inverse are not well defined. However, we can trivially demonstrate that the matrix products $\V_{n_{x}} \S$ and $\W_{n_{x}} \S^{-1}$ are well defined and finite in the limits $\xi \to 1$ and $\omega \to 1$, respectively, if $s^{-} = \xi$. The following discussion, therefore, holds for all $\xi, \omega \in \mathbb{R}^{+}$, as required~\eqref{eq:probability-conditions}. From here, we are free to set $s^{+} = \xi$ such that the matrix products trivialise,
\begin{equation}\label{eq:matrix-product-simplification}
    \V_{n_{x}} \S = \W_{n_{x}}, \qquad \W_{n_{x}} \S^{-1} = \V_{n_{x}}.
\end{equation}
For the special case where $\xi = \omega = 1$, the states $\p$ and $\pp$ converge to the \textit{maximum entropy state}: the state for which the probabilities of every configuration are equally likely. In this limit, the MPS representation for the ESS simplifies, as detailed in Appendix~\ref{app:maximum-entropy-state}.

Akin to the situation for the PSA, the stationary state $\pp$, corresponding to the odd time step, takes an identical form to the even time step stationary state $\p$, but with the spectral parameters exchanged $\xi \leftrightarrow \omega$ which equates to exchanging the physical space vectors $\v_{x} \leftrightarrow \w_{x}$,
\begin{equation}\label{eq:mps-tensor-primed}
    \pp = \frac{1}{Z} \text{Tr} \big[ \w_{1} \v_{2} \cdots \w_{2N - 1} \v_{2N} \big].
\end{equation}
Explicitly, the components $p^{\prime}_{n}$ of the ESS $\pp$ read
\begin{equation}
    p^{\prime}_{n} = \frac{1}{Z} \text{Tr} \big( \W_{n_{1}} \V_{n_{2}} \cdots \W_{n_{2N - 1}} \V_{n_{2N}} \big).
\end{equation}
The stationarity conditions~\eqref{eq:stationarity-condition} then follow directly from the algebraic relations in Eqs.~\eqref{eq:mps-algebraic-relation-tensor} and \eqref{eq:mps-algebraic-relation-tensor-primed}.

To prove the first of the conditions~\eqref{eq:stationarity-condition}, we insert $\S^{}\S^{-1}$ between the matrices $\V_{n_{1}}$ and $\W_{n_{2}}$ and apply the local time evolution operator $\U_{2N}$ whilst utilising~\eqref{eq:mps-algebraic-relation},
\begin{equation}
\begin{aligned}
    \!\! \M{E} \p{} & = \U_{2} \cdots \U_{2N} \text{Tr} \big[ \v_{1} \S \S^{-1} \w_{2} \cdots \v_{2N - 1} \w_{2N} \big], \\
    & = \U_{2} \cdots \U_{2N - 2}\text{Tr} \big[ \w_{1} \S^{-1} \w_{2} \cdots \v_{2N - 1} \S \v_{2N} \big]. \!\!\!
\end{aligned}
\end{equation}
We then continually apply the local time evolution operators $\U_{x}$, in order, each shifting the delimiter matrix $\S$ two sites to the left, until we are left with the following,
\begin{equation}
\begin{aligned}
    \M{E} \p{} & = \U_{2} \text{Tr} \big[ \w_{1} \S^{-1} \w_{2} \v^{}_{3} \S \v_{4} \cdots \w_{2N - 1} \v_{2N} \big], \\
    & = \text{Tr} \big[ \w_{1} \v_{2} \w_{3} \S^{-1} \S \v_{4} \cdots \w_{2N - 1} \v_{2N} \big],
\end{aligned}
\end{equation}
where, to obtain the second equality, we utilised the dual-relation in Eq.~\eqref{eq:mps-algebraic-relation-tensor-primed}, together with the property that the time evolution operators are involutory (i.e., $\U^{2} = \bm{I}^{\otimes 3}$). Noting that extracting the product $\S^{-1} \S$ yields the ESS $\pp$ proves the stationarity in Eq.~\eqref{eq:stationarity-condition}. The second condition then follows directly from the first by taking advantage of Eqs.~\eqref{eq:delimiter-matrix-inverse} and~\eqref{eq:mps-parameter-exchange}. 

\subsection{Partition function}\label{sec:partition-function}

As demonstrated in Sec.~\ref{sec:patch-state-ansatz}, the components of the stationary states $p_{n}$ are distributed according to a simple grand canonical ensemble,
\begin{equation}\label{eq:quasiparticle-gibbs-state}
    p_{n} = \frac{1}{Z} \exp( -\mu^{+} \q_{n}^{+} - \mu^{-} \q_{n}^{-}),
\end{equation}
where the spectral parameters $\xi$ and $\omega$ are given in terms of the chemical potentials $\mu^{\pm}$ associated to the numbers of quasiparticles $\q_{n}^{\pm}$ in the configuration $\n$~\eqref{eq:chemical-potentials}. It then follows directly from the normalization of the MPS representation of the state $\p$, that the corresponding grand canonical partition function can be written as a sum over the trace of the product of the MPS auxiliary matrices. That is,
\begin{equation}\label{eq:mps-partition-function}
    Z = \sum_{n} \Tr \big( \V_{n_{1}} \W_{n_{2}} \cdots \W_{n_{2N}} \big) \equiv \Tr \big( \T^{N} \big),
\end{equation}
where, to obtain the second expression, we have used the linearity of the trace, and for ease of notation, introduced the \textit{transfer matrix} $\T$, defined as the sum of all products of auxiliary matrices on two adjacent sites,
\begin{equation}\label{eq:transfer-matrix}
    \T = (\V_{0} + \V_{1}) (\W_{0} + \W_{1}) =
    \begin{bmatrix}
        1 + \xi \omega & \omega + \xi \\
        \omega + \xi & 1 + \xi \omega \\
    \end{bmatrix}.
\end{equation}

Similarly, it follows directly from the normalization of Eq.~\eqref{eq:grand-canonical-ensemble} that $Z$ can equivalently be expressed explicitly in terms of a sum over the spectral parameters exponentiated by their respective quasiparticle numbers,
\begin{equation}\label{eq:combinatoric-partition-function}
    Z = \sum_{n} \xi^{\q^{+}_{n}} \omega^{\q^{-}_{n}} \equiv \sum_{\q^{\pm}} 
    \Omega(N, \q^{+}, \q^{-})
 \xi^{\q^{+}} \omega^{\q^{-}},
\end{equation}
where, in the second expression, we have introduced the counting function $ \Omega$ which counts the number of distinct configurations with $\q^{+}$ positive and $\q^{-}$ negative quasiparticles. More precisely, $\Omega$ takes the following combinatoric form,
\begin{equation}\label{eq:partition-function-entropy}
    \Omega(N, \q^{+}, \q^{-}) = 2 \binom{N}{\q^{+}} \binom{N}{\q^{-}}.
\end{equation}
Additionally, we have introduced the shorthand notation for the index of summation $\q^{\pm}$ to denote the set of pairs of numbers of positive and negative quasiparticles that satisfy the constraint~\eqref{eq:quasiparticle-constraint}, imposed by the even system size and PBC, and the following inequalities manifesting from the finite size of the quasiparticles,
\begin{equation}\label{eq:quasiparticle-size-constraint}
    0 \leq \q^{\pm} \leq N,
\end{equation}
which are implicitly given by the following binomial identity, $\binom{n < k}{k} = 0$. To prove that Eq.~\eqref{eq:partition-function-entropy} really counts the total number of configurations of even size $2N$ with $\q^{+}$ positive and $\q^{-}$ negative quasiparticles, it is sufficient to show that the two forms of the grand canonical partition function~\eqref{eq:mps-partition-function} and~\eqref{eq:combinatoric-partition-function} are equivalent. An explicit proof of this equivalence, as well as a qualitative derivation of the counting function from physical arguments, is given in Appendix~\ref{app:partition-function}.

In the thermodynamic limit (i.e., $N \to \infty$), the expression for the grand canonical partition function in Eq.~\eqref{eq:combinatoric-partition-function} can be rewritten in terms of an integral over the \textit{densities} of positive and negative quasiparticles,
\begin{equation}
    n^{\pm} \equiv \lim_{N \to \infty} \frac{\q^{\pm}}{N},
\end{equation}
such that it reads
\begin{equation}
    \mathcal{Z} \equiv \lim_{N \to \infty} Z = \iint_{0}^{1} \text{d} n^{+} \text{d} n^{-} \exp(- N \mathcal{F}),
\end{equation}
where $\mathcal{F}$ can be interpreted as a \textit{free energy density}. More precisely, the free energy density is defined as
\begin{equation}
     \mathcal{F} = \mu^{+} n^{+} + \mu^{-} n^{-} - \mathcal{S},
\end{equation}
where the entropic term $\mathcal{S}$ corresponds to an entropy density, which comes from the counting of degenerate configurations (i.e., states with equivalent numbers of positive and negative quasiparticles) and is obtained by applying the Stirling approximation to Eq.~\eqref{eq:partition-function-entropy}. Explicitly,
\begin{equation}
\begin{aligned}
    \mathcal{S} = - \big( n^{+} \ln n^{+} + (1 - n^{+}) \ln{} (1 - n^{+}) & \\
    & \hspace{-108pt} + n^{-} \ln n^{-} + (1 - n^{-}) \ln{} (1 - n^{-}) \big),
\end{aligned}
\end{equation}
which has the form of an entropy density of mixing of the quasiparticles, subject to the constraints~\eqref{eq:quasiparticle-constraint} and~\eqref{eq:quasiparticle-size-constraint}.

\section{Nonequilibrium Stationary States for stochatic boundary conditions}\label{sec:non-equilibrium-stationary-state}

As demonstrated in Sec.~\ref{sec:equilibrium-stationary-states}, the dynamics of the model with PBC is entirely deterministic and reversible, and is \textit{integrable} (i.e., the system exhibits conserved quantities, possesses an algebraic geometry, and is exactly solvable), which necessarily implies that the system is \textit{nonergodic}. The configuration space is \textit{reducible} under the dynamics and is composed of dynamically disconnected subspaces (i.e., the orbits, or \textit{trajectories}, of the dynamical system). The number of ESS of the periodic system is, therefore, numerous and highly degenerate. To make the dynamics ergodic we impose stochastic boundary conditions (SBC) by considering a chain of finite size coupled to stochastic reservoirs that inject and eject quasiparticles, as was done for Rule 54 (see Refs.~\cite{Prosen2016, Prosen2017, Inoue2018}) and Rule 201 (see Ref.~\cite{Wilkinson2020}). We start by taking the MPS representation of the ESS for a system with PBC and use it to express the probability distribution of a finite subsection of the chain in the large system size, or \textit{thermodynamic}, limit (i.e., $N \to \infty$). We demonstrate that the resulting state can be understood as a \textit{nonequilibrium stationary state} (NESS) of the finite Markov chain with stochastic boundaries that create and destroy the quasiparticles with rates compatible with the chemical potentials $\mu^{\pm}$ of the Gibbs state in Sec~\ref{sec:equilibrium-stationary-states}. We proceed to show that the generator of the dynamics (i.e., the Markov operator) is \textit{irreducible} and \textit{aperiodic}, which implies the uniqueness of the NESS, and the asymptotic approach towards it from any initial state. The dynamics is, therefore, \textit{ergodic} and \textit{mixing}.

\subsection{Asymptotic states}\label{sec:asymptotic-states}

We consider a closed system of even size $2M$ with PBC that is assumed to be in an ESS given by the parameters $\xi$ and $\omega$ as in Sec.~\ref{sec:equilibrium-stationary-states}. The stationary probabilities of a subsection of the chain of even length $2N \leq 2M$ are then given by summing over the probabilities corresponding to the configurations with the same $2N$ sites,
\begin{equation}
    p_{n_{1}, \ldots, n_{2N}}^{(2M)} = \sum_{n_{2N + 1}, \ldots, n_{2M}} \frac{1}{Z} \Tr \big( \V_{n_{1}} \cdots \W_{n_{2M}} \big).
\end{equation}
Utilising the transfer matrix $\T$, defined as the sum of all products of matrices on two adjacent sites [see Eq.~\eqref{eq:transfer-matrix}], the state vectors $\p^{(2M)}$ can be written succinctly as
\begin{equation}
    \p^{(2M)} = \frac{\Tr \big( \v_{1} \w_{2} \cdots \v_{2N - 1} \w_{2N} \T^{M - N} \big)}{\Tr \big( \T^{M} \big)}.
\end{equation}
We then define the state of the subsystem, of fixed even size $2N$, as the large system size limit (i.e., $M \to \infty$) of the probability distribution $\p^{(2M)}$,
\begin{equation}\label{eq:stochastic-probability-state}
    \p \equiv \lim_{M \to \infty} \p^{(2M)} = \frac{\l \v_{1} \w_{2} \cdots \v_{2N - 1} \w_{2N} \r}{\chi^{N} \braket*{l}{r}},
\end{equation}
where $\p$ denotes the \textit{asymptotic probability distribution} of the open subsystem of size $2N$. Here, we have introduced $\chi$ which denotes the leading eigenvalue of $\T$ with $\r$ and $\l$ the corresponding right and left eigenvectors,
\begin{equation}\label{eq:transfer-matrix-eigenvectors}
    \T \r = \chi \r, \qquad \l \T = \chi \l.
\end{equation}
Explicitly, the leading eigenvalue is given by
\begin{equation}
    \chi = (1 + \xi) (1 + \omega),
\end{equation}
while the associated right and left eigenvectors read
\begin{equation}\label{eq:stochastic-eigenvectors}
    \r = r
    \begin{bmatrix}
        1 \\
        1 \\
    \end{bmatrix}, \qquad
    \l = l
    \begin{bmatrix}
        1 & 1 \\
    \end{bmatrix},
\end{equation}
where $r$ and $l$ are scalars determined by the normalization [n.b., the transfer matrix is \textit{symmetric} (i.e., $\T \equiv \T^{\text{T}}$), so the leading right and left eigenvectors are equivalent up to an arbitrary scalar]. Note that the leading eigenvalue is the largest solution of the characteristic polynomial,
\begin{equation}
    \chi^{2} - 2 (1 + \xi \omega) \chi + (1 - \xi^{2})(1 - \omega^{2}) = 0,
\end{equation}
which for $\xi, \omega \in \mathbb{R}^{+}$ is the only real root greater than 1. 

We can similarly define the odd state $\pp$ as the asymptotic form of the primed probability distribution, which takes the same form as $\p$, but with the spectral parameters exchanged (i.e., $\xi \leftrightarrow \omega$). In particular,
\begin{equation}\label{eq:stochastic-probability-state-primed}
    \pp = \frac{\lp \w_{1} \v_{2} \cdots \w_{2N - 1} \v_{2N} \rp}{\chi^{N} \braket*{l^{\prime}}{r^{\prime}}},
\end{equation}
where $\rp$ and $\lp$ are the (leading) right and left eigenvectors of the primed transfer matrix $\T^{\prime}(\xi, \omega) = \T(\omega, \xi)$, respectively, defined as
\begin{equation}\label{eq:boundary-parameter-exchange}
    \ket*{r^{\prime}(\xi, \omega)} = \ket*{r(\omega, \xi)}, \qquad \bra*{l^{\prime}(\xi, \omega)} = \bra*{l(\omega, \xi)}.
\end{equation}
Explicitly,
\begin{equation}\label{eq:stochastic-eigenvectors-primed}
    \rp = r^{\prime}
    \begin{bmatrix}
        1 \\
        1 \\
    \end{bmatrix}, \qquad
    \lp = l^{\prime}
    \begin{bmatrix}
        1 & 1 \\
    \end{bmatrix},
\end{equation}
where $r^{\prime}(\xi, \omega) = r(\omega, \xi)$ and $l^{\prime}(\xi, \omega) = l(\omega, \xi)$. Note that the transfer matrix is invariant under the exchange of the parameters $\xi \leftrightarrow \omega$, namely, $\T \equiv \T^{\prime}$ and, therefore, so are the leading eigenvalue and eigenvectors (similarly, up to an arbitrary scalar). We remark that the expressions for the asymptotic probability distributions, $\p$ and $\pp$, hold for all finite subsections of the periodic chain that start at odd sites, at even and odd times, respectively. For the case where the first site of the subsection is even, we need to exchange the spectral parameters, which, as shown in Sec.~\ref{sec:matrix-product-state}, is equivalent to exchanging the physical space vectors $\v_{x} \leftrightarrow \w_{x}$.

\subsection{Compatible boundaries}\label{sec:compatible-boundaries}

\begin{figure}[t]
    \centering
    \includegraphics{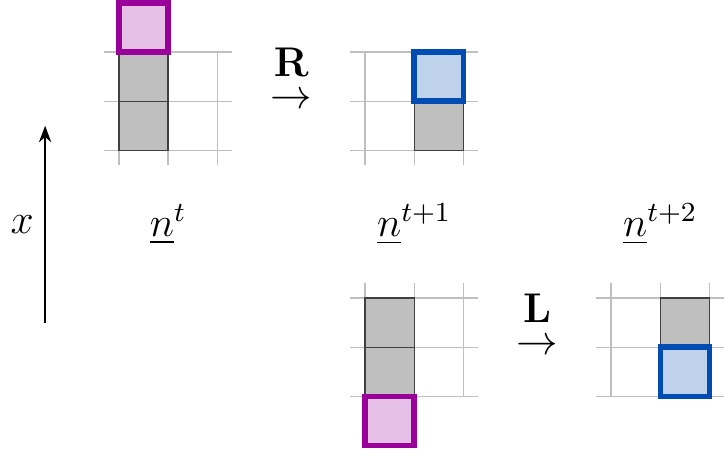}
    \caption{\textbf{Boundary propagators.} The action of $\bm{R}$ and $\bm{L}$ can be understood by appending a \textit{virtual site} to the edge of the lattice, whose state is dependent on the subconfiguration of the pair of adjacent sites, and then deterministically evolving the boundary site according to Eq.~\eqref{eq:rule-150}. Virtual sites are denoted by purple, while updated sites are colored blue.}
    \label{fig:boundary-propagators}
\end{figure}

Alternatively, the asymptotic probability distributions $\p$ and $\pp$ can be understood as the NESS of a boundary driven system whereby time evolution is deterministic in the bulk and stochastic at the boundaries. In particular, during the even time step, the sites $n_{1}, n_{2}, \ldots, n_{2N - 1}$ are updated deterministically by the bulk matrices $\U_{x}$, while the site $n_{2N}$ is updated stochastically by the \textit{right boundary matrix} $\R$,
\begin{equation}\label{eq:stochastic-even-operator}
    \M{E} = \R \prod_{x = 1}^{\mathclap{N - 1}} \U_{2x}.
\end{equation}
Similarly, for the odd time step, the evolution of the sites $n_{2}, n_{3}, \ldots, n_{2N}$ is deterministic, whilst site $n_{1}$ is updated stochastically by the \textit{left boundary matrix} $\L$,
\begin{equation}\label{eq:stochastic-odd-operator}
    \M{O} = \L \prod_{x = 1}^{\mathclap{N - 1}} \U_{2x + 1}.
\end{equation}
To ensure that only sites $n_{1}$ and $n_{2N}$ are updated stochastically by $\L$ and $\R$, the boundary matrices must satisfy the following compatibility conditions,
\begin{equation}\label{eq:boundary-compatibility-condition}
    [\L, \U_{3}] = 0, \qquad [\R, \U_{2N - 2}] = 0,
\end{equation}
which are analogous to the conditions in Eq.~\eqref{eq:bulk-compatibility-condition}. We can interpret the action of the boundary propagators equivalently, by imagining we temporarily append a \textit{virtual} site to the edge of the lattice, in a state that depends on the configuration of the boundary site and its neighbour, and then updating the three sites deterministically according to Eq.~\eqref{eq:rule-150}, as demonstrated in Figure~\ref{fig:boundary-propagators}. Explicitly, the components of the local boundary propagators $\bm{R}$ and $\bm{L}$, which are given by
\begin{equation}\label{eq:ness-boundary-propagator-definitions}
    \R = \bm{I}^{\otimes (2N - 2)} \otimes \bm{R}, \qquad \L = \bm{L} \otimes \bm{I}^{\otimes (2N - 2)},
\end{equation}
can be parametrized as
\begin{equation}\label{eq:boundary-propagator-ansatz}
\begin{aligned}
    \bm{R}_{(m_{3}, m_{4}),(n_{3}, n_{4})} & = \sum_{\mathclap{n_{5} = 0}}^{1} \delta_{m_{3}, n_{3}} \delta_{m_{4}, f_{4}} R_{n_{3}, n_{4}, n_{5}}, \\
    \bm{L}_{(m_{1}, m_{2}),(n_{1}, n_{2})} & = \sum_{\mathclap{n_{0} = 0}}^{1} \delta_{m_{1}, f_{1}} \delta_{m_{2}, n_{2}} L_{n_{0}, n_{1}, n_{2}},
\end{aligned}
\end{equation}
where, to improve readability, we have set $N = 2$ for $\bm{R}$. The boundary matrices, therefore, read
\begin{equation}\label{eq:local-boundary-time-evolution-operators}
\begin{aligned}
    \bm{R} & =
    \begin{bmatrix}
        R_{000} & R_{011} & & \\
        R_{001} & R_{010} & & \\
        & & R_{101} & R_{110} \\
        & & R_{100} & R_{111} \\
    \end{bmatrix}, \\
    \bm{L} & =
    \begin{bmatrix}
        L_{000} & & L_{110} & \\
        & L_{101} & & L_{011} \\
        L_{100} & & L_{010} & \\
        & L_{001} & & L_{111} \\
    \end{bmatrix},
\end{aligned}
\end{equation}
with the scalar quantities $R_{n_{3}, n_{4}, n_{5}}, L_{n_{0}, n_{1}, n_{2}} \in (0, 1)$ the \textit{conditional probabilities} of the virtual sites being $n_{5}$ and $n_{0}$, respectively, given that the sites at the right and left boundaries are $(n_{3}, n_{4})$ and $(n_{1}, n_{2})$. We can equivalently interpret the components of $\bm{R}$ and $\bm{L}$ as the conditional probabilities of either creating or destroying negative and positive quasiparticles at the boundaries, given the state of sites $(n_{3}, n_{4})$ and $(n_{1}, n_{2})$, respectively. For example, $R_{001}$ can be understood to be the conditional probability of creating a negative quasiparticle at the right boundary given that the pair of sites $(n_{3}, n_{4}) = (0, 0)$, while $L_{110}$ is the conditional probability of destroying a negative quasiparticle, or equivalently not creating a positive quasiparticle, at the left boundary given that $(n_{1}, n_{2}) = (1, 0)$.

To ensure that the asymptotic probability distribution vectors $\p$ and $\pp$ are indeed stationary states under the stochastic time evolution, the conditions for stationarity in Eq.~\eqref{eq:stationarity-condition} must hold. Specifically,
\begin{equation}\label{eq:stochastic-stationarity-condition}
    \M{E} \p = \pp, \qquad \M{O} \pp = \p.
\end{equation}
In addition to the bulk algebraic relations~\eqref{eq:mps-algebraic-relation-tensor} and \eqref{eq:mps-algebraic-relation-tensor-primed}, the probability states~\eqref{eq:stochastic-probability-state} and \eqref{eq:stochastic-probability-state-primed} must also satisfy appropriate boundary relations to guarantee that Eq.~\eqref{eq:stochastic-stationarity-condition} is met. In particular, for the even time step, the following boundary relations must hold,
\begin{equation}\label{eq:even-boundary-equations-tensor}
\begin{aligned}
    \l \v_{1} \S & = \frac{1}{\Gamma_{\text{R}}} \lp \w_{1}, \\
    \R \big[ \v_{2N - 1} \w_{2N} \r \big] & = \Gamma_{\text{R}} \v_{2N - 1} \S \v_{2N} \rp,
\end{aligned}
\end{equation}
while for the odd time step, we have
\begin{equation}\label{eq:odd-boundary-equations-tensor}
\begin{aligned}
    \L \big[ \lp \w_{1} \v_{2} \big] & = \Gamma_{\text{L}} \l \v_{1} \w_{2} \S^{-1}, \\
    \w_{2N} \S^{-1} \rp & = \frac{1}{\Gamma_{\text{L}}} \w_{2N} \r,
\end{aligned}
\end{equation}
where the scalar parameters $\Gamma_{\text{R}}$ and $\Gamma_{\text{L}}$ ensure the MPS is normalized and satisfies the fixed point condition~\eqref{eq:stochastic-stationarity-condition}. Immediately, we impose that the right and left boundary matrices must be \textit{left stochastic}, more precisely, each and every column of $\bm{R}$ and $\bm{L}$ must sum to unity, implying
\begin{equation}\label{eq:conditional-probability-normalization}
    \sum_{\mathclap{n_{5} = 0}}^{1} R_{n_{3}, n_{4}, n_{5}} = 
    \sum_{\mathclap{n_{0} = 0}}^{1} L_{n_{0}, n_{1}, n_{2}} = 1,
\end{equation}
which reduces the $4 \times 4$ stochastic matrices $\bm{R}$ and $\bm{L}$ to two nondeterministic $2 \times 2$ blocks of two parameters per boundary propagator.

Substituting the boundary ansatz~\eqref{eq:boundary-propagator-ansatz} into the system of equations for the even time step~\eqref{eq:even-boundary-equations-tensor} yields the following matrix product identities,
\begin{align}
    \l \V_{n_{1}} \S & = \frac{1}{\Gamma_{\text{R}}} \lp \W_{n_{1}}, \label{eq:even-left-boundary-equation} \\
    \sum_{\mathclap{n_{5} = 0}}^{1} R_{n_{3}, f_{4}, n_{5}} \V_{n_{3}} \W_{f_{4}} \r & = \Gamma_{\text{R}} \V_{n_{3}} \S \V_{n_{4}} \rp, \label{eq:even-right-boundary-equation}
\end{align}
where, for readability, we have again set $N = 2$. Solving separately these equations, whilst taking into account the normalization~\eqref{eq:conditional-probability-normalization}, returns the following expressions for the components of the right boundary propagator,
\begin{equation}\label{eq:right-boundary-probabilities}
\begin{aligned}
    R_{001} & = \xi \theta_{0}, \qquad
    & R_{011} & = \frac{\xi - \omega}{\xi (1 + \omega)} + \theta_{0}, \\
    R_{110} & = \xi \theta_{1},
    & R_{100} & = \frac{\xi - \omega}{\xi (1 + \omega)} + \theta_{1},
\end{aligned}
\end{equation}
where $\theta_{0}$ and $\theta_{1}$ are the free parameters corresponding to the two nondeterministic blocks of $\R$, with the boundary vector normalization given by
\begin{equation}\label{eq:right-boundary-parameters}
    \frac{r}{r^{\prime}} = \Gamma_{\text{R}}, \qquad \frac{l}{l^{\prime}} = \frac{1}{\Gamma_{\text{R}}}.
\end{equation}
Similarly, substituting the ansatz~\eqref{eq:boundary-propagator-ansatz} into the equations for the odd time step~\eqref{eq:odd-boundary-equations-tensor} gives the following identities,
\begin{align}
    \sum_{\mathclap{n_{0} = 0}}^{1} L_{n_{0}, f_{1}, n_{2}} \lp \W_{f_{1}} \V_{n_{2}} & = \Gamma_{\text{L}} \l \V_{n_{1}} \W_{n_{2}} \S^{-1}, \label{eq:odd-left-boundary-equation} \\
    \W_{n_{2N}} \S^{-1} \rp & = \frac{1}{\Gamma_{\text{L}}} \W_{n_{2N}} \r, \label{eq:odd-right-boundary-equation}
\end{align}
which, after solving, return the following expressions for the left boundary propagator components,
\begin{equation}\label{eq:left-boundary-probabilities}
\begin{aligned}
    L_{100} & = \omega \vartheta_{0}, \qquad
    & L_{110} & = \frac{\omega - \xi}{\omega (1 + \xi)} + \vartheta_{0}, \\
    L_{011} & = \omega \vartheta_{1},
    & L_{001} & = \frac{\omega - \xi}{\omega (1 + \xi)} + \vartheta_{1},
\end{aligned} 
\end{equation}
where $\vartheta_{0}$ and $\vartheta_{1}$ are the corresponding left boundary free parameters, with the normalization reading
\begin{equation}\label{eq:left-boundary-parameters}
    \frac{r}{r^{\prime}} = \Gamma_{\text{L}} \frac{1 + \xi}{1 + \omega}, \qquad \frac{l}{l^{\prime}} = \frac{1}{\Gamma_{\text{L}}} \frac{1 + \omega}{1 + \xi}.
\end{equation}
Equating the expressions for the boundary parameters in Eq.~\eqref{eq:right-boundary-parameters} and~\eqref{eq:left-boundary-parameters} then necessarily implies that
\begin{equation}\label{eq:ness-normalization-solution}
    \frac{\Gamma_{\text{R}}}{\Gamma_{\text{L}}} = \frac{1 + \xi}{1 + \omega}.
\end{equation}
At this point, we are free to choose specific values for the normalization parameters that satisfy Eq.~\eqref{eq:boundary-parameter-exchange} and set
\begin{equation}
    \Gamma_{\text{R}} = 1, \qquad \Gamma_{\text{L}} = \frac{1 + \omega}{1 + \xi},
\end{equation}
such that the right and left boundary vectors read,
\begin{equation}\label{eq:ness-boundary-vector-solutions}
    \r \equiv \rp =
    \begin{bmatrix}
        1 \\
        1 \\
    \end{bmatrix}, \qquad
    \l \equiv \lp =
    \begin{bmatrix}
        1 & 1 \\
    \end{bmatrix}.
\end{equation}
The solutions in Eqs.~\eqref{eq:right-boundary-probabilities} and~\eqref{eq:left-boundary-probabilities} constitute the most general form for the boundary propagators $\R$ and $\L$, where the asymptotic probability distributions $\p$ and $\pp$ in Eqs.~\eqref{eq:stochastic-probability-state} and~\eqref{eq:stochastic-probability-state-primed} are exactly the fixed points. Notice, however, that the stochastic parameters $\theta_{0}$, $\theta_{1}$, $\vartheta_{0}$, $\vartheta_{1}$ are not completely arbitrary as the elements of the boundary matrices must be appropriately bounded and the spectral parameters must be strictly nonnegative and equal at the right and left boundary. A particularly convenient choice for the parametrization is achieved by setting
\begin{equation}\label{eq:equilibrium-free-parameter-choice}
    \theta_{0} \equiv \theta_{1} = \frac{\omega}{\xi(1 + \omega)}, \qquad \vartheta_{0} \equiv \vartheta_{1} = \frac{\xi}{\omega (1 + \xi)},
\end{equation}
as it facilitates the following summary for the conditional probabilities at the boundaries,
\begin{equation}\label{eq:equilibrium-boundary-conditional-probability-parametrization}
\begin{aligned}
    R_{n_{3}, n_{4}, n_{5}} & = \frac{p_{n_{3}, n_{4}, n_{5}, 0} + p_{n_{3}, n_{4}, n_{5}, 1}}{p_{n_{3}, n_{4}}}, \\
    L_{n_{0}, n_{1}, n_{2}} & = \frac{p_{0, n_{0}, n_{1}, n_{2}}^{\prime} + p_{1, n_{0}, n_{1}, n_{2}}^{\prime}}{p_{n_{1}, n_{2}}^{\prime}}, \\
\end{aligned}
\end{equation}
which is comparable to the identities obtained for Rule 54 (see, e.g., Refs.~\cite{Klobas2020a, Buca2021}) and, similarly, for Rule 201 (see Ref.~\cite{Wilkinson2020}). Explicitly, the probability of finding the virtual sites at the right and left boundaries in the states $n_{5}$ and $n_{0}$, respectively, given that the pairs of adjacent spins are in the configurations $(n_{3}, n_{4})$ and $(n_{1}, n_{2})$, that is $R_{n_{3}, n_{4}, n_{5}}$ and $L_{n_{0}, n_{1}, n_{2}}$ is equivalent to the conditional probability of finding the three sites in the configurations $(n_{3}, n_{4}, n_{5})$ and $(n_{0}, n_{1}, n_{2})$, given the states of the sites $(n_{3}, n_{4})$ and $(n_{1}, n_{2})$. The asymptotic distributions $\p$ and $\pp$ can then equally be interpreted as the nonequilibrium stationary states of a boundary driven system.

While the solutions in Eqs.~\eqref{eq:right-boundary-probabilities} and~\eqref{eq:left-boundary-probabilities} are general, they are not completely arbitrary. By this, we mean that the parameters $\theta_{0}, \theta_{1}, \vartheta_{0}, \vartheta_{1}$ cannot take arbitrary values, in particular, for given values of the spectral parameters $\xi, \omega \in \mathbb{R}^{+}$, the parameters $\theta_{0}, \theta_{1}, \vartheta_{0}, \vartheta_{1}$ must take values such that the conditional probabilities are appropriately bounded, namely, $R_{n_{3}, n_{4}, n_{5}}, L_{n_{0}, n_{1}, n_{2}} \in (0, 1)$. Requiring this puts additional constraints on the boundary matrices $\bm{R}$ and $\bm{L}$. Explicitly, it demands that the matrix elements $R_{n_{3}, n_{4}, n_{5}}$ and $L_{n_{0}, n_{1}, n_{2}}$ obey the particle-hole symmetry of the model (see Sec~\ref{sec:model} for details and Appendix~\ref{app:boundary-matrices-constraint} for a proof),
\begin{equation}\label{eq:conditional-probability-symmetry}
\begin{aligned}
    R_{n_{3}, n_{4}, n_{5}} & = R_{1 - n_{3}, 1 - n_{4}, 1 - n_{5}}, \\
    L_{n_{0}, n_{1}, n_{2}} & = L_{1 - n_{0}, 1 - n_{1}, 1 - n_{2}},
\end{aligned}
\end{equation}
which immediately implies the equivalence of the free parameters of the right and left boundaries,
\begin{equation}
    \theta_{0} = \theta_{1}, \qquad \vartheta_{0} = \vartheta_{1}.
\end{equation}
In order to guarantee the consistency of the solutions in Eqs.~\eqref{eq:right-boundary-probabilities} and~\eqref{eq:left-boundary-probabilities} (i.e., the equivalence of the spectral parameters $\xi$ and $\omega$ at the right and left boundaries), we eliminate the free parameters $\theta_{0}$ and $\vartheta_{0}$ by equating the expressions at the right and left boundaries, respectively, and subsequently solve for the spectral parameters which yields the following unique nontrivial solution,
\begin{equation}\label{eq:ness-spectral-parameters}
\begin{aligned}
    \xi & = \frac{R_{001} (1 - L_{110}) + (1 - R_{001}) L_{100}}{R_{011} (1 - L_{100}) + (1 - R_{011}) L_{110}}, \\
    \omega & = \frac{L_{100} (1 - R_{011}) + (1 - L_{100}) R_{001}}{L_{110} (1 - R_{001}) + (1 - L_{110}) R_{011}},
\end{aligned}
\end{equation}
which can be easily verified to be appropriately bounded, that is, $\xi, \omega \in \mathbb{R}^{+}$ for any $R_{n_{3}, n_{4}, n_{5}}, L_{n_{0}, n_{1}, n_{2}} \in (0, 1)$, as required. Remarkably, this solution is equivalent to that obtained by the parametrization introduced in Eq.~\eqref{eq:equilibrium-free-parameter-choice}. This can be proven straightforwardly by substituting the conditional probabilities~\eqref{eq:equilibrium-boundary-conditional-probability-parametrization} directly into the solutions for the spectral parameters~\eqref{eq:ness-spectral-parameters}.

\subsection{Statistical independence}\label{sec:statistical-independence}

The asymptotic probability distributions~\eqref{eq:stochastic-probability-state} and~\eqref{eq:stochastic-probability-state-primed} admit a remarkable factorization property similar to that of Rule 54~\cite{Klobas2020a}. In particular, the conditional probability of observing site $2N$ in the state $n_{2N}$, given the previous $2N - 1$ sites $(n_{1}, \ldots, n_{2N - 1})$, depends only on the state of the last two sites $(n_{2N - 1}, n_{2N})$. Explicitly, 
\begin{equation}\label{eq:right-conditional-probability-relations}
\begin{aligned}
    \frac{p_{n_{1}, \ldots, n_{2N}}}{p_{n_{1}, \ldots, n_{2N - 1}}} & = \frac{p_{n_{2N - 1}, n_{2N}}}{p_{n_{2N - 1}}}, \\
    \frac{p_{n_{1}, \ldots, n_{2N}}^{\prime}}{p_{n_{1}, \ldots, n_{2N - 1}}^{\prime}} & = \frac{p_{n_{2N - 1}, n_{2N}}^{\prime}}{p_{n_{2N - 1}}^{\prime}}. \\
\end{aligned}
\end{equation}
Analogously, the conditional probability of finding site $1$ in the state $n_{1}$, given the next $2N - 1$ sites $(n_{2}, \ldots, n_{2N})$, depends only on sites $(n_{1}, n_{2})$. Namely,
\begin{equation}\label{eq:left-conditional-probability-relations}
\begin{aligned}
    \frac{p_{n_{1}, \ldots, n_{2N}}}{p_{n_{2}, \ldots, n_{2N}}} & = \frac{p_{n_{1}, n_{2}}}{p_{n_{2}}}, \\
    \frac{p_{n_{1}, \ldots, n_{2N}}^{\prime}}{p_{n_{2}, \ldots, n_{2N}}^{\prime}} & = \frac{p_{n_{1}, n_{2}}^{\prime}}{p_{n_{2}}^{\prime}}.
\end{aligned}
\end{equation}
An explicit proof of these equalities, which follow directly from the definitions of the MPS matrices $\V_{n_{x}}$ and $\W_{n_{x}}$, as well as formal definitions of the asymptotic conditional probabilities are presented in Appendix~\ref{app:conditional-probabilities}.

An important consequence of this factorization of the asymptotic conditional probabilities~\eqref{eq:stochastic-probability-state} and~\eqref{eq:stochastic-probability-state-primed} is the \textit{statistical independence of quasiparticles}. Namely, in the stationary state, the probability of observing a quasiparticle at any given site of the lattice is the same at every site, independent of the positions of other quasiparticles. Let the conditional probability of encountering a positive or negative quasiparticle at any given pair of sites, given the state of either site, be denoted by $p^{+}$ and $p^{-}$, respectively. Then, in terms of the asymptotic probabilities, we can express these now well-defined quantities as
\begin{align}
    p^{+} & = \frac{p_{10}}{p_{0}} = \frac{p_{01}}{p_{1}} = \frac{p_{01}}{p_{0}} = \frac{p_{10}}{p_{1}} = \frac{\xi}{1 + \xi}, \label{eq:positive-quasiparticle-conditional-probabilities} \\
    p^{-} & = \frac{p_{01}^{\prime}}{p_{0}^{\prime}} = \frac{p_{10}^{\prime}}{p_{1}^{\prime}} = \frac{p_{10}^{\prime}}{p_{0}^{\prime}} = \frac{p_{01}^{\prime}}{p_{1}^{\prime}} = \frac{\omega}{1 + \omega}, \label{eq:negative-quasiparticle-conditional-probabilities}
\end{align}
which are identical to the expressions in Eq.~\eqref{eq:equilibrium-boundary-conditional-probability-parametrization} for the conditional probabilities of encountering quasiparticles at the left and right boundaries, respectively. In particular, let us denote the conditional probability of introducing a positive quasiparticle at the left boundary given the state of site $n_{0}$ by $L^{+}$, specifically,
\begin{equation}
    L^{+} = L_{100} = 1 - L_{110},
\end{equation}
and that of a negative quasiparticle at the right boundary given $n_{2N + 1}$ by $R^{-}$, that is,
\begin{equation}
    R^{-} = R_{001} = 1 - R_{011}.
\end{equation}
It then follows directly from~\eqref{eq:equilibrium-boundary-conditional-probability-parametrization} that
\begin{equation}
    L^{+} = \frac{\xi}{1 + \xi} = p^{+}, \qquad R^{-} = \frac{\omega}{1 + \omega} = p^{-}.
\end{equation}
Note that the conditional probabilities $p^{+}$ and $p^{-}$ provide an equivalent parametrization for the stationary states as their relation to the spectral parameters can be inverted. Explicitly,
\begin{equation}
    \xi = \frac{p^{+}}{1 - p^{+}}, \qquad \omega = \frac{p^{-}}{1 - p^{-}}.
\end{equation}
In addition, $p^{+}$ and $p^{-}$ exhibit a notable thermodynamic property, which is obtained by substituting the relations for the spectral parameters in Eq.~\eqref{eq:chemical-potentials}, in terms of their associated chemical potentials, into Eqs.~\eqref{eq:positive-quasiparticle-conditional-probabilities} and~\eqref{eq:negative-quasiparticle-conditional-probabilities}. Doing so yields
\begin{equation}
    p^{\pm} = \frac{1}{\exp(\mu^{\pm}) + 1},
\end{equation}
which can be immediately identified as being exactly the \textit{Fermi-Dirac distributions} of the quasiparticles.

\subsection{Irreducibility and aperiodicity}\label{sec:irreducibility-and-aperiodicity}

To prove that the NESS~\eqref{eq:stochastic-probability-state} is unique and asymptotically approached from any initial state requires we show that the Markov operator $\M{}$ is \textit{irreducible} and \textit{aperiodic} (cf. Theorem 1 in Ref.~\cite{Prosen2016}). As per the Perron-Frobenius theorem~\cite{Serfozo2009}, this amounts to demonstrating that, firstly, for any two basis states $\bm{e}_{n}$ and $\bm{e}_{m}$ (i.e., configurations $\n$ and $\underline{m}$) there exists a nonnegative integer $T$ such that
\begin{equation}
    \bm{e}_{m} \cdot \M{}^{T} \bm{e}_{n} > 0,
\end{equation}
and, secondly, for the case where $\bm{e}_{m} \equiv \bm{e}_{n}$ that the greatest common divisor of the set of $T$ is unity.

To prove the irreducibility, we recall that the dynamics in the bulk is deterministic. Therefore, every positive and negative quasiparticle in the system propagates towards the right and left boundary, respectively. In contrast, the boundary dynamics is stochastic and so we are effectively free to choose the values of the sites $n_{1}$ and $n_{2N}$ for every state between $\bm{e}_{n}$ and $\bm{e}_{m}$. Now, consider the sequence of configurational states,
\begin{equation}
    \bm{e}_{n}^{0} \to \bm{e}_{n}^{1} \to \cdots \to \bm{e}_{n}^{2 T - 1} \to \bm{e}_{n}^{2 T},
    \label{seq}
\end{equation}
connected by the Markov operator
$\bm{e}_{n}^{t + 1} \cdot \M{} \bm{e}_{n}^{t} > 0$ where $\bm{e}_{n}^{0} \equiv \bm{e}_{n}$ and $\bm{e}_{n}^{2T} \equiv \bm{e}_{m}$, and with $T$ counting the number of \textit{full time steps} between states $\bm{e}_{n}$ and $\bm{e}_{m}$. For the first part of the sequence, we argue that we can set the values of the virtual sites $n_{0}$ and $n_{2N + 1}$ so that they \textit{eject} each and every quasiparticle from the initial state $\bm{e}_{n}$. Indeed, by recalling that the quasiparticles propagate ballistically with velocities of $v^{\pm} = \pm 1$ and interact trivially without scattering (i.e., are noninteracting), then after an integer number of full time steps $t_{+} \leq 2N$ we are guaranteed to be in the vacuum state [i.e., either the state $(0, \ldots, 0)$ or $(1, \ldots, 1)$], irrespective of the initial state $\bm{e}_{n}$. We do so with the following rules for the virtual sites,
\begin{equation}
    n_{0}^{t} = n_{1}^{t - 1}, \qquad n_{2N + 1}^{t} = n_{2N}^{t - 1}.
\end{equation}
For the second part of the sequence in Eq.~\eqref{seq}, we need to show that we can set the values of the virtual sites such that the quasiparticles are \textit{injected}, so that after an integer number of full time steps $t_{-} \leq 2N$ we obtain the state $\bm{e}_{m}$. To achieve this we exploit the \textit{time reversibility} of the bulk dynamics and site freedom of the boundaries to get to the vacuum state from the final state $\bm{e}_{m}$, but with time evolution inverted. In particular, we apply the following rules,
\begin{equation}
    n_{0}^{t} = n_{1}^{t + 1}, \qquad n_{2N + 1}^{t} = n_{2N}^{t + 1}.
\end{equation}
Due to the nonnegativity of the Markov matrix elements, and the sequence that connects the initial and final states $\bm{e}_{n}$ and $\bm{e}_{m}$ in $T = t_{+} + t_{-} \leq 4N$ full time steps, we have that $\bm{e}_{m} \cdot \M{}^{T} \bm{e}_{n}$ is nonvanishing for any arbitrary states $\bm{e}_{n}$ and $\bm{e}_{m}$, thus, proving the irreducibility of $\M{}$.

\begin{figure*}[t]
    \centering
    \includegraphics{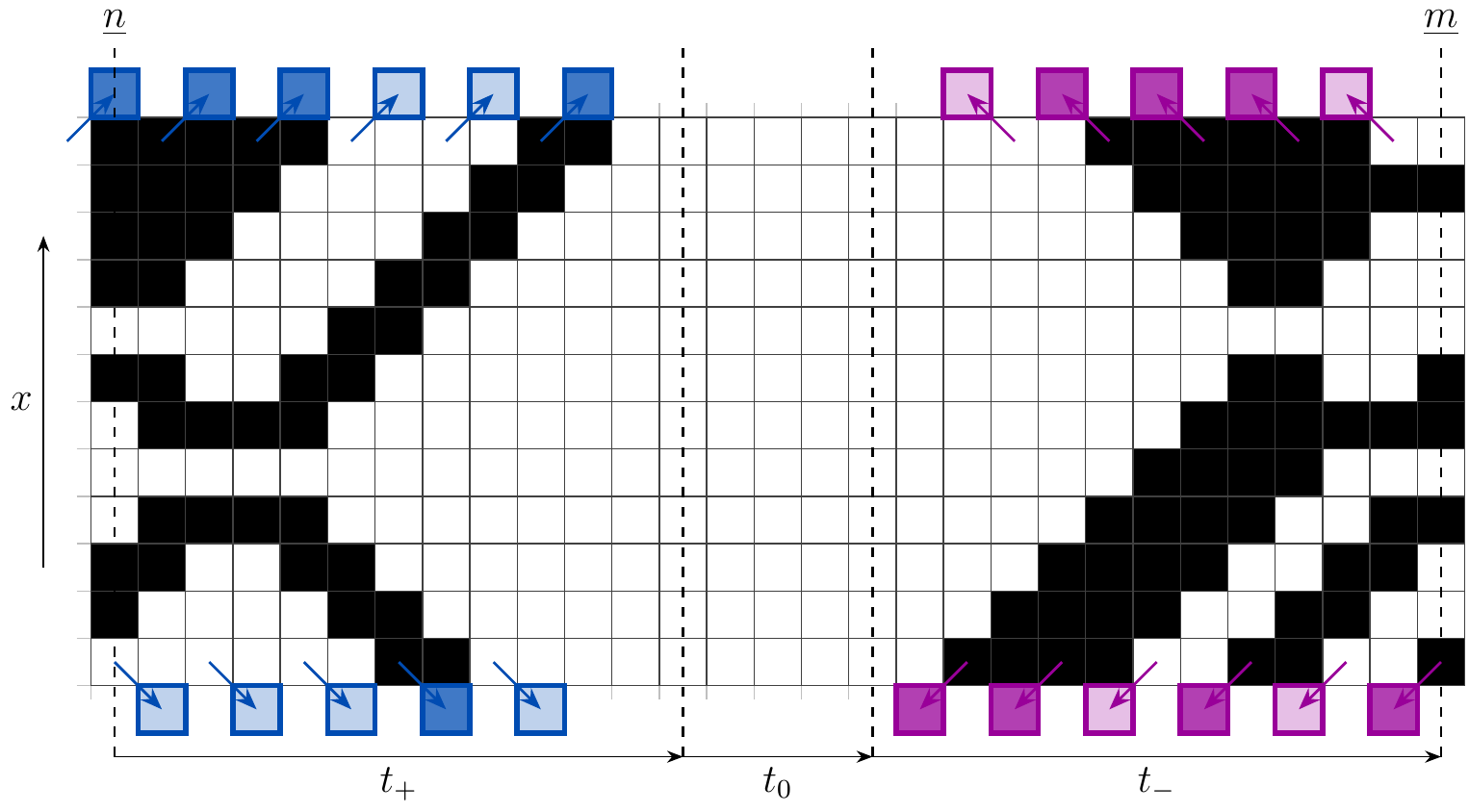}
    \caption{\textbf{Irreducibility and aperiodicity.} Illustrative explanation of the idea of the proof of irreducibility and aperiodicity of the Markov operator $\M{}$. Each and every configuration is connected via a walk of at least $T = t_{+} + t_{0} + t_{-}$ time steps where $t_{+}$, $t_{0}$, and $t_{-}$ denote the number of full time steps taken to reach the vacuum from $\bm{e}_{n} \equiv \bm{e}_{n}^{0}$, reach $\bm{e}_{m} \equiv \bm{e}_{n}^{2T}$ from the vacuum, and waited in the vacuum, respectively. In the example given, the states $\bm{e}_{n}^{0}$ and $\bm{e}_{n}^{2T}$ are connected in \textit{at least} $T = 12$ full time steps. At the start of the walk (i.e., $t = 0, \ldots, t_{+}$) the virtual sites, shown in blue, are set to causally \textit{destroy} quasiparticles by taking the values of \textit{past} boundary sites, specifically, with the rule $\smash{n_{0}^{t} = n_{1}^{t - 1}}$ and $\smash{n_{2N + 1}^{t} = n_{2N}^{t - 1}}$. In contrast, at the end of the walk (i.e., $t = t_{+} + t_{0}, \ldots, t_{+} + t_{0} + t_{-})$ the virtual sites, colored purple, are chosen to causally \textit{create} quasiparticles by taking the values of \textit{future} sites, that is, $\smash{n_{0}^{t} = n_{1}^{t + 1}}$ and $\smash{n_{2N + 1}^{t} = n_{2N}^{t + 1}}$. Consequently, there always exists an integer $T = t_{+} + t_{-}$ such that $\bm{e}_{m}\cdot\M{}^{T} \bm{e}_{n} > 0$ for any arbitrary states $\bm{e}_{n}$ and $\bm{e}_{m}$, hence, the Markov operator $\M{}$ is irreducible. Finally, we consider the middle of the walk (i.e., $t = t_{+}, \ldots, t_{+} + t_{0}$) where the virtual sites take the values of \textit{present} boundary sites, explicitly, $\smash{n_{0}^{t} = n_{1}^{t}}$ and $\smash{n_{2N + 1}^{t} = n_{2N}^{t}}$. This necessarily implies that the system can remain in either vacuum state for any integer number of full time steps $t_{0} \in \mathbb{N}$ which guarantees that $\gcd(\{t_{+} + t_{0} + t_{-}\}) = 1$ and, therefore, proves that the Markov operator $\M{}$ is aperiodic.}
    \label{fig:irreducibility}
\end{figure*}

To show the aperiodicity, we recall that we are free to remain in either vacuum state for an indefinite number of full time steps $t_{0}$. Consequently, $T$ can take any integer value in the closed interval $[t_{+} + t_{-}, t_{+} + t_{0} + t_{-}]$ which implies that the greatest common divisor of the set of $T$ has to be unity. That is,
\begin{equation}
    \gcd(\{t_{+} + t_{0} + t_{-}\}) = 1,
\end{equation}
for $t_{0} \in \mathbb{N}$. This, therefore, proves the aperiodicity of $\M{}$. For an illustrative explanation of the proof see Figure~\ref{fig:irreducibility}.

\section{Spectrum and relaxation dynamics}\label{sec:decay-modes}

\begin{figure*}[t]
    \centering
    \includegraphics{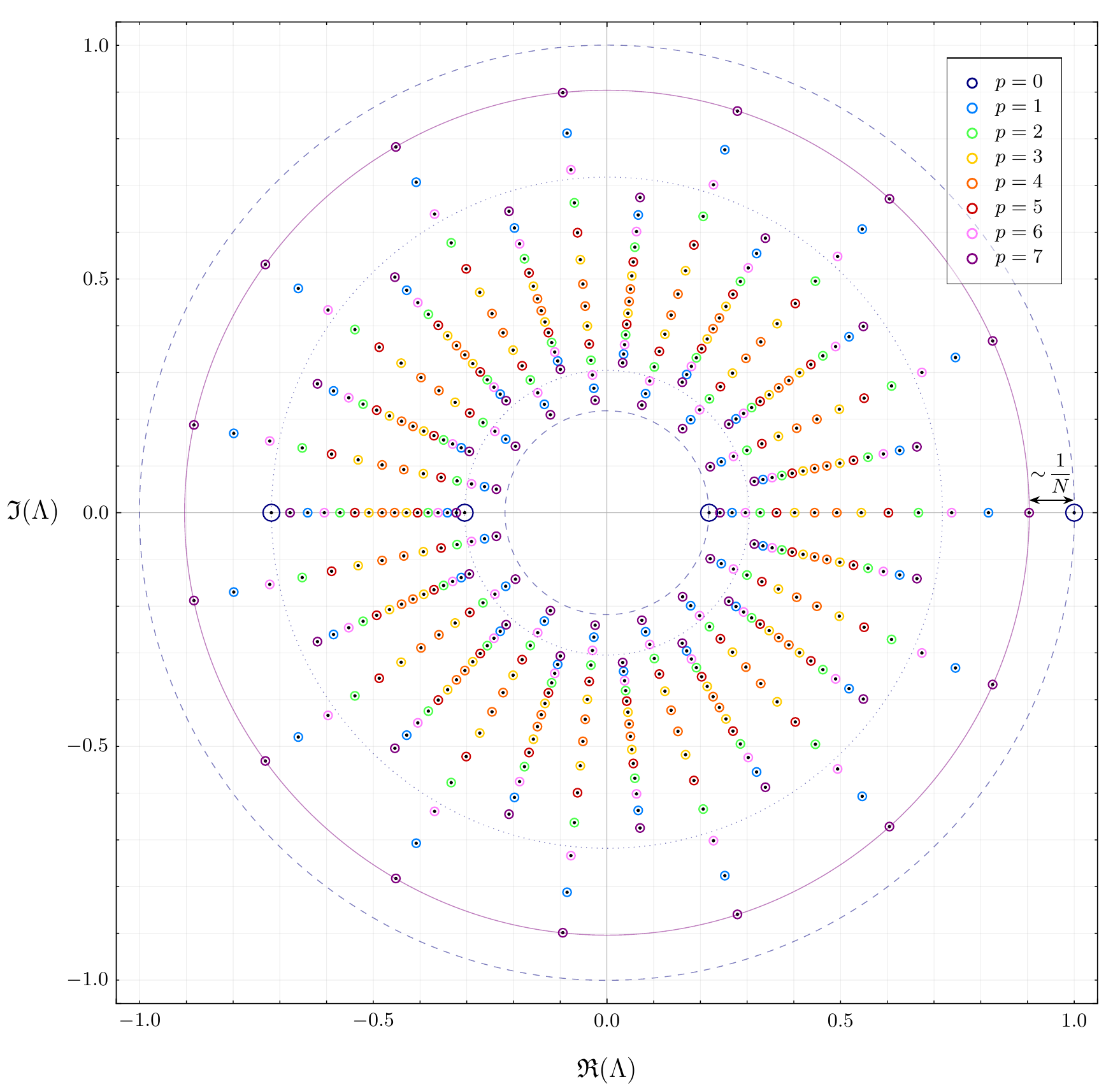}
    \caption{\textbf{Orbitals.} Spectrum of the Markov operator $\M{}$ for a system of even size $2N = 16$ with $\alpha = 3/5$, $\beta = 7/8$, $\gamma = 8/9$, and $\delta = 4/7$. The black dots mark the numerical solutions computed by exact diagonalization. The colored circles (see legend) then denote the analytic results for the orbital $p$ eigenvalues $\Lambda$ obtained from the conjectured expressions~\eqref{eq:eigenvalue-solutions}. The dark blue circles represent the roots $\lambda$ of the quadratic characteristic polynomials~\eqref{eq:decay-eigenvalues}, which are precisely the eigenvalues of the zeroth orbital. The dashed blue curves denote the circles with radii $\smash{r = 1}$ and $\smash{r = \abs*{\mu}}$ which, together with the dotted blue curves of radii $\smash{r = \abs*{\eta \pm \sqrt{\eta^{2} - \mu}}}$, bound the sets of eigenvalues generated by the momentum parameter $z$~\eqref{eq:momentum-solutions}. In the thermodynamic limit $N \to \infty$, the purple curve, which corresponds to the algebraic curve bounding the eigenvalues of the orbital associated to the leading decay modes, converges to the unit circle, where the spectral gap that characterizes the relaxation dynamics of the system scales with $1/N$.}
    \label{fig:orbitals}
\end{figure*}

In Sec.~\ref{sec:non-equilibrium-stationary-state}, we demonstrated that the deterministic and reversible dynamics with PBC could be made ergodic by considering a finite subsection of the chain in the infinite size limit, which effectively imposed SBC. The resulting state could then be understood as a NESS. In this section we generalise the results above to study the \textit{full relaxation dynamics} of the model, that is, to resolve the spectrum of the Markov operator $\M{}$. As was observed in Refs.~\cite{Prosen2016, Prosen2017}, we find that the spectrum is composed of \textit{orbitals}, that is, subsets of the set of eigenvalues which are roots of simple polynomial factors of the characteristic polynomial of the Markov operator $\M{}$, as illustrated in Figure~\ref{fig:orbitals}. We show that the eigenvectors of the simplest orbital, that we refer to as the \textit{zeroth orbital}, which contains the NESS derived in Sec.~\ref{sec:non-equilibrium-stationary-state} and a triplet of decay modes whose associated eigenvalues are size invariant, can be expressed explicitly in terms of an MPS similar to that of the stationary states $\p$ and $\pp$ in Sec.~\ref{sec:asymptotic-states}. We then propose a conjecture for the Bethe-like equations for the entire spectrum (i.e., the distinct eigenvalues and the corresponding degeneracies), which follows directly as a consequence of the consistency conditions imposed, and that generalizes the expressions for the NESS. In addition, we study the thermodynamic limit and demonstrate that the leading decay modes, that is, the eigenvectors of the Markov operator $\M{}$ associated to the eigenvalues with the largest real parts not equal to unity, that characterize the spectral gap and determine the relaxation rate of the system in the asymptotic limit, scale with $1/N$.

\subsection{Markov operator}\label{sec:eigenvalue-equation} 

We are interested in obtaining exact analytic solutions to the eigenvalue equation for the Markov operator,
\begin{equation}\label{eq:markov-operator-eigenvalue-equation}
    \M{} \p = \Lambda \p,
\end{equation}
which we can conveniently separate into a pair of coupled linear equations for the even and odd time steps,
\begin{equation}\label{eq:even-odd-operator-eigenvalue-equation}
    \M{E} \p = \Lambda_{\text{R}} \pp, \qquad \M{O} \pp = \Lambda_{\text{L}} \p,
\end{equation}
with the eigenvalue of the Markov operator $\M{}$ factorizing as $\Lambda = \Lambda_{\text{L}} \Lambda_{\text{R}}$. Here, the stochastic matrices $\M{E}$ and $\M{O}$ are defined as in~\eqref{eq:stochastic-even-operator} and~\eqref{eq:stochastic-odd-operator}, respectively, however, for simplicity and without loss of generality, we assume that the constraints imposed on the boundary matrices $\bm{R}$ and $\bm{L}$, by the normalization~\eqref{eq:conditional-probability-normalization} and symmetry~\eqref{eq:conditional-probability-symmetry}, apply implicitly, such that we have
\begin{equation}
\begin{aligned}
    \bm{R} & =
    \begin{bmatrix}
        1 - \alpha & \beta & & \\
        \alpha & 1 - \beta & & \\
        & & 1 - \beta & \alpha \\
        & & \beta & 1 - \alpha \\
    \end{bmatrix}, \\
    \bm{L} & =
    \begin{bmatrix}
        1 - \gamma & & \delta & \\
        & 1 - \delta & & \gamma \\
        \gamma & & 1 - \delta & \\
        & \delta & & 1 - \gamma \\
    \end{bmatrix},
\end{aligned}
\end{equation}
where $\alpha, \beta, \gamma, \delta \in (0, 1)$ are boundary driving parameters (i.e., conditional probabilities) that determine the rate at which the quasiparticles are either created or destroyed. For example, $\alpha$ and $\delta$ respectively denote the conditional probability that a negative quasiparticle is injected at the right boundary and ejected at the left boundary.

It can be straightforwardly demonstrated that solving the eigenvalue equation~\eqref{eq:markov-operator-eigenvalue-equation} provides access to the full relaxation dynamics of the model, as the probability for a given state $\p^{t}$ at time $t$ can be written explicitly in terms of the eigenvalues $\Lambda_{j}$ and corresponding eigenvectors $\p_{j}$. In particular, we can write
\begin{equation}
    \p^{t} = \sum_{j = 0}^{\mathclap{2^{2N} - 1}} c_{j} \Lambda_{j}^{t} \p_{j},
\end{equation}
where $c_{j}$ are coefficients that depend on the initial state. In Sec.~\ref{sec:irreducibility-and-aperiodicity}, we proved that the Markov operator $\M{}$ is \textit{irreducible} and \textit{aperiodic} for arbitrary nontrivial driving parameters $0 < \alpha, \beta, \gamma, \delta < 1$. The Perron-Frobenius theorem~\cite{Serfozo2009}, therefore, guarantees that the unique eigenvector $\p_{0}$, associated to the eigenvalue $\Lambda_{0} = 1$, namely, the NESS, does not decay in time while the eigenvectors $\p_{j}$ for $j > 0$ exponentially decay as their associated eigenvalues are bounded within the unit circle by $\abs*{\Lambda_{j}} < 1$. We refer to these eigenvectors as \textit{decay modes} as they encode the time evolution of any initial state towards the NESS in the asymptotic limit.

\subsection{Decay modes}\label{sec:zeroth-orbital}

We begin by presenting the ansatz for the eigenvectors of the zeroth orbital of the Markov operator $\M{}$, in terms of a simple staggered MPS which reads
\begin{align}
    \p & = \bra*{L} \v_{1} \w_{2} \cdots \v_{2N - 1} \w_{2N} \ket*{R}, \label{eq:decay-eigenvector-mps} \\
    \pp & = \bra*{L^{\prime}} \w_{1} \v_{2} \cdots \w_{2N - 1} \v_{2N} \ket*{R^{\prime}}, \label{eq:decay-eigenvector-mps-primed}
\end{align}
where $\v_{x}$ and $\w_{x}$ are the vectors of matrices~\eqref{eq:mps-matrices-tensor-product-notation} that we showed satisfy the bulk algebraic relations~\eqref{eq:mps-algebraic-relation-tensor} and~\eqref{eq:mps-algebraic-relation-tensor-primed}. To ensure that the states $\p$ and $\pp$ in Eqs.~\eqref{eq:decay-eigenvector-mps} and~\eqref{eq:decay-eigenvector-mps-primed} satisfy the coupled eigenvalue equations in Eq.~\eqref{eq:even-odd-operator-eigenvalue-equation}, we additionally require that the following modified boundary algebraic relations hold for the row vectors $\bra*{L}$ and $\bra*{L^{\prime}}$, and column vectors $\ket*{R}$ and $\ket*{R^{\prime}}$,
\begin{align}
    \bra*{L} \v_{1} \S & = \bra*{L^{\prime}} \w_{1}, \label{eq:decay-left-even-boundary-relation} \\
    \R \big[ \v_{2N - 1} \w_{2N} \ket*{R} \big] & = \Lambda_{\text{R}} \v_{2N - 1} \S \v_{2N} \ket*{R^{\prime}}, \label{eq:decay-right-even-boundary-relation} \\
    \L \big[ \bra*{L^{\prime}} \w_{1} \v_{2} \big] & = \Lambda_{\text{L}} \bra*{L} \v_{1} \w_{2} \S^{-1}, \label{eq:decay-left-odd-boundary-relation} \\
    \w_{2N} \S^{-1} \ket*{R^{\prime}} & = \w_{2N} \ket*{R}, \label{eq:decay-right-odd-boundary-relation}
\end{align}
where $\S$ is the delimiter matrix~\eqref{eq:delimiter-matrix} and $\S^{-1}$ its inverse. We can readily verify that these algebraic relations solve the staggered eigenvalue equations~\eqref{eq:even-odd-operator-eigenvalue-equation} by substituting the ansatz into either of the equations and applying the appropriate relations to transform $\p \leftrightarrow \pp$. In particular, to obtain $\p$ from $\pp$, we first write out $\M{O} \pp$ in terms of the matrix product~\eqref{eq:stochastic-odd-operator} and ansatz~\eqref{eq:decay-eigenvector-mps-primed}. Applying the operator $\L$ and utilising the boundary relation~\eqref{eq:decay-left-odd-boundary-relation}, we introduce the delimiter matrix inverse $\S^{-1}$ on the left, as well as the parameter $\Lambda_{\text{L}}$. We then repeatedly apply $\U_{x}$ to the odd sites of the chain (i.e., sites $n_{3}, n_{5}, \ldots, n_{2N - 1}$) using the bulk relation~\eqref{eq:mps-algebraic-relation-tensor-primed}, which shifts $\S^{-1}$ to the right, two sites at a time. Finally, we eliminate $\S^{-1}$ with~\eqref{eq:decay-right-odd-boundary-relation} to yield $\Lambda_{\text{L}} \p$. The other condition for the even time step then follows analogously. As an example, we consider the transformation $\p \to \pp$ for $N = 3$,
\begin{equation}
\begin{aligned}
    \M{E} \p & = \U_{2} \U_{4} \bm{R}_{6} \bra*{L} \v_{1} \w_{2} \v_{3} \w_{4} \v_{5} \w_{6} \ket*{R} \\
    & = \Lambda_{\text{R}} \U_{2} \U_{4} \bra*{L} \v_{1} \w_{2} \v_{3} \w_{4} \v_{5} \S \v_{6} \ket*{R^{\prime}} \\
    & = \Lambda_{\text{R}} \U_{2} \bra*{L} \v_{1} \w_{2} \v_{3} \S \v_{4} \w_{5} \v_{6} \ket*{R^{\prime}} \\
    & = \Lambda_{\text{R}} \bra*{L} \v_{1} \S \v_{2} \w_{3} \v_{4} \w_{5} \v_{6} \ket*{R^{\prime}} \\
    & = \Lambda_{\text{R}} \bra*{L^{\prime}} \w_{1} \v_{2} \w_{3} \v_{4} \w_{5} \v_{6} \ket*{R^{\prime}} \\
    & = \Lambda_{\text{R}} \pp.
\end{aligned}
\end{equation}

Solving separately the equations for the right boundary \eqref{eq:decay-right-even-boundary-relation} and~\eqref{eq:decay-right-odd-boundary-relation}, we obtain the following pair of solutions, identical up to a sign, for the spectral parameters,
\begin{align}
    \xi & = \sigma \frac{\Lambda_{\text{R}} - (1 - \alpha)}{\beta}, \label{eq:decay-right-positive-spectral-parameter} \\
    \omega & = \sigma \frac{\Lambda_{\text{R}} (1 - \beta) - (1 - \alpha - \beta)}{\Lambda_{\text{R}} \beta}, \label{eq:decay-right-negative-spectral-parameter}
\end{align}
with $\sigma = \pm 1$ and associated right boundary vectors,
\begin{equation}\label{eq:decay-right-boundary-vectors}
    \ket*{R} = R
    \begin{bmatrix}
        1 \\
        \sigma \\
    \end{bmatrix}, \qquad
    \ket*{R^{\prime}} = R \frac{1 + \sigma \omega}{1 + \sigma \xi}
    \begin{bmatrix}
        1 \\
        \sigma \\
    \end{bmatrix},
\end{equation}
where $R$ is a scalar that determines the normalization of the solutions of the right boundary. Similarly solving the left boundary equations~\eqref{eq:decay-left-even-boundary-relation} and~\eqref{eq:decay-left-odd-boundary-relation} then returns an equivalent pair of solutions for the spectral parameters,
\begin{align}
    \xi & = \tau \frac{\Lambda_{\text{L}} (1 - \delta) - (1 - \gamma - \delta)}{\Lambda_{\text{L}} \delta}, \label{eq:decay-left-positive-spectral-parameter} \\
    \omega & = \tau \frac{\Lambda_{\text{L}} - (1 - \gamma)}{\delta}, \label{eq:decay-left-negative-spectral-parameter}
\end{align}
with $\tau = \pm 1$ and left boundary vectors,
\begin{equation}\label{eq:decay-left-boundary-vectors}
    \bra*{L} = L
    \begin{bmatrix}
        1 & \tau \\
    \end{bmatrix}, \qquad
    \bra*{L^{\prime}} = L
    \begin{bmatrix}
        1 & \tau \\
    \end{bmatrix},
\end{equation}
with $L$ the corresponding scalar determining the normalization of the left boundary solutions. In order to obtain solutions that are consistent with the results in Secs.~\ref{sec:equilibrium-stationary-states} and~\ref{sec:non-equilibrium-stationary-state}, we choose to set
\begin{equation}
    R = \frac{1}{1 + \sigma \omega}, \qquad L = 1,
\end{equation}
such that the components $p_{n}$ of the eigenvectors $\p$ of the Markov operator $\M{}$ take a form reminiscent of the grand canonical ensemble~\eqref{eq:grand-canonical-ensemble}. Specifically,
\begin{equation}
    p_{n} = \tau^{n_{1}} \xi^{\q^{+}_{n}} \omega^{\q^{-}_{n}},
\end{equation}
where $\tau$ corresponds to the choice of solutions for the left boundary equations~\eqref{eq:decay-left-positive-spectral-parameter},~\eqref{eq:decay-left-negative-spectral-parameter}, and~\eqref{eq:decay-left-boundary-vectors}.

To guarantee that the solutions at the boundaries that were obtained independently of each other are consistent necessarily requires that we demand that the expressions for the spectral parameters $\xi$ in Eqs.~\eqref{eq:decay-right-positive-spectral-parameter} and~\eqref{eq:decay-left-positive-spectral-parameter} and $\omega$ in Eqs.~\eqref{eq:decay-right-negative-spectral-parameter} and~\eqref{eq:decay-left-negative-spectral-parameter} are, respectively, equal. Notice, however, that the signs of the solutions at the right and left boundaries are independent and, therefore, pairwise equating all possible combinations of expressions for the spectral parameters $\xi$ and $\omega$ returns a doubly degenerate closed pair of equations for the eigenvalue parameters $\Lambda_{\text{R}}$ and $\Lambda_{\text{L}}$ that we interpret as \textit{Bethe equations}, imposed by the consistency conditions at the boundaries. Explicitly,
\begin{align}
    \frac{\Lambda_{\text{R}} - (1 - \alpha)}{\beta} & = \tau \frac{\Lambda_{\text{L}} (1 - \delta) - (1 - \gamma - \delta)}{\Lambda_{\text{L}} \delta}, \label{eq:decay-positive-consistency-equation} \\
    \frac{\Lambda_{\text{L}} - (1 - \gamma)}{\delta} & = \tau \frac{\Lambda_{\text{R}} (1 - \beta) - (1 - \alpha - \beta)}{\Lambda_{\text{R}} \beta}, \label{eq:decay-negative-consistency-equation}
\end{align}
where, for simplicity, we have taken the positive solutions at the right boundary. Eliminating either $\Lambda_{\text{R}}$ or $\Lambda_{\text{L}}$ using the eigenvalue $\Lambda = \Lambda_{\text{L}} \Lambda_{\text{R}}$ and subsequently solving yields the following quadratic characteristic polynomial,
\begin{equation}\label{eq:decay-characteristic-polynomial}
    \Lambda^{2} - (1 + \mu - \nu + \tau \nu) \Lambda + \mu = 0,
\end{equation}
where, for readability, we have introduced the coefficients $\mu \in (-1, 1)$ and $\nu \in (0, 2)$, which are defined by
\begin{equation}
    \mu = (1 - \alpha - \beta)(1 - \gamma - \delta), \qquad
    \nu = \alpha \delta + \beta \gamma.
\end{equation}
It follows straightforwardly that as Eq.~\eqref{eq:decay-characteristic-polynomial} is a \textit{pair} of quadratic equations it has, in general, four distinct roots that can be written succinctly as
\begin{equation}\label{eq:decay-eigenvalues}
    \Lambda = 1, \mu, \eta \pm \sqrt{\eta^{2} - \mu},
\end{equation}
where the coefficient $\eta \in (-1, 1)$ is given by
\begin{equation}
    \eta = \frac{1 + \mu - 2 \nu}{2}.
\end{equation}
Clearly, $\Lambda \equiv \Lambda_{0} = 1$ is always guaranteed to be a solution with the corresponding eigenvector being the NESS. The remaining solutions $\Lambda \equiv \Lambda_{j}$ for $j \neq 0$ then correspond to three decay modes whose eigenvalues are independent of the system size, that is, they are \textit{size invariant}. We refer to this set of four eigenvalues as the \textit{zeroth orbital}.

\subsection{Quasiparticle excitations}\label{sec:complete-spectrum} 

Despite the fact that we are unable to find an explicit MPS expression for eigenvectors of the Markov operator $\M{}$ beyond the zeroth orbital, exact numerical diagonalization for small systems suggest that the remaining eigenvalues also organize into orbitals, see  Figure~\ref{fig:orbitals}. This is simlar to what occurs in Rule 54~\cite{Prosen2017}, with the number of orbitals scaling linearly with the size of the system and the degeneracy of the eigenvalues increasing exponentially with the orbital level. 

Using these observations, together with similar conjectures as in Ref.~\cite{Prosen2017}, we are able to construct exact analytic forms for the Bethe equations [cf. Eqs.~\eqref{eq:decay-positive-consistency-equation} and~\eqref{eq:decay-negative-consistency-equation}] that completely reproduce the entire spectrum of the Markov operator $\M{}$. To start, we introduce some additional parameters required for the conjecture, specifically, the nonnegative integer $p$ that counts the orbital level, $z \in \mathbb{C}$ which we interpret as the momentum associated to quasiparticle excitations of the NESS, which in turn is intuitively understood as the vacuum state of the Markovian dynamics, and $A_{\pm} \in \mathbb{C}$, a pair of complex amplitudes associated to the operators that create the aforementioned quasiparticle excitations. 

Having introduced the necessary prerequisites, we now postulate the following generalized expressions for $\xi$ and $\omega$ at the right boundary [cf. Eqs.~\eqref{eq:decay-right-positive-spectral-parameter} and~\eqref{eq:decay-right-negative-spectral-parameter}],
\begin{align}
    \xi & = \sigma \frac{\Lambda_{\text{R}} - z (1 - \alpha)}{z \beta}, \label{eq:general-decay-right-positive-spectral-parameter} \\
    \omega & = \sigma \frac{\Lambda_{\text{R}} (1 - \beta) - z (1 - \alpha - \beta)}{\Lambda_{\text{R}} \beta}, \label{eq:general-decay-right-negative-spectral-parameter}
\end{align}
while at the left boundary [cf. Eqs.~\eqref{eq:decay-left-positive-spectral-parameter} and~\eqref{eq:decay-left-negative-spectral-parameter}],
\begin{align}
    \xi & = \tau \frac{\Lambda_{\text{L}} (1 - \delta) - z (1 - \gamma - \delta)}{\Lambda_{\text{L}} \delta}, \label{eq:general-decay-left-positive-spectral-parameter} \\
    \omega & = \tau \frac{\Lambda_{\text{L}} - z (1 - \gamma)}{z \delta} \label{eq:general-decay-left-negative-spectral-parameter},
\end{align}
where $\sigma, \tau = \pm 1$. In addition, we require that the pair of amplitude parameters $A_{\pm}$ satisfy the following identities at the right and left boundary, respectively,
\begin{align}
    \frac{A_{+}}{A_{-}} & = \frac{\Lambda_{\text{R}}^{2 p} z^{2N - 1}}{z^{2 p} (1 - \alpha - \beta)^{p}}, \label{eq:general-decay-right-complex-amplitude-ratio} \\
    \frac{A_{-}}{A_{+}} & = \frac{\Lambda_{\text{L}}^{2 p}}{z^{2 p} (1 - \gamma - \delta)^{p}}. \label{eq:general-decay-left-complex-amplitude-ratio}
\end{align}
Imposing the consistency condition, i.e., demanding that the expressions for the spectral parameters $\xi$ and $\omega$, and amplitude parameters $A_{+}$ and $A_{-}$ are pairwise equivalent then returns the following closed set of generalized Bethe equations for $\Lambda_{\text{R}}$, $\Lambda_{\text{L}}$, and $z$,
\begin{align}
    \frac{\Lambda_{\text{R}} - z (1 - \alpha)}{z \beta} & = \tau \frac{\Lambda_{\text{L}} (1 - \delta) - z (1 - \gamma - \delta)}{\Lambda_{\text{L}} \delta}, \\
    \frac{\Lambda_{\text{L}} - z (1 - \gamma)}{z \delta} & = \tau \frac{\Lambda_{\text{R}} (1 - \beta) - z (1 - \alpha - \beta)}{\Lambda_{\text{R}} \beta}, \\
    \frac{\Lambda_{\text{R}}^{2 p} z^{2N - 1}}{z^{2 p} (1 - \alpha - \beta)^{p}} & = \frac{z^{2 p} (1 - \gamma - \delta)^{p}}{\Lambda_{\text{L}}^{2 p}},
\end{align}
where, as for the zeroth orbital, the $\pm$ signs are obtained by equating expressions for the spectral parameters with positive signs for the right boundary with both solutions of the left boundary. Replacing either $\Lambda_{\text{R}}$ or $\Lambda_{\text{L}}$ with the eigenvalue $\Lambda = \Lambda_{\text{L}} \Lambda_{\text{R}}$ and eventually solving transforms the set of equations into a pair of identities for $\Lambda$ and $z$. The first, which reads
\begin{equation}\label{eq:nonequilibrium-dispersion-relation}
    \Lambda^{2} - (1 + \mu - \nu + \tau \nu) \Lambda z^{2} + \mu z^{4} = 0, 
\end{equation}
can be interpreted as a \textit{nonequilibrium dispersion relation} that connects the eigenvalues and momentum parameter and can be straightforwardly shown to be a direct generalization of the quadratic characteristic polynomial in Eq.~\eqref{eq:decay-characteristic-polynomial}, for which $z = 1$. The second identity,
\begin{equation}\label{eq:momentum-conservation-relation}
    \Lambda^{2 p} z^{2N - 4p - 1} - \mu^{p} = 0,
\end{equation}
can, instead, be understood as a \textit{momentum conservation relation}. Indeed, remarking that the solution to Eq.~\eqref{eq:nonequilibrium-dispersion-relation} can be compactly written as
\begin{equation}\label{eq:decay-modes-eigenvalue-simplification}
    \Lambda = \lambda z^{2},
\end{equation}
where we have introduced the parameter $\lambda$, which can be straightforwardly demonstrated to be equivalent to the $\Lambda$ in Eq.~\eqref{eq:decay-eigenvalues} (i.e., the eigenvalues of the zeroth orbital),
\begin{equation}\label{eq:decay-modes-eigenvalue-parameter}
    \lambda = 1, \mu, \eta \pm \sqrt{\eta^{2} - \mu},
\end{equation}
then allows us to rewrite Eq.~\eqref{eq:momentum-conservation-relation} as
\begin{equation}
    z^{2N - 1} = \frac{\mu^{p}}{\lambda^{2 p}}.
\end{equation}
Therefore, for a given orbital $p$ and arbitrary parameters $\alpha, \beta, \gamma, \delta$, the magnitude of the momentum $z$ is conserved. Specifically, the solutions to the momentum conservation relation are the $2N - 1$ distinct roots, that read
\begin{equation}\label{eq:momentum-solutions}
    z = \exp\!\Bigg(\!\frac{p \ln \rho}{2N - 1} + i \bigg(\frac{p \phi + 2 \pi p q}{2N - 1}\bigg)\!\!\Bigg),
\end{equation}
where we have introduced the polar parameters,
\begin{equation}
    \rho = \abs{\frac{\mu}{\lambda^{2}}}, \qquad \phi = \mathrm{Arg}\left(\frac{\mu}{\lambda^{2}}\right),
\end{equation}
with $q = 0, \ldots, 2N - 2$. The eigenvalues then read
\begin{equation}\label{eq:eigenvalue-solutions}
    \Lambda = \exp\!\Bigg(\!\ln \varrho + \frac{2 p \ln \rho}{2N - 1} + i \bigg(\varphi + \frac{2 p \phi + 2 \pi p q}{2N - 1}\bigg)\!\!\Bigg),
\end{equation}
where, additionally, we have defined,
\begin{equation}
    \varrho = \abs{\lambda}, \qquad \varphi = \mathrm{Arg}(\lambda),
\end{equation}
with the orbital number $p = 0, \ldots, N - 1$. We conjecture that the multivalued function~\eqref{eq:eigenvalue-solutions} completely describes the entire spectrum of $\M{}$. Indeed, comparing the results calculated analytically with numerical values obtained by exact diagonalization of the Markov matrix $\M{}$ for $N \leq 8$ we see perfect agreement as demonstrated in Figure~\ref{fig:orbitals}. In contrast to the typical Bethe ansatz~\cite{Bethe1931}, this conjecture implies that the entire spectrum is characterized by just one \textit{universal} momentum parameter $z$, irrespective of the number of quasiparticle excitations (cf. Rule 54~\cite{Friedman2019, Buca2021}). This can be seen as following directly from the dispersion relation, in that each and every quasiparticle propagates with constant (group) velocity $v^{\pm} = \pm 1$ (i.e., each species is \textit{nondispersive}). 

Additionally, we present a conjecture for the associated degeneracy $g$ of the eigenvalue $\Lambda$. Explicitly,
\begin{equation}\label{eq:decay-eigenvalue-degeneracy}
    g = \sum_{\mathclap{d | D}} \frac{d}{2N - 1} \sum_{\mathclap{d^{\prime} | D^{\prime}}} \mu(d^{\prime}) \binom{\frac{2N - 1}{d d^{\prime}}}{\frac{p}{d d^{\prime}}},
\end{equation}
where $\mu(\ \cdot\ )$ denotes the M\"{o}bius function~\cite{Hardy2008} and $j | k$ the set of positive integer divisors $j$ of the integer $k$, with
\begin{equation}
    D = \gcd(2N - 1, p, q), \qquad D^{\prime} = \frac{\gcd(2N - 1, p)}{q},
\end{equation}
where $\gcd(\ \cdot\ )$ denotes the greatest common divisor. This conjecture can be confirmed numerically for small system sizes (see Appendix~\ref{app:eigenvalue-degeneracy} for details).

\subsection{Thermodynamic limit}

In the thermodynamic limit $N \to \infty$, the series expansion of the momentum conservation relation~\eqref{eq:momentum-conservation-relation}, in the small parameter $1/N$, to leading order reads
\begin{equation}
    z \simeq 1 + \frac{1}{N} z^{\prime}, \qquad z^{\prime} = \frac{p \ln \rho + i (p \phi + 2 \pi p q)}{2},
\end{equation}
which immediately implies that, in the asymptotic limit, the momentum parameter $z$ is given by
\begin{equation}\label{eq:decay-modes-unimodular-momentum}
    z(\kappa, \epsilon) = \exp(\epsilon \ln\rho + i \kappa),
\end{equation}
where we have introduced the \textit{momentum} $\kappa \in [0, 2\pi)$ and \textit{decay} $\epsilon \in [0, \tfrac{1}{2})$, defined by
\begin{equation}
     \kappa = \lim_{N \to \infty} \frac{p \phi + 2 \pi p q}{2N - 1}, \qquad \epsilon = \lim_{N \to \infty} \frac{p}{2N - 1}.
\end{equation}
A direct consequence of this is that, in the limit $N \to \infty$, the eigenvalues of each and every orbital $p$ converge to a set of algebraic curves, specifically, circles $\Lambda(\kappa, \epsilon)$, that are given by inserting 
Eq.~\eqref{eq:decay-modes-unimodular-momentum} into Eq.~\eqref{eq:decay-modes-eigenvalue-simplification}. Explicitly,
\begin{equation}
    \Lambda(\kappa, \epsilon) = \exp(\ln\varrho + 2 \epsilon \ln\rho + i \kappa).
\end{equation}
Writing the series expansion of the eigenvalue $\Lambda$ as
\begin{equation}
    \Lambda \simeq \varrho\bigg(1 - \frac{1}{N} \Lambda^{\prime}\bigg),
\end{equation}
and substituting into the nonequilibrium dispersion relation~\eqref{eq:nonequilibrium-dispersion-relation}, we obtain
\begin{equation}
    \Lambda^{\prime} = - 2 z^{\prime},
\end{equation}
which is consistent with the interpretation of the dynamics in terms of the ballistic propagation of quasiparticles.

In the long time limit, the asymptotic relaxation rate of the system is determined by the leading decay mode, defined as the eigenvector $\p_{1}$ of the Markov operator $\M{}$, associated to the eigenvalue $\Lambda_{1}$ satisfying
\begin{equation}\label{eq:decay-mode-leading-decay-mode-condition}
    \abs*{\mathfrak{R}(\Lambda_{1})} = \max_{j > 0}\!\big(\abs*{\mathfrak{R}(\Lambda_{j})}\!\big),
\end{equation}
that is, the eigenvalue with the largest real part not equal to unity. In contrast to Rule 54 (see, e.g., Refs.~\cite{Prosen2016, Prosen2017, Buca2021}), the leading decay modes, that determine the spectral gap of the Markov operator $\M{}$, are associated to eigenvalues with orbital number $p = N - 1$, as opposed to $p = 1$. To prove this, we begin by rewriting the condition~\eqref{eq:decay-mode-leading-decay-mode-condition} as
\begin{widetext}
\begin{equation}
    \abs*{\mathfrak{R}(\Lambda_{1})} = \max_{\lambda, p}\!\Bigg(\abs{\exp\!\bigg(\!\frac{(2N - 2p - 1) \ln\abs*{\lambda} + 2p \ln\abs{\frac{\mu}{\lambda}}}{2N - 1}
    \!\bigg) \cos\!\bigg(\!\frac{(2N - 2p - 1) \mathrm{Arg}(\lambda) + 2p \mathrm{Arg}\left(\frac{\mu}{\lambda}\right) + 2 \pi p q}{2N - 1}\bigg)}\Bigg),
\end{equation}
\end{widetext}
for $\lambda \in \{1, \mu, \eta \pm \sqrt{\eta^{2} - \mu}\}$ and $p \in \{1, \ldots, N - 1\}$ where, to obtain the equality, we have used the properties of the logarithm, absolute value, and principle argument. From here, we remark that $\abs*{\lambda} \leq 1$ and $p \leq N - 1$, which imply that the first term of the exponential is nonpositive, and that $\abs*{\lambda} \geq \abs*{\mu}$ and $p \geq 1$, which similarly imply that the second term is nonpositive. Together, with the constraint that $\abs*{\mu} < 1$, these observations ensure that the exponent is strictly negative and must, therefore, be minimized to maximize the exponential. Similarly, the cosine function is maximized by minimizing the modulus of its argument, which, since $\mu \in \mathbb{R}$, $\lambda \in \mathbb{C}$, and $q \in \{0, \ldots, 2N - 2\}$, can be achieved by setting $\lambda \in \mathbb{R}$ (i.e., $\lambda \in \{1, \mu\}$) and $q = 0$. For the case with $\lambda = 1$, it follows straightforwardly that $\abs*{\mathfrak{R}(\Lambda_{j})}$ is maximized by choosing $p = 1$, while for $\lambda = \mu$, it is maximized by selecting $p = N - 1$. Comparing both cases, and recalling that $\abs*{\mu} < 1$, we immediately realize that the leading decay modes are guaranteed to be in the orbital $p = N - 1$, with
\begin{equation}
    \Lambda_{1} = \exp\!\bigg(\frac{1}{2N - 1} \ln\abs*{\mu}\!\bigg),
\end{equation}
for all $\alpha, \beta, \gamma, \delta \in (0, 1)$ and $N \in \mathbb{N}^{+}$.

Whilst, naively, one would expect that the boundaries would become irrelevant in the thermodynamic limit and, therefore, force each and every eigenvalue to collapse onto the unit circle, as was the case for the closed system with periodic boundaries, this does not happen here. Instead, we observe that the eigenvalues distribute themselves over an infinite set of circles $\Lambda(\kappa, \epsilon)$, that are parametrized radially by $\epsilon$ and angularly by $\kappa$,
\begin{equation}
    \lim_{N \to \infty} \abs*{\Lambda} = \abs*{\mu}^{2 \epsilon} \abs*{\lambda}^{1 - 4 \epsilon}, \qquad \lim_{N \to \infty} \mathrm{Arg}(\Lambda) = \kappa,
\end{equation}
for $\epsilon \in [0, \tfrac{1}{2})$ and $\kappa \in [0, 2 \pi)$, which then implies that the thermodynamic $N \to \infty$ and long time $t \to \infty$ limits are \textit{distinct} (i.e, the stationary state $\p \equiv \p_{0}$ is the only state in the asymptotic time limit for any even system size $2N$, but with the time taken to reach it increasing with $N$).

\subsection{Observables and correlations}

We now consider computing observables in the NESS. To do so, we define the partition function for the open system out of equilibrium as we did for the closed system with periodic boundaries, namely, via normalization of the MPS probabilities,
\begin{equation}
    Z = \sum_{n} p_{n} = \sum_{n} \bra*{L} \V_{n_{1}} \cdots \W_{n_{2N}} \ket*{R} = \bra*{L} \T^{N} \ket*{R},
\end{equation}
which, using the transfer matrix eigenvalue equation, can be simplified to
\begin{equation}
    Z = \chi^{N} \braket*{L}{R} = 2 (1 + \xi)^{N} (1 + \omega)^{N - 1}.
\end{equation}

The average density function for the NESS is 
\begin{equation}
    \langle n_{x} \rangle = \frac{1}{Z} \sum_{n} n_{x} p_{n_{1}, \ldots, n_{2N}}.
\end{equation}
A direct computation shows that we can rewrite this as
\begin{equation}
\begin{aligned}
    \langle n_{2x} \rangle & = \frac{1}{Z} \bra*{L} \T^{x - 1} \V \bm{D}_{2x} \T^{N - x} \ket*{R}, \\
    \langle n_{2x - 1} \rangle & = \frac{1}{Z} \bra*{L} \T^{x - 1} \bm{D}_{2x - 1} \W \T^{N - x} \ket*{R},
\end{aligned}
\end{equation}
where we have introduced the site density operator,
\begin{equation}
    \bm{D}_{x} =
    \begin{cases}
        \W_{1}, & x = 0 \pmod{2}, \\
        \V_{1}, & x = 1 \pmod{2}, \\
    \end{cases}
\end{equation}
with the shorthand notations,
\begin{equation}
    \V \equiv \V_{0} + \V_{1}, \qquad \W \equiv \W_{0} + \W_{1}.
\end{equation}
Using the eigenvalue equation for the transfer matrix we get,
\begin{equation}
\begin{aligned}
    \langle n_{2x} \rangle & = \frac{\bra*{L} \V \bm{D}_{2x} \ket*{R}}{\chi \braket*{L}{R}} = \frac{1}{2}, \\
    \langle n_{2x - 1} \rangle & = \frac{\bra*{L} \bm{D}_{2x - 1} \W \ket*{R}}{\chi \braket*{L}{R}} = \frac{1}{2}.
\end{aligned}
\end{equation}

We can similarly calculate multi-point correlation functions for arbitrary products $\smash{n_{2x_{1} - y_{1}}, \ldots, n_{2x_{k} - y_{k}}}$, with $x_{j} = 1, \ldots, N$ and $y_{j} = 0, 1$, for $j = 1, \ldots, k$ where $x_{0} \equiv 0$ and $x_j \geq x_{j-1}$. Assuming $x_{j}>x_{j-1}$, we write
\begin{widetext}
\begin{equation}
    \bigg\langle\! \prod_{j} n_{2x_{j} - y_{j}} \!\bigg\rangle = \frac{1}{Z} \sum_{n} \bigg(\! \prod_{j} n_{2x_{j} - y_{j}} \!\bigg) p_{n_{1}, \ldots, n_{2N}} = \frac{1}{Z} \bra*{L} \prod_{j}\bigg(\!  \T^{x_{j} - x_{j - 1} - 1} \V^{1 - y_{j}} \bm{D}_{2x_{j} - y_{j}} \W^{y_{j}} \!\bigg) \T^{N - x_{k}} \ket*{R}.
\end{equation}
\end{widetext}
Specifically, the two-point correlator, for example for sites $2x_{1} - 1$ and $2x_{2}$, reads
\begin{equation}
    \langle n_{2x_{1} - 1} n_{2x_{2}} \rangle = \frac{\bra*{L} \bm{D}_{2x_{1} - 1} \T^{x_{2} - x_{1}} \bm{D}_{2x_{2}} \ket*{R}}{\chi^{x_{2} - x_{1} + 1} \braket*{L}{R}}.
\end{equation}

Defining the connected correlation,
\begin{equation}
    C_{x_{1}, x_{2}} = \langle n_{x_{1}} n_{x_{2}} \rangle - \langle n_{x_{1}} \rangle \langle n_{x_{2}} \rangle,
\end{equation}
and using the decomposition of the transfer matrix $\T$,
\begin{equation}
    \T = \sum_{j = 1}^{2} \chi_{j} \dyad{R_{j}}{L_{j}},
\end{equation}
where the normalized eigenvectors are 
\begin{equation}
\begin{aligned}
    \ket*{R_{1}} & = \frac{1}{\sqrt{2}}
    \begin{bmatrix}
        1 \\
        1 \\
    \end{bmatrix} = \bra*{L_{1}}^\dagger, &
    \ket*{R_{2}} & = \frac{1}{\sqrt{2}}
    \begin{bmatrix}
        1 \\
        - 1 \\
    \end{bmatrix} = \bra*{L_{2}}^\dagger,  
    \nonumber
\end{aligned}
\end{equation}
and corresponding eigenvalues,
\begin{equation}
    \chi_{1} = (1 + \xi)(1 + \omega), \qquad \chi_{2} = (1 - \xi)(1 - \omega),
\end{equation}
can be rewritten compactly for arbitrary sites as
\begin{widetext}
\begin{equation}
    C_{2x_{1} - y_{1}, 2x_{2} - y_{2}} = \frac{\bra*{L_{1}} \V^{1 - y_{1}} \bm{D}_{2x_{1} - y_{1}} \W^{y_{1}} \dyad{R_{2}}{L_{2}} \V^{1 - y_{2}} \bm{D}_{2x_{2} - y_{2}} \W^{y_{2}} \ket*{R_{1}}}{\chi_{1} \chi_{2}} \bigg(\frac{\chi_{2}}{\chi_{1}}\bigg)^{x_{2} - x_{1}}.
\end{equation}
\end{widetext}
As expected, the correlation function depends only on the distance between the sites and decays exponentially as
\begin{equation}
    C_{2x_{1} - y_{1}, 2x_{2} - y_{2}} \sim \exp(- \frac{\abs*{x_{2} - x_{1}}}{\ell}),
\end{equation}
with correlation length $\ell$
\begin{equation}
\ell = \ln\abs{\frac{\chi_{1}}{\chi_{2}}}.
\end{equation}

Finally, we consider the ensemble average quasiparticle current in the nonequilibrium stationary state, defined as the difference between the densities of the quasiparticles. Explicitly, the density of positive  quasiparticles $j^{+} \equiv j^{+}_{x}$, which is independent of site $x$ in the NESS, is given by
\begin{equation}
    j^{+} = \frac{1}{Z} \sum_{n} \big(n_{2x - 1} (1 - n_{2x}) + (1 - n_{2x - 1}) n_{2x})\big) p_{n},
\end{equation}
whilst the density of negative quasiparticles $j^{-} \equiv j^{-}_{x}$ is
\begin{equation}
    j^{-} = \frac{1}{Z} \sum_{n} \big(n_{2x} (1 - n_{2x + 1}) + (1 - n_{2x}) n_{2x + 1}\big) p_{n}.
\end{equation}
Computing these expressions, we find that they read
\begin{equation}
    j^{+} = \frac{\xi}{1 + \xi}, \qquad j^{-} = \frac{\omega}{1 + \omega},
\end{equation}
which we notice are exactly equivalent to the conditional probabilities of detecting quasiparticles in the NESS, $p^{+}$ and $p^{-}$, in Eqs.~\eqref{eq:positive-quasiparticle-conditional-probabilities} and~\eqref{eq:negative-quasiparticle-conditional-probabilities}, respectively. Therefore, the ensemble average quasiparticle current,
\begin{equation}
    j \equiv j^{+} - j^{-} = \frac{\xi - \omega}{(1 + \xi)(1 + \omega)},
\end{equation}
which, we remark, is linear in the small parameter regime (i.e., $\xi, \omega \ll 1$), as expected,
\begin{equation}
    j \sim \xi - \omega.
\end{equation}

\section{Large Deviations}\label{sec:large-deviations}

A central feature of stochastic KCMs is
the existence of {\em trajectory phase transitions} 
\cite{Garrahan2007,Garrahan2009} (see also~\cite{Lecomte2007,Appert-Rolland2008,Espigares2013,Karevski2017,Helms2019,Monthus2021} and~\cite{Garrahan2018} for a review). 
This refers to the singular change displayed by trajectories with dynamical behaviour that is very different from typical. 
Specifically, the XOR-FA model~\cite{Causer2020}, which has the same constraint as Rule 150, was shown to have an active-inactive trajectory phase transition, demonstrated by studying the large deviation (LD) statistics of an appropriate trajectory observable (the total number of configuration changes, or dynamical activity~\cite{Garrahan2007,Lecomte2007,Maes2019}). 
We now show that the dynamics of the boundary driven Rule 150 also displays such transitions. We do so by computing the exact LD functions that determine the long-time statistics of a large class of trajectory observables.

\subsection{Time integrated observables}

We consider general time (and space) additive observables of the form
\begin{equation}
    K(N, T) = \sum_{t = 0}^{T - 1} \sum_{x = 1}^{2N - 1} (a_{x}^{2t} + b_{x}^{2t + 1}),
\end{equation}
where $a$ and $b$ are functions of the occupation on two consecutive sites at given times in a trajectory,
\begin{equation}\label{eq:large-deviation-functions}
    a_{x}^{2t} \equiv a_x(n_{x}^{2t}, n_{x + 1}^{2t}), \qquad \! b_{x}^{2t + 1} \equiv b_x(n_{x}^{2t + 1}, n_{x + 1}^{2t + 1}).
\end{equation}
We refer to observables of this type as \textit{dynamical} as they depend on the full time history of the state of the system, namely, the \textit{trajectory} $(\n^{0}, \n^{1}, \ldots, \n^{2T - 1})$. For example, one could consider the time integrated number of excited sites given by $a_{x}^{2t} = \smash{\frac{1}{2}(n_{x}^{2t} + n_{x + 1}^{2t})}$ and $b_{x}^{2t + 1} = 0$.

In the long time limit, $T \to \infty$, the probability distribution of $K$ has a large deviation (LD) form~\cite{Touchette2009},
\begin{equation}
    P_{T}(K) \asymp \exp\big(-T \varphi_{N}(k)\big),
\end{equation}
where $\varphi_{N}(k) \equiv \varphi_{N}(K/T)$ is the \textit{rate function}. Similarly, it can be shown that the \textit{moment generating function} has a LD form too,
\begin{equation}
    M_{T}(s) \asymp \exp\big(T \theta_{N}(s)\big),
\end{equation}
where we refer to $\theta_{N}(s)$ as the \textit{scaled cumulant generating function} (SCGF) as its derivatives at $s = 0$ are related to the cumulants of $k \equiv K/T$. The LD functions are connected through a Legendre transform,
\begin{equation}\label{eq:large-deviation-legendre-transform}
    \theta_{N}(s) = - \min_{k}\big(s k + \varphi_{N}(k)\big),
\end{equation}
which implies that they can intuitively be interpreted as corresponding to the free energy and entropy density of the trajectory ensemble.

In order to obtain an analytic form for the SCGF, we follow the approach of Refs.~\cite{Buca2019, Buca2021} whereby we deform, or \textit{tilt} the Markov operator~\cite{Touchette2009}. As will be demonstrated, we then have that
\begin{equation}\label{eq:large-deviation-scaled-cumulant-generating-function}
    \theta_{N}(s) = \ln \tilde{\Lambda}(s),
\end{equation}
where $\smash{\tilde{\Lambda}(s)}$ is the eigenvalue of the tilted Markov operator with the largest real part. Finding $\smash{\tilde{\Lambda}(s)}$, therefore, allows us to study the statistics of $K$ and its cumulants.

\subsection{Tilted Markov operator}\label{sec:tilted-markov-operator}

The tilted Markov operator $\tM{}(s)$ is defined as
\begin{equation}
    \tM{}(s) \equiv \tM{O}(s) \tM{E}(s),
\end{equation}
where $\tM{E}(s)$ and $\tM{O}(s)$ are the tilted propagators that act on the even and odd time steps, respectively,
\begin{equation}\label{eq:large-deviation-tilted-even-odd-operators}
    \tM{E}(s) = \M{E} \A(s), \qquad \tM{O} (s) = \M{O} \B(s),
\end{equation}
with $\A(s)$ and $\B(s)$ the diagonal operators introduced to apply the deformation. It follows that these extensive tilt operators can be expressed as products of local operators acting on pairs of adjacent sites,
\begin{equation}\label{eq:large-deviation-operator-product}
\begin{aligned}
    \A(s) & = \A_{1, 2}^{(1)} \A_{2, 3}^{(2)} \cdots \A_{2N - 1, 2N}^{(2N - 1)}, \\
    \B(s) & = \B_{1, 2}^{(1)} \B_{2, 3}^{(2)} \cdots \B_{2N - 1, 2N}^{(2N - 1)},
\end{aligned}
\end{equation}
where the subscript index denotes the sites of the lattice on which the operators act nontrivially,
\begin{equation}\label{eq:large-deviation-local-operators}
\begin{aligned}
    \A_{x, x + 1}^{(x)} & = \bm{I}^{\otimes (x - 1)} \otimes \A^{(x)} \otimes \bm{I}^{\otimes (2N - x - 1)}, \\
    \B_{x, x + 1}^{(x)} & = \bm{I}^{\otimes (x - 1)} \otimes \B^{(x)} \otimes \bm{I}^{\otimes (2N - x - 1)},
\end{aligned}
\end{equation}
while the superscript index denotes that the matrices are \textit{site dependent}. Specifically, the operators $\A^{(x)}$ and $\B^{(x)}$ are given by the following local $4 \times 4$ diagonal matrices,
\begin{equation}\label{eq:large-deviation-diagonal-matrices}
\begin{aligned}
    \A^{(x)} & =
    \begin{bmatrix}
        a_{00}^{(x)} & & & \\
        & a_{01}^{(x)} & & \\
        & & a_{10}^{(x)} & \\
        & & & a_{11}^{(x)} \\
    \end{bmatrix}, \\
    \B^{(x)} & =
    \begin{bmatrix}
        b_{00}^{(x)} & & & \\
        & b_{01}^{(x)} & & \\
        & & b_{10}^{(x)} & \\
        & & & b_{11}^{(x)} \\
    \end{bmatrix}, 
\end{aligned}
\end{equation}
where we have introduced the shorthand notations,
\begin{equation}\label{eq:large-deviation-functions-shorthand}
\begin{aligned}
    a_{n_{x}, n_{x + 1}}^{(x)} & \equiv \exp(-s a_{x}(n_x,n_{x+1})), \\
    b_{n_{x}, n_{x + 1}}^{(x)} & \equiv \exp(-s b_{x}(n_x,n_{x+1})),
\end{aligned}
\end{equation}
to denote the exponents of the local functions~\eqref{eq:large-deviation-functions}.

It follows directly from computation that the local tilt operators~\eqref{eq:large-deviation-diagonal-matrices} can be distributed between the local time evolution operators~\eqref{eq:local-time-evolution-operator} and~\eqref{eq:local-boundary-time-evolution-operators} in such a way that the tilted propagators~\eqref{eq:large-deviation-tilted-even-odd-operators} can be expressed as
\begin{equation}
\begin{aligned}
    \tM{E}(s) & = \tU^{(2)}_{2} \tU^{(4)}_{4} \cdots \tU^{(2N - 2)}_{2N - 2} \tR^{(2N)}, \\
    \tM{O}(s) & = \tL^{(1)} \tU^{(3)}_{3} \cdots \tU^{(2N - 3)}_{2N - 3} \tU^{(2N - 1)}_{2N - 1}, 
\end{aligned}
\end{equation}
where the tilted bulk matrices read
\begin{equation}
\begin{aligned}
    \tU_{2x}^{(2x)} & = \U_{2x} \A^{(2x - 1)}_{2x - 1, 2x} \A^{(2x)}_{2x, 2x + 1}, \\
    \tU_{2x - 1}^{(2x - 1)} & = \U_{2x - 1} \B^{(2x - 2)}_{2x - 2, 2x - 1} \B^{(2x - 1)}_{2x - 1, 2x},
\end{aligned}
\end{equation}
while the tilted boundary matrices are given by
\begin{equation}
    \tR^{(2N)} = \R \A^{(2N - 1)}_{2N - 1, 2N}, \qquad \tL^{(1)} = \L \B^{(1)}_{1, 2}.
\end{equation}

\subsection{Dominant eigenvalue}

We now look to construct an explicit expression for the leading eigenvector of the tilted Markov operator $\tM{}(s)$, namely, the eigenvector associated to the eigenvalue $\smash{\tilde{\Lambda}(s)}$ with the largest real part. Specifically, we seek a pair of vectors $\tp$ and $\tpp$, that satisfy the coupled equations,
\begin{equation}\label{eq:large-deviation-coupled-eigenvalue-equations}
    \tM{E}(s) \tp = \tilde{\Lambda}_{\text{R}}(s) \tpp, \qquad \tM{O}(s) \tpp = \tilde{\Lambda}_{\text{L}}(s) \tp,
\end{equation}
where $\tilde{\Lambda}(s) = \tilde{\Lambda}_{\text{R}}(s) \tilde{\Lambda}_{\text{L}}(s)$, which indeed implies that
\begin{equation}
    \tM{}(s) \tp = \tilde{\Lambda}(s) \tp.
\end{equation}
We now postulate a simple staggered MPS ansatz for the components of the eigenvectors similar to the ansatz used for the vectors in Sec.~\ref{sec:decay-modes}, that reads
\begin{equation}
\begin{aligned}
    \tilde{p}_{n} & = \tl \tV^{(1)}_{n_{1}} \tW^{(2)}_{n_{2}} \cdots \tV^{(2N - 1)}_{n_{2N - 1}} \tW^{(2N)}_{n_{2N}} \tr, \\
    \tilde{p}^{\prime}_{n} & = \tlp \tW^{(1)}_{n_{1}} \tV^{(2)}_{n_{2}} \cdots \tW^{(2N - 1)}_{n_{2N - 1}} \tV^{(2N)}_{n_{2N}} \trp,
\end{aligned}
\end{equation}
where the matrices $\tV^{(x)}_{n_{x}}$ and $\tW^{(x)}_{n_{x}}$ acting in the auxiliary space $\mathbb{C}^{2}$ are now \textit{site dependent}. It then follows that we can efficiently write the pair of vectors, using the compact tensor product notation, as
\begin{equation}
\begin{aligned}
    \tp & = \tl \tv^{(1)}_{1} \tw^{(2)}_{2} \cdots \tv^{(2N - 1)}_{2N - 1} \tw^{(2N)}_{2N} \tr, \\
    \tpp & = \tlp \tw^{(1)}_{1} \tv^{(2)}_{2} \cdots \tw^{(2N - 1)}_{2N - 1} \tv^{(2N)}_{2N} \trp,
\end{aligned}
\end{equation}
where, explicitly, the vectors of tilted matrices read
\begin{equation}
    \tv^{(x)}_{x} =
    \begin{bmatrix}
        \tV^{(x)}_{0} \\
        \tV^{(x)}_{1} \\
    \end{bmatrix}, \qquad
    \tw^{(x)}_{x} =
    \begin{bmatrix}
        \tW^{(x)}_{0} \\
        \tW^{(x)}_{1} \\
    \end{bmatrix}.
\end{equation}

We demand that these vectors of tilted matrices satisfy the following inhomogeneous bulk relations, that generalizes the homogeneous bulk algebraic cancellation scheme in Eqs.~\eqref{eq:mps-algebraic-relation} and~\eqref{eq:mps-algebraic-relation-primed}. Explicitly, we require that
\begin{equation}
\begin{aligned}
    \tU^{(2x)}_{2x} \big[ \tv^{(2x - 1)}_{2x - 1} \tw^{(2x)}_{2x} \tz^{(2x + 1)}_{2x + 1} \big] & \\
    & \hspace{-18pt} = \tz^{(2x - 1)}_{2x - 1} \tv^{(2x)}_{2x} \tw^{(2x + 1)}_{2x + 1}, \\
    \tU^{(2x + 1)}_{2x + 1} \big[ \tz^{(2x)}_{2x} \tw^{(2x + 1)}_{2x + 1} \tv^{(2x + 2)}_{2x + 2} \big] & \\
    & \hspace{-18pt} = \tw^{(2x)}_{2x} \tv^{(2x + 1)}_{2x + 1} \tz^{(2x + 2)}_{2x + 2},
\end{aligned}
\end{equation}
which encodes the matrix product equations,
\begin{equation}\label{eq:large-deviation-bulk-algebraic-relations}
\begin{aligned}
    a^{(2x - 1)}_{n_{2x - 1}, f_{2x}} a^{(2x)}_{f_{2x}, n_{2x + 1}} \tV^{(2x - 1)}_{n_{2x - 1}} \tW^{(2x)}_{f_{2x}} \tZ^{(2x + 1)}_{n_{2x + 1}} & \\
    & \hspace{-84pt} = \tZ^{(2x - 1)}_{n_{2x - 1}} \tV^{(2x)}_{n_{2x}} \tW^{(2x + 1)}_{n_{2x + 1}}, \\
    b^{(2x)}_{n_{2x}, f_{2x + 1}} b^{(2x + 1)}_{f_{2x + 1}, n_{2x + 2}} \tZ^{(2x)}_{n_{2x}} \tW^{(2x + 1)}_{f_{2x + 1}} \tV^{(2x + 2)}_{n_{2x + 2}} & \\
    & \hspace{-84pt} = \tW^{(2x)}_{n_{2x}} \tV^{(2x + 1)}_{n_{2x + 1}} \tZ^{(2x + 2)}_{n_{2x + 2}},
\end{aligned}
\end{equation}
where, for convenience, we have introduced the \textit{exchange matrices} $\smash{\tZ_{n_{x}}^{(x)}}$ [i.e., site dependent generalizations of the delimiter matrix $\S$~\eqref{eq:delimiter-matrix}], with the associated vector,
\begin{equation}
    \tz^{(x)}_{x} =
    \begin{bmatrix}
        \tZ^{(x)}_{0} \\
        \tZ^{(x)}_{1} \\
    \end{bmatrix}.
\end{equation}

We now postulate the following ansatz for the matrices of the inhomogeneous algebra that generalizes Eq.~\eqref{eq:mps-matrices},
\begin{equation}\label{eq:large-deviation-auxiliary-matrix-ansatz}
\begin{aligned}
    \tV_{0}^{(x)} & =
    \begin{bmatrix}
        1 & \tilde{v}_{01}^{(x)} \\
        0 & 0 \\
    \end{bmatrix}, \\
    \tV_{1}^{(x)} & =
    \begin{bmatrix}
        0 & 0 \\
        \tilde{v}_{10}^{(x)} & 1 \\
    \end{bmatrix}, \qquad
\end{aligned}
\begin{aligned}
    \tW_{0}^{(x)} & =
    \begin{bmatrix}
        1 & \tilde{w}_{01}^{(x)} \\
        0 & 0 \\
    \end{bmatrix}, \\
    \tW_{1}^{(x)} & =
    \begin{bmatrix}
        0 & 0 \\
        \tilde{w}_{10}^{(x)} & 1 \\
    \end{bmatrix},
\end{aligned}
\end{equation}
while the ansatz for the exchange matrices reads
\begin{equation}\label{eq:large-deviation-exchange-matrix-ansatz}
    \tZ_{0}^{(x)} =
    \begin{bmatrix}
        \tilde{z}_{00}^{(x)} & \tilde{z}_{01}^{(x)} \\
        0 & 0 \\
    \end{bmatrix}, \qquad
    \tZ_{1}^{(x)} =
    \begin{bmatrix}
        0 & 0 \\
        \tilde{z}_{10}^{(x)} & \tilde{z}_{11}^{(x)} \\
    \end{bmatrix}.
\end{equation}
Requiring that the inhomogeneous algebraic relations in Eq.~\eqref{eq:large-deviation-bulk-algebraic-relations} can be exactly solved using the generalized site dependent matrix ansatz postulated imposes constraints on the tilt operators $\A(s)$ and $\B(s)$, reminiscent of those placed on the boundary operators $\bm{R}$ and $\bm{L}$ in Sec.~\ref{sec:compatible-boundaries}. A particularly convenient choice of parametrization, that has a remarkably simple physical interpretation, can be obtained by setting
\begin{equation}\label{eq:large-deviation-particle-hole-symmetry}
\begin{aligned}
    a_{n_{x}, n_{x + 1}}^{(x)} & = a_{1 - n_{x}, 1 - n_{x + 1}}^{(x)}, \\
    b_{n_{x}, n_{x + 1}}^{(x)} & = b_{1 - n_{x}, 1 - n_{x + 1}}^{(x)},
\end{aligned}
\end{equation}
which is nothing but the requirement that $\A(s)$ and $\B(s)$ obey the \textit{particle-hole} symmetry of the model. Under this set of conditions, the inhomogeneous bulk algebra yields the following two-parameter family of solutions,
\begin{equation}\label{eq:large-deviation-bulk-algebra-solutions}
\begin{aligned}
    \tilde{v}_{01}^{(2x)} = \tilde{v}_{10}^{(2x)} & = \xi \prod_{j = 0}^{x - 1} \frac{b_{01}^{(2j)} a_{01}^{(2j + 1)}}{b_{00}^{(2j)} a_{00}^{(2j + 1)}}, \\
    \tilde{w}_{01}^{(2x)} = \tilde{w}_{10}^{(2x)} & = \omega \prod_{j = 1}^{x} \frac{b_{00}^{(2j - 1)} a_{00}^{(2j)}}{b_{01}^{(2j - 1)} a_{01}^{(2j)}}, \\
    \tilde{z}_{00}^{(2x)} = \tilde{z}_{11}^{(2x)} & = \prod_{j = 1}^{\mathclap{2x - 1}} b_{00}^{(j)}, \\
    \tilde{z}_{01}^{(2x)} = \tilde{z}_{10}^{(2x)} & = \xi \prod_{j = 0}^{x - 1} \frac{b_{01}^{(2j)} a_{01}^{(2j + 1)} b_{00}^{(2j + 1)}}{a_{00}^{(2j + 1)}}, \\
    \tilde{v}_{01}^{(2x + 1)} = \tilde{v}_{10}^{(2x + 1)} & = \xi \prod_{j = 1}^{x} \frac{a_{01}^{(2j - 1)} b_{01}^{(2j)}}{a_{00}^{(2j - 1)} b_{00}^{(2j)}}, \\
    \tilde{w}_{01}^{(2x + 1)} = \tilde{w}_{10}^{(2x + 1)} & = \omega \prod_{j = 0}^{x} \frac{a_{00}^{(2j)} b_{00}^{(2j + 1)}}{a_{01}^{(2j)} b_{01}^{(2j + 1)}}, \\
    \tilde{z}_{00}^{(2x + 1)} = \tilde{z}_{11}^{(2x + 1)} & = \prod_{j = 1}^{\mathclap{2x}} \frac{1}{a_{00}^{(j)}}, \\
    \tilde{z}_{01}^{(2x + 1)} = \tilde{z}_{10}^{(2x + 1)} & = \omega \prod_{j = 0}^{x} \frac{b_{00}^{(2j + 1)}}{a_{01}^{(2j)} b_{01}^{(2j + 1)} a_{00}^{(2j + 1)}},
\end{aligned}
\end{equation}
where we have used the convention that
\begin{equation}
\begin{aligned}
    a_{n_{-1}, n_{0}}^{(-1)} = a_{n_{0}, n_{1}}^{(0)} = 1, \\
    b_{n_{-1}, n_{0}}^{(-1)} = b_{n_{0}, n_{1}}^{(0)} = 1.
\end{aligned}
\end{equation}

Analogous to the treatment of the NESS in Sec.~\ref{sec:decay-modes}, we additionally require that the tilted row vectors of the left boundary $\smash{\tl}$ and $\smash{\tlp}$ and tilted column vectors of the right boundary $\smash{\tr}$ and $\smash{\trp}$, satisfy inhomogeneous site dependent boundary algebraic relations generalizing the homogeneous identities~\eqref{eq:decay-left-even-boundary-relation},~\eqref{eq:decay-right-even-boundary-relation},~\eqref{eq:decay-left-odd-boundary-relation}, and~\eqref{eq:decay-right-odd-boundary-relation}. In particular, we demand the following relations hold,
\begin{align}
    \tl \tz_{1}^{(1)} & = \tlp \tw_{1}^{(1)}, \\
    \! \tR \big[ \tv_{2N - 1}^{(2N - 1)} \tw_{2N}^{(2N)} \tr \big] & = \tilde{\Lambda}_{\text{R}} \tz_{2N - 1}^{(2N - 1)} \tv_{2N}^{(2N)} \trp, \\
    \tL \big[ \tlp \tw_{1}^{(1)} \tv_{2}^{(2)} \big] & = \tilde{\Lambda}_{\text{L}} \tl \tv_{1}^{(1)} \tz_{2}^{(2)}, \\
    \tz_{2N}^{(2N)} \trp & = \tw_{2N}^{(2N)} \tr,
\end{align}
which, in terms of the auxiliary matrices, read
\begin{align}
    \tl \tZ_{n_{1}}^{(1)} & = \tlp \tW_{n_{1}}^{(1)}, \label{eq:large-deviation-left-even-boundary-relations} \\
    \smash{\sum_{\mathclap{n_{5} = 0}}^{1}} R_{n_{3}, f_{4}, n_{5}} a_{n_{3}, f_{4}}^{(3)} \tV_{n_{3}}^{(3)} \tW_{f_{4}}^{(4)} \tr \hspace{-12pt} & \nonumber \\
    & = \tilde{\Lambda}_{\text{R}} \tZ_{n_{3}}^{(3)} \tV_{n_{4}}^{(4)} \trp, \label{eq:large-deviation-right-even-boundary-relations} \\
    \smash{\sum_{\mathclap{n_{0} = 0}}^{1}} L_{n_{0}, f_{1}, n_{2}} b_{f_{1}, n_{2}}^{(1)} \tlp \tW_{f_{1}}^{(1)} \tV_{n_{2}}^{(2)} \hspace{-12pt} & \nonumber \\
    & = \tilde{\Lambda}_{\text{L}} \tl \tV_{n_{1}}^{(1)} \tZ_{n_{2}}^{(2)}, \label{eq:large-deviation-left-odd-boundary-relations} \\
    \tZ_{n_{4}}^{(4)} \trp & = \tW_{n_{4}}^{(4)} \tr, \label{eq:large-deviation-right-odd-boundary-relations}
\end{align}
where, to save space, we set $N = 2$ at the right boundary. It can be straightforwardly shown by direct computation that if these inhomogeneous bulk and boundary relations are satisfied, then the coupled eigenvalue equations~\eqref{eq:large-deviation-coupled-eigenvalue-equations} are solved. As an example, for a chain of size $N = 2$, the second relation follows as 
\begin{equation}
\begin{aligned}
    \tM{O} \tpp & = \tL \tU_{3}^{(3)} \tlp \tw_{1}^{(1)} \tv_{2}^{(2)} \tw_{3}^{(3)} \tv_{4}^{(4)} \trp \\
    & = \tilde{\Lambda}_{\text{L}} \tU_{3}^{(3)} \tl \tv_{1}^{(1)} \tz_{2}^{(2)} \tw_{3}^{(3)} \tv_{4}^{(4)}  \trp \\
    & = \tilde{\Lambda}_{\text{L}} \tl \tv_{1}^{(1)} \tw_{2}^{(2)} \tv_{3}^{(3)} \tz_{4}^{(4)} \trp \\
    & = \tilde{\Lambda}_{\text{L}} \tl \tv_{1}^{(1)} \tw_{2}^{(2)} \tv_{3}^{(3)} \tw_{4}^{(4)} \tr \\
    & = \tilde{\Lambda}_{\text{L}} \tp,
\end{aligned}
\end{equation}
with the first equation of Eq.~\eqref{eq:large-deviation-coupled-eigenvalue-equations} following analogously. Solving the pair of inhomogeneous algebraic relations for the right boundary, Eqs.~\eqref{eq:large-deviation-right-even-boundary-relations} and~\eqref{eq:large-deviation-right-odd-boundary-relations}, yields a pair of expressions for the spectral parameters,
\begin{align}
    \xi & = \sigma \frac{\tilde{\Lambda}_{\text{R}} \tilde{z}_{00}^{(2N + 1)} - \tilde{z}_{00}^{(2N)} (1 - \alpha)}{\tilde{z}_{01}^{(2N)} \beta}, \label{eq:large-deviation-right-even-spectral-parameter} \\
    \omega & = \sigma \frac{\tilde{\Lambda}_{\text{R}} \tilde{z}_{00}^{(2N + 1)} (1 - \beta) - \tilde{z}_{00}^{(2N)} (1 - \alpha - \beta)}{\tilde{\Lambda}_{\text{R}} \tilde{z}_{01}^{(2N + 1)} \beta}, \label{eq:large-deviation-right-odd-spectral-parameter}
\end{align}
with $\sigma = \pm 1$ and right boundary vectors,
\begin{equation}
    \tr = \tilde{R}
    \begin{bmatrix}
        1 \\
        \sigma \\
    \end{bmatrix}, \quad
    \trp = \sigma \frac{\tilde{R}}{\tilde{z}_{00}^{(2N)}}
    \frac{1 + \sigma \tilde{w}^{(2N)}_{01}}{1 + \sigma \tilde{v}^{(2N)}_{01}}
    \begin{bmatrix}
        1 \\
        \sigma \\
    \end{bmatrix},
\end{equation}
where $\tilde{R}$ is a scalar that determines the normalization of the right boundary vector and with the convention that
\begin{equation}
\begin{aligned}
    a_{n_{2N}, n_{2N + 1}}^{(2N)} & = a_{n_{2N + 1}, n_{2N + 2}}^{(2N + 1)} = 1, \\
    b_{n_{2N}, n_{2N + 1}}^{(2N)} & = b_{n_{2N + 1}, n_{2N + 2}}^{(2N + 1)} = 1.
\end{aligned}
\end{equation}
Similarly, for the left boundary relations~\eqref{eq:large-deviation-left-even-boundary-relations} and~\eqref{eq:large-deviation-left-odd-boundary-relations}, we obtain a pair of solutions for the spectral parameters,
\begin{align}
    \xi & = \tau \frac{\tilde{\Lambda}_{\text{L}} (1 - \delta) - (1 - \gamma - \delta)}{\tilde{\Lambda}_{\text{L}} \delta}, \label{eq:large-deviation-left-even-spectral-parameter} \\
    \omega & = \tau \frac{\tilde{\Lambda}_{\text{L}} - (1 - \gamma)}{\delta}, \label{eq:large-deviation-left-odd-spectral-parameter}
\end{align}
with $\tau = \pm 1$ and left boundary vectors,
\begin{equation}
    \tl = \tilde{L}
    \begin{bmatrix}
        1 & \tau \\
    \end{bmatrix}, \qquad
    \tlp = \tilde{L}
    \begin{bmatrix}
        1 & \tau \\
    \end{bmatrix},
\end{equation}
where $\tilde{L}$ is the corresponding scalar that determines the normalization of the left boundary vector. A particularly convenient parametrization for the eigenvector $\tp$, that is consistent with the expression for the vector $\p$ in Sec.~\ref{sec:decay-modes}, is obtained by setting
\begin{equation}
    \tilde{R} = \frac{1}{1 + \sigma \tilde{w}_{01}^{(2N)}}, \qquad \tilde{L} = 1.
\end{equation}
It then follows from the ansatz~\eqref{eq:large-deviation-auxiliary-matrix-ansatz} and~\eqref{eq:large-deviation-exchange-matrix-ansatz}, that the components $\smash{\tilde{p}_{n}}$ of the eigenvectors $\smash{\tp}$ take the form of a generalized \textit{site dependent} grand canonical ensemble,
\begin{equation}\label{eq:large-deviation-site-dependent-grand-canonical-ensemble}
    \tilde{p}_{n} = \tau^{n_{1}} \prod_{x^{\pm}} \tilde{v}_{01}^{(x^{+})} \tilde{w}_{01}^{(x^{-})} \propto \xi^{\q^{+}_{n}} \omega^{\q^{-}_{n}}, 
\end{equation}
where, as before, the $\tau$ corresponds to the choice of solution at the left boundary while the sets $x^{\pm}$ denote the sets of sites which are occupied, respectively, by the positive and negative quasiparticles (e.g., $\smash{\tilde{p}_{0010} = \tilde{w}_{01}^{(2)} \tilde{v}_{01}^{(3)} \propto \xi \omega}$).

As was done in Sec.~\ref{sec:decay-modes}, we now impose equality between the pair of expressions for the spectral parameters at the right and left boundaries. Rearranging and subsequently solving for the eigenvalue $\smash{\tilde{\Lambda}(s) = \tilde{\Lambda}_{\text{L}}(s) \tilde{\Lambda}_{\text{R}}(s)}$ returns the following pair of quadratic characteristic polynomials,
\begin{equation}\label{eq:large-deviation-characteristic-polynomial}
    \tilde{\Lambda}^{2} - \Bigg( \sum_{j = 0}^{1} \sum_{k = 0}^{1} \tau^{j + k} \psi_{j, k} \tilde{Z}_{j, k} \Bigg) \tilde{\Lambda} + \psi \tilde{Z} = 0,
\end{equation}
where we have introduced,
\begin{equation}
\begin{aligned}
    \tilde{Z}_{00} & = \prod_{j = 1}^{2N} a_{00}^{(j)} b_{00}^{(j)}, \\
    \tilde{Z}_{01} & = \prod_{j = 1}^{N} a_{00}^{(2j - 1)} b_{01}^{(2j - 1)} a_{01}^{(2j)} b_{00}^{(2j)}, \\
    \tilde{Z}_{10} & = \prod_{j = 1}^{N} a_{01}^{(2j - 1)} b_{00}^{(2j - 1)} a_{00}^{(2j)} b_{01}^{(2j)}, \\
    \tilde{Z}_{11} & = \prod_{j = 1}^{2N} a_{01}^{(j)} b_{01}^{(j)},
\end{aligned}
\end{equation}
that satisfy the following identity,
\begin{equation}
    \tilde{Z} \equiv \tilde{Z}_{00} \tilde{Z}_{11} \equiv \tilde{Z}_{01} \tilde{Z}_{10} = \prod_{j = 1}^{2N} a_{00}^{(j)} a_{01}^{(j)} b_{00}^{(j)} b_{01}^{(j)},
\end{equation}
and the coefficients,
\begin{equation}
\begin{aligned}
    \psi_{00} & = (1 - \alpha)(1 - \gamma), \\
    \psi_{01} & = \alpha \delta, \\
    \psi_{10} & = \beta \gamma, \\
    \psi_{11} & = (1 - \beta)(1 - \delta),
\end{aligned}
\end{equation}
which satisfy the equality,
\begin{equation}
    \psi \equiv \sum_{j = 0}^{1} \sum_{\smash{k = 0}}^{1} \psi_{j, k} - 1 = (1 - \alpha - \beta)(1 - \gamma - \delta).
\end{equation}
As a consistency check, when $s = 0$ we recover the quadratic characteristic polynomials~\eqref{eq:decay-characteristic-polynomial} in Sec.~\ref{sec:decay-modes}, for which the corresponding dominant eigenvalue $\smash{\tilde{\Lambda} = \Lambda = 1}$, as expected for a stochastic operator.

\subsection{Dynamical phase transition}\label{eq:dynamical-phase-transition}

We can straightforwardly solve the quadratic equation in Eq.~\eqref{eq:large-deviation-characteristic-polynomial} to obtain an explicit expression for $\smash{\tilde{\Lambda}(s)}$, for any arbitrary observables satisfying the constraint~\eqref{eq:large-deviation-particle-hole-symmetry}. Specifically, the dominant eigenvalue reads 
\begin{equation}
    \tilde{\Lambda}(s) = \tilde{\eta}(s) + \sqrt{\tilde{\eta}^{2}(s) - \tilde{\mu}(s)},
\end{equation}
with,
\begin{equation}
    \tilde{\eta}(s) = \frac{1}{2} \sum_{j = 0}^{1} \sum_{\smash{k = 0}}^{1} \psi_{j, k} \tilde{Z}_{j, k}, \qquad
    \tilde{\mu}(s) = \psi \tilde{Z}.
\end{equation}
As the observables~\eqref{eq:large-deviation-functions} are extensive in the system size, we can define the tilting functions as
\begin{equation}
    \tilde{Z}_{j, k} = \exp(-N s \zeta_{j, k}), 
\end{equation}
such that the following limits exist and are finite,
\begin{equation}
\begin{aligned}
    \zeta_{j, k} = - \lim_{\mathclap{N \to \infty}} \ \frac{\ln \tilde{Z}_{j, k}}{N s}.
\end{aligned}
\end{equation}
The SCGF can then be expressed in the scaling form,
\begin{equation}\label{eq:large-deviation-scgf-scaling-form}
    \theta_{N}(s) \equiv \vartheta(N s).
\end{equation}
Immediately, this scaling form~\eqref{eq:large-deviation-scgf-scaling-form} provides us with the cumulants of $K$ in the long time limit $T \to \infty$, namely,
\begin{equation}
    \kappa_{j} \equiv \lim_{T \to \infty} \frac{1}{T} \langle \! \langle K^{j} \rangle \! \rangle = (-1)^{j} \frac{\d^{j}}{\d s^{j}} \theta_{N}(s) \bigg\rvert_{s = 0} \propto N^{j},
\end{equation}
where $\langle \! \langle K^{j} \rangle \! \rangle$ denotes the $j$th cumulant of the observable $K$. Note that in the thermodynamic limit (i.e., $N \to \infty$), the long time cumulants of $K$ for $j \geq 2$ diverge for $s = 0$, therefore, indicating the existence of a singularity (i.e., a \textit{dynamical phase transition} in the trajectory statistics).

We can construct explicitly the exact form of all cumulants $\langle \! \langle K^{j} \rangle \! \rangle$ for all even system sizes $2N$ in the long time limit $T \to \infty$, as detailed in Appendix~\ref{app:cumulants-of-long-time-observables}. For $j = 1$, we obtain the \textit{mean} of $K$ per unit time,
\begin{equation}
    \kappa_{1} = - \frac{4 \tilde{\eta}' - \tilde{\mu}'}{2(1 - \tilde{\mu})},
\end{equation}
while for $j = 2$, we get the \textit{variance} of $K$ per unit time,
\begin{equation}
    \kappa_{2} = \frac{4 \tilde{\eta}'' - \tilde{\mu}''}{2(1 - \tilde{\mu})} + \frac{(4 \tilde{\eta}' - \tilde{\mu}') \tilde{\mu}'}{4(1 - \tilde{\mu})^{2}} - \frac{(4 \tilde{\eta}' - \tilde{\mu}')^{2}}{4(1 - \tilde{\mu})^{3}},
\end{equation}
where the derivatives are given by,
\begin{equation}
\begin{aligned}
    \tilde{\eta}^{(m)} \equiv \frac{\d^{m}}{\d s^{m}} \tilde{\eta}(s) \bigg\rvert_{s = 0} & = \frac{1}{2} \sum_{j = 0}^{1} \sum_{\smash{k = 0}}^{1} \psi_{j, k} (-N \zeta_{j, k})^{m}, \\
    \tilde{\mu}^{(m)} \equiv \frac{\d^{m}}{\d s^{m}} \tilde{\mu}(s) \bigg\rvert_{s = 0} & = \psi (-N \zeta)^{m}.
\end{aligned}
\end{equation}

We now consider the asymptotic behaviour of the scaling function $\vartheta(Ns)$, which under the assumption of positive tilting functions $\zeta_{11} > \cdots > \zeta_{00}$, takes the form,
\begin{equation}
    \vartheta(Ns) =
    \begin{cases}
        - N s \zeta_{11} + \ln \psi_{11} + \cdots & N s \to \infty \\
        - N s \zeta_{00} + \ln \psi_{00} + \cdots & N s \to - \infty \\
    \end{cases}
\end{equation}
It can then be deduced that the SCGF converges to,
\begin{equation}\label{eq:large-deviation-scgf-asymptotic-limit}
    \lim_{N \to \infty} \frac{1}{N} \theta_{N}(s) =
    \begin{cases}
        - s \zeta_{11} & s > 0 \\
        - s \zeta_{00} & s < 0 \\
    \end{cases}
\end{equation}
where the singularity at $s = 0$ corresponds to a \textit{first order phase transition}. 

\begin{figure*}[t]
    \centering
    \includegraphics{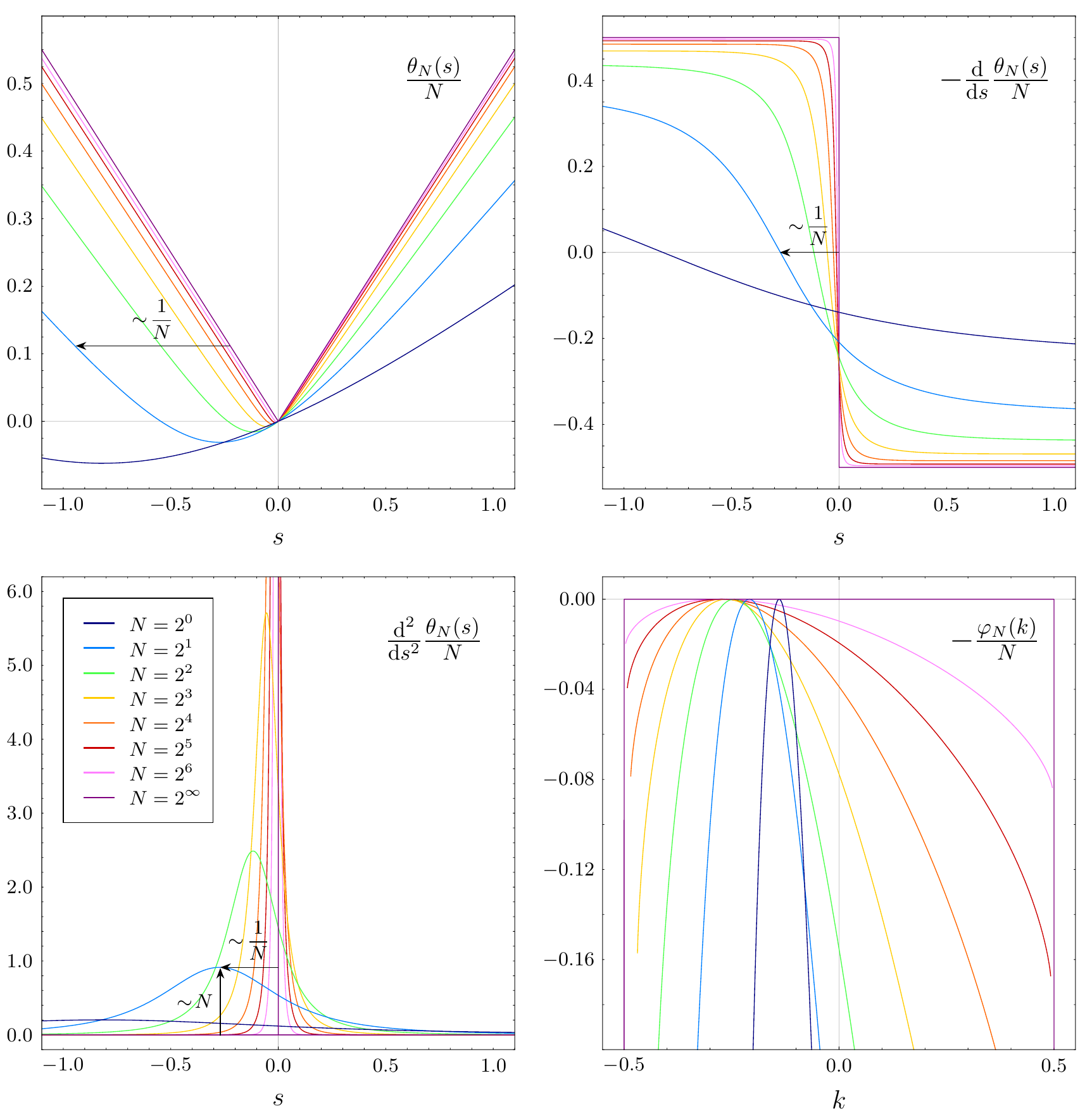}
    \caption{\textbf{Dynamical first order phase transition.} SCGF $\theta_{N}(s)/N$ [top left], mean $-\theta_{N}^{\prime}(s)/N$ [top right], variance $\theta_{N}^{\prime\prime}(s)/N$ [bottom left], and rate function $\varphi_{N}(k)/N$ [bottom right] for the current $\zeta_{00} = \zeta_{11} = 0$, $\zeta_{01} = -\zeta_{10} = 1/2$, explicitly, the time integrated number of quasiparticles, for systems of even sizes $2N$ (see legend) with boundary conditional probabilities $\alpha = 3/5$, $\beta = 7/8$, $\gamma = 8/9$, and $\delta = 4/7$. In the thermodynamic limit $N \to \infty$, the SCGF approaches the asymptotic form in Eq.~\eqref{eq:large-deviation-scgf-asymptotic-limit} while the mean exhibits a first order phase transition about $s = 0$. Correspondingly, the variance diverges as $N \to \infty$, with the  singularity converging towards $s = 0$ as $1/N$. Additionally shown is the rate function which broadens towards a square well in the asymptotic limit.}
    \label{fig:phase-transition}
\end{figure*}

In Figure~\ref{fig:phase-transition} we plot the SCGF $\theta_{N}(s)$ and its cumulants, specifically, the mean $-\theta^{\prime}_{N}(s)$ and the variance $\theta^{\prime\prime}_{N}(s)$, for a particular observable, namely, the current (i.e., the time integrated number of quasiparticles). From inspection, it is clear that the SCGF converges towards the asymptotic form in Eq.~\eqref{eq:large-deviation-scgf-asymptotic-limit} as $N \to \infty$, with the discontinuity (i.e., the critical point) occurring at $s = 0$. Similarly, the mean transitions from being positive for $s < 0$, to negative for $s > 0$ around the critical point at $s = 0$, with the change becoming discontinuous in the thermodynamic limit (i.e., $N \to \infty$). This transformation in the shape of the mean, characterizing the dynamical phase transition, manifests in the variance as its maximum scales with $N$, whilst the corresponding value of $s$, which indicates the singularity, scales with $1/N$. We also show the rate function $\varphi_{N}(k)$, which we obtain by taking the Legendre transform of the SCGF~\eqref{eq:large-deviation-legendre-transform},
\begin{equation}
    \varphi_{N}(k) = - \min_{s}\big(s k + \theta_{N}(s)\big).
\end{equation}
As can be seen from inspection, the rate function $\varphi_{N}(k)$ broadens with increasing finite system size $N$, indicating large fluctuations in the dynamics. In the limit $N \to \infty$, $\varphi_{N}(k)$ converges towards a square well, with the extrema $\max{k} = \zeta_{00}$ and $\min{k} = \zeta_{11}$ associated to the coexisting dynamically active phases.

These results are reminiscent of those obtained for the Rule 54 RCA~\cite{Buca2019}. Dynamics of the Rule 150 RCA sits at the coexistence point between two dynamical phases, one of high activity where $K$ is large, and one of low activity with $K$ vanishing in the large size limit. Fluctuations in each of these phases are highly suppressed with the main source coming from the coexistence (cf. Figure~\ref{fig:phase-transition}). Much like the case of Rule 54 (i.e., the RCA counterpart to the FA model), we find that the main ingredient facilitating the active-inactive transitions are the kinetic constraints. The simplicity in the form of $\theta_N(s)$, as compared to other KCM, is a consequence of the deterministic dynamics in the bulk, as all fluctuations originate from the stochastic boundaries. Since the probabilistic cost of realizing a rare fluctuation in the boundary (e.g., not emitting any given quasiparticle if doing so yields the empty state) does not scale with the system size, rare trajectories can be easily realized. Boundary control over the bulk is a hallmark of a phase transition, and here, by construction, we have an immediate realization of this phenomenon in space-time.

\subsection{\label{sec:doob-transformation} Doob transformation}

Having the exact form of the leading eigenstate of the tilted operator also allows us to find the exact {\em generalised Doob transform}~\cite{Borkar2003,Jack2010,Chetrite2015,Garrahan2016}. This refers to the construction of a stochastic operator whose trajectories are the atypical trajectories described by the non-stochastic tilted operator. In other words, to derive the operator that gives the optimal dynamics with which to sample the exponentially rare trajectories of the original dynamics associated with counting field $s$. 
For long times, this construction only requires the leading eigenvalue and eigenvector.

We can obtain an explicit expression for the long time Doob operator from the MPS representation of the leading eigenvector $\tp$ of the tilted Markov operator $\tM{}(s)$
through~\cite{Borkar2003,Jack2010,Chetrite2015,Garrahan2016},
\begin{equation}\label{eq:large-deviation-doob-transformation}
    \tilde{\bm{D}}(s) \equiv \frac{1}{\tilde{\Lambda}(s)} \tilde{\bm{Q}}(s) \tM{}(s) \tilde{\bm{Q}}^{-1}(s),
\end{equation}
where $\tilde{\Lambda}(s)$ is the dominant eigenvalue of $\tM{}(s)$ and $\smash{\tilde{\bm{Q}}(s)}$ is a diagonal operator defined in terms of the components of the corresponding leading \text{left} eigenvector $\tq$ of $\tM{}(s)$,
\begin{equation}\label{eq:large-deviation-diagonal-left-eigenvector}
    \tilde{\bm{Q}}(s) = \sum_{n} 
    \bm{e}_{n} \, \bm{e}_{n}^{\text{T}} \, \bm{e}_{n} \cdot \tq \,,
\end{equation}
where $\bm{e}_{n}$ denotes a standard basis (column) vector (cf. Sec~\ref{sec:model}) and
\begin{equation}\label{eq:large-deviation-left-eigenequation}
    \tq \tM{}(s) = \tilde{\Lambda}(s) \tq.
\end{equation}
To reduce the computation required to obtain an analytic expression for the leading left eigenvector $\tq$ we utilise the technique used for Rule 54 in Ref.~\cite{Buca2019}. Namely, define
\begin{equation}\label{eq:large-deviation-left-eigenvector-similarity-transformation}
    \tilde{\bm{q}} \equiv \big( \bm{A}(s) \hat{\bm{q}} \big)^{\text{T}} = \hat{\bm{q}}^{\text{T}} \bm{A}(s),
\end{equation}
where $\hat{\bm{q}}$ can be straightforwardly shown to be the leading right eigenvector of the newly introduced operator $\hat{\bm{M}}(s)$, with $\smash{\hat{\Lambda}(s) \equiv \tilde{\Lambda}(s)}$ the associated eigenvalue. Specifically, taking the transpose of the left eigenvalue equation~\eqref{eq:large-deviation-left-eigenequation}, whilst utilising the similarity transformation for the left eigenvector~\eqref{eq:large-deviation-left-eigenvector-similarity-transformation}, we have for the left hand side,
\begin{equation}
\begin{aligned}
    \big( \hat{\bm{q}}^{\text{T}} \bm{A}(s) \tM{}(s) \big)^{\text{T}} & =
    \big( \hat{\bm{q}}^{\text{T}} \bm{A}(s) \M{O} \bm{B}(s) \M{E} \bm{A}(s) \big)^{\text{T}} \\
    & = \bm{A}^{\text{T}}(s) \M{E}^{\text{T}} \bm{B}^{\text{T}}(s) \M{O}^{\text{T}} \bm{A}^{\text{T}}(s) \hat{\bm{q}} \\
    & = \bm{A}(s) \M{E}^{\text{T}} \bm{B}(s) \M{O}^{\text{T}} \bm{A}(s) \hat{\bm{q}},
\end{aligned}
\end{equation}
where, to obtain the final equality, we used the property that the tilt operators $\bm{A}(s)$ and $\bm{B}(s)$ are diagonal. The corresponding right side of the equation then reads
\begin{equation}
    \big( \hat{\Lambda}(s) \hat{\bm{q}}^{\text{T}} \bm{A}(s) \big)^{\text{T}} = \hat{\Lambda}(s) \bm{A}^{\text{T}}(s) \hat{\bm{q}} = \hat{\Lambda}(s) \bm{A}(s) \hat{\bm{q}}.
\end{equation}
Multiplying both sides on the left by $\bm{A}^{-1}(s)$ then gives,
\begin{equation}
    \hat{\bm{M}}(s) \hat{\bm{q}} \equiv \M{E}^{\text{T}} \bm{B}(s) \M{O}^{\text{T}} \bm{A}(s) \hat{\bm{q}} = \hat{\Lambda}(s) \hat{\bm{q}},
\end{equation}
with the transposed even and odd time step operators,
\begin{equation}
    \M{E}^{\text{T}} = \R^{\text{T}} \prod_{x = 1}^{N - 1} \U_{2x}, \qquad \M{O}^{\text{T}} = \L^{\text{T}} \prod_{x = 1}^{N - 1} \U_{2x + 1}.
\end{equation}
As we did for the leading right eigenvector $\tp$, we now construct a pair of vectors $\hat{\bm{q}}$ and $\hat{\bm{q}}^{\prime}$ that satisfy the coupled eigenvalue equations,
\begin{equation}
    \M{O}^{\text{T}} \bm{A}(s) \hat{\bm{q}} = \hat{\Lambda}_{\text{L}}(s) \hat{\bm{q}}^{\prime}, \qquad \M{E}^{\text{T}} \bm{B}(s) \hat{\bm{q}}^{\prime} = \hat{\Lambda}_{\text{R}}(s) \hat{\bm{q}},
\end{equation}
where $\hat{\bm{q}}$ and $\hat{\bm{q}}^{\prime}$ take a matrix product form, reminiscent of the ansatz of the vectors $\tpp$ and $\tp$, respectively. More precisely, their components read
\begin{equation}
\begin{aligned}
    \hat{q}_{n} & = \bra*{\hat{L}} \hat{\bm{W}}_{n_{1}}^{(1)} \hat{\bm{V}}_{n_{2}}^{(2)} \cdots \hat{\bm{W}}_{n_{2N - 1}}^{(2N - 1)} \hat{\bm{V}}_{n_{2N}}^{(2N)} \ket*{\hat{R}}, \\
    \hat{q}_{n}^{\prime} & = \bra*{\hat{L}^{\prime}} \hat{\bm{V}}_{n_{1}}^{(1)} \hat{\bm{W}}_{n_{2}}^{(2)} \cdots \hat{\bm{V}}_{n_{2N - 1}}^{(2N - 1)} \hat{\bm{W}}_{n_{2N}}^{(2N)} \ket*{\hat{R}^{\prime}}.
\end{aligned}
\end{equation}
Analogous to the right eigenvectors $\tp$ and $\tpp$, the vectors $\hat{\bm{q}}$ and $\hat{\bm{q}}^{\prime}$ satisfy inhomogeneous bulk algebraic relations that can be written as
\begin{equation}
\begin{aligned}
    b^{(2x - 1)}_{n_{2x - 1}, f_{2x}} b^{(2x)}_{f_{2x}, n_{2x + 1}} \hat{\bm{V}}^{(2x - 1)}_{n_{2x - 1}} \hat{\bm{W}}^{(2x)}_{f_{2x}} \hat{\bm{Z}}^{(2x + 1)}_{n_{2x + 1}} & \\
    & \hspace{-84pt} = \hat{\bm{Z}}^{(2x - 1)}_{n_{2x - 1}} \hat{\bm{V}}^{(2x)}_{n_{2x}} \hat{\bm{W}}^{(2x + 1)}_{n_{2x + 1}}, \\
    a^{(2x)}_{n_{2x}, f_{2x + 1}} a^{(2x + 1)}_{f_{2x + 1}, n_{2x + 2}} \hat{\bm{Z}}^{(2x)}_{n_{2x}} \hat{\bm{W}}^{(2x + 1)}_{f_{2x + 1}} \hat{\bm{V}}^{(2x + 2)}_{n_{2x + 2}} & \\
    & \hspace{-84pt} = \hat{\bm{W}}^{(2x)}_{n_{2x}} \hat{\bm{V}}^{(2x + 1)}_{n_{2x + 1}} \hat{\bm{Z}}^{(2x + 2)}_{n_{2x + 2}},
\end{aligned}
\end{equation}
which are identical to Eq.~\eqref{eq:large-deviation-bulk-algebraic-relations}, but with the local tilting functions exchanged $\smash{a^{(x)}_{n_{x}, n_{x + 1}}} \leftrightarrow \smash{b^{(x)}_{n_{x}, n_{x + 1}}}$. It then follows directly that the explicit solutions to these equations are of the same form as Eq.~\eqref{eq:large-deviation-bulk-algebra-solutions}, but with
\begin{equation}
\begin{aligned}
    \hat{\bm{V}}_{n_{x}}^{(x)} \big(a^{(x)}_{n_{x}, n_{x + 1}}, b^{(x)}_{n_{x}, n_{x + 1}}\big) & \\
    & \hspace{-24pt} \equiv \tV_{n_{x}}^{(x)} \big(b^{(x)}_{n_{x}, n_{x + 1}}, a^{(x)}_{n_{x}, n_{x + 1}}\big), \\
    \hat{\bm{W}}_{n_{x}}^{(x)} \big(a^{(x)}_{n_{x}, n_{x + 1}}, b^{(x)}_{n_{x}, n_{x + 1}}\big) & \\
    & \hspace{-24pt} \equiv \tW_{n_{x}}^{(x)} \big(b^{(x)}_{n_{x}, n_{x + 1}}, a^{(x)}_{n_{x}, n_{x + 1}}\big), \\
    \hat{\bm{Z}}_{n_{x}}^{(x)} \big(a^{(x)}_{n_{x}, n_{x + 1}}, b^{(x)}_{n_{x}, n_{x + 1}}\big) & \\
    & \hspace{-24pt} \equiv \tZ_{n_{x}}^{(x)} \big(b^{(x)}_{n_{x}, n_{x + 1}}, a^{(x)}_{n_{x}, n_{x + 1}}\big).
\end{aligned}
\end{equation}
The corresponding boundary relations then read
\begin{equation}
\begin{aligned}
    \bra*{\hat{L}^{\prime}} \hat{\bm{Z}}_{n_{1}}^{(1)} & = \bra*{\hat{L}} \hat{\bm{W}}_{n_{1}}^{(1)}, \\
    \smash{\sum_{\mathclap{n_{5} = 0}}^{1}} R_{n_{3}, n_{4}, n_{5}} b_{n_{3}, f_{4}}^{(3)} \hat{\bm{V}}_{n_{3}}^{(3)} \hat{\bm{W}}_{f_{4}}^{(4)} \ket*{\hat{R}^{\prime}} \hspace{-18pt} & \\
    & = \hat{\Lambda}_{\text{R}} \hat{\bm{Z}}_{n_{3}}^{(3)} \hat{\bm{V}}_{n_{4}}^{(4)} \ket*{\hat{R}}, \\
    \smash{\sum_{\mathclap{n_{0} = 0}}^{1}} L_{n_{0}, n_{1}, n_{2}} a_{f_{1}, n_{2}}^{(1)} \bra*{\hat{L}} \hat{\bm{W}}_{f_{1}}^{(1)} \hat{\bm{V}}_{n_{2}}^{(2)} \hspace{-18pt} & \\
    & = \hat{\Lambda}_{\text{L}} \bra*{\hat{L}^{\prime}} \hat{\bm{V}}_{n_{1}}^{(1)} \hat{\bm{Z}}_{n_{2}}^{(2)}, \\
    \hat{\bm{Z}}_{n_{4}}^{(4)} \ket*{\hat{R}} & = \hat{\bm{W}}_{n_{4}}^{(4)} \ket*{\hat{R}^{\prime}},
\end{aligned}
\end{equation}
which are the same as Eqs.~\eqref{eq:large-deviation-left-even-boundary-relations},~\eqref{eq:large-deviation-right-even-boundary-relations},~\eqref{eq:large-deviation-left-odd-boundary-relations}, and~\eqref{eq:large-deviation-right-odd-boundary-relations}, but the functions $\smash{a^{(x)}_{n_{x}, n_{x + 1}}}$ and $\smash{b^{(x)}_{n_{x}, n_{x + 1}}}$ interchanged and with the stochastic boundary matrices $\bm{R}$ and $\bm{L}$ replaced by their transposes, which is equivalent to exchanging the elements $R_{n_{3}, f_{4}, n_{5}} \leftrightarrow R_{n_{3}, n_{4}, n_{5}}$, $L_{n_{0}, f_{1}, n_{2}} \leftrightarrow L_{n_{0}, n_{1}, n_{2}}$. If we solve the right boundary equations, as we did for the right eigenvector, we obtain the following expressions for the spectral parameters,
\begin{equation}
\begin{aligned}
    \xi & = \sigma \frac{\hat{\Lambda}_{\text{R}} \hat{z}_{00}^{(2N + 1)} - \hat{z}_{00}^{(2N)} (1 - \alpha)}{\hat{z}_{01}^{(2N)} \alpha}, \\
    \omega & = \sigma \frac{\hat{\Lambda}_{\text{R}} \hat{z}_{00}^{(2N + 1)} (1 - \beta) - \hat{z}_{00}^{(2N)} (1 - \alpha - \beta)}{\hat{\Lambda}_{\text{R}} \hat{z}_{01}^{(2N + 1)} \alpha},
\end{aligned}
\end{equation}
where the right boundary vectors are given by,
\begin{equation}
    \ket*{\hat{R}} = \hat{R}
    \begin{bmatrix}
        1 \\
        \sigma \\
    \end{bmatrix}, \quad
    \ket*{\hat{R}^{\prime}} = \sigma \hat{R} \tilde{z}^{(2N)}_{00} \frac{1 + \sigma \tilde{v}^{(2N)}_{01}}{1 + \sigma \tilde{w}^{(2N)}_{01}}
    \begin{bmatrix}
        1 \\
        \sigma \\
    \end{bmatrix}. \!\!
\end{equation}
Similarly solving the left boundary equations gives,
\begin{equation}
\begin{aligned}
    \xi & = \tau \frac{\hat{\Lambda}_{\text{L}} (1 - \delta) - (1 - \gamma - \delta)}{\hat{\Lambda}_{\text{L}} \gamma}, \\
    \omega & = \tau \frac{\hat{\Lambda}_{\text{L}} - (1 - \gamma)}{\gamma},
\end{aligned}
\end{equation}
for the spectral parameters, with the boundary vectors,
\begin{equation}
    \bra*{\hat{L}} = \hat{L}
    \begin{bmatrix}
        1 & \tau \\
    \end{bmatrix}, \qquad
    \bra*{\hat{L}^{\prime}} = \hat{L}
    \begin{bmatrix}
        1 & \tau \\
    \end{bmatrix}.
\end{equation}
It can be easily verified that equating the expressions for the spectral parameters $\xi$ and $\omega$ at either boundary and then solving for the eigenvalue $\hat{\Lambda}(s) = \hat{\Lambda}_{\text{L}} \hat{\Lambda}_{\text{R}}$ returns the quadratic characteristic polynomials~\eqref{eq:large-deviation-characteristic-polynomial}, as expected. Furthermore, it follows that by setting,
\begin{equation}
    \hat{R} = \frac{1}{1 + \sigma \hat{v}_{01}^{(2N)}}, \qquad \hat{L} = 1,
\end{equation}
the components $\hat{q}_{n}$ of the eigenvectors $\hat{\bm{q}}$ of the operator $\hat{\bm{M}}(s)$ take a form reminiscent to that of the vectors $\tp$ [cf. Eq.~\eqref{eq:large-deviation-site-dependent-grand-canonical-ensemble}]. Specifically,
\begin{equation}
    \hat{q}_{n} = \tau^{n_{1}} \prod_{x^{\pm}} \hat{v}_{01}^{(x^{+})} \hat{w}_{01}^{(x^{-})} \propto \xi^{\q^{+}_{n}} \omega^{\q^{-}_{n}}. 
\end{equation}
Explicit expressions for the left eigenvector $\smash{\tq}$ of the tilted Markov operator $\smash{\tM{}(s)}$, the diagonal operator $\smash{\tilde{\bm{Q}}(s)}$, and, most importantly, the long time Doob operator $\smash{\tilde{\bm{D}}(s)}$ then follow trivially from Eqs.~\eqref{eq:large-deviation-left-eigenvector-similarity-transformation}, \eqref{eq:large-deviation-diagonal-left-eigenvector}, and~\eqref{eq:large-deviation-doob-transformation}.

\bigskip

\section{Conclusions}\label{sec:conclusions}

The aim of this paper was to present a comprehensive  study into the dynamics of a simple integrable cellular automaton. The model we studied here, the Rule 150 RCA, is integrable~\cite{Gombor2021}, but in contrast to other recently studied integrable RCA, its quasiparticles are noninteracting~~\cite{Gopalakrishnan2018b}. This allowed us to present within the paper a significant number of exact results, including the stationary states for both closed and open boundaries, the full spectrum of the evolution operator, and the large deviation dynamical phase diagram. Our work here adds to the growing number of exact results on statistical mechanics of classical deterministic RCA, which is of interest to a number of other fields as explained in the introduction. There are also many interesting extensions and generalisations, including the dynamics of integrable RCA with stochastic or unitary dynamics~\cite{Prosen2021, Gombor2021}. We expect our paper which studies the simplest of these models can also serve as a useful introduction to this rich field. 

\bigskip

\acknowledgments

JWPW and JPG acknowledge support of The Leverhulme Trust through Grant number RPG-2018-181.
TP acknowledges support of European Research Council (ERC) through the Advanced Grant No.\ 694544 -- OMNES, and of Slovenian Research Agency (ARRS)
through the Programme P1-0402.

\renewcommand{\v}[1]{\defaultv{#1}}
\bibliography{bibfile}
\renewcommand{\v}{\bm{v}}

\appendix

\section{Discrete symmetries}\label{app:symmetries}

The discrete local symmetries of the model are given in Sec.~\ref{sec:states} by the commutation relations between the local time evolution operator $\U$ and the symmetry generators. Explicitly [cf. Eq.~\eqref{eq:model-local-symmetry-conditions}],
\begin{equation}\label{eq:symmetries-local-symmetries}
    [\U, \bm{J}_{\mathrm{S}}] = 0, \qquad [\U, \bm{J}_{\mathrm{T}}] = 0, \qquad [\U, \bm{J}_{\mathrm{P}}] = 0,
\end{equation}
where $\bm{J}_{\mathrm{S}}$, $\bm{J}_{\mathrm{T}}$, and $\bm{J}_{\mathrm{P}}$ are $8 \times 8$ matrix representations of the respective symmetry generators which are defined by their action on either the operator $\U$ or vector $\p^{t}$. First, we consider the generator of spatial-inversions $\bm{J}_{\mathrm{S}}$, whose action inverts the spatial indices of the sites $x - 1, x + 1$ about the site $x$. Specifically,
\begin{equation}\label{eq:symmetries-local-spatial-inversion-symmetry}
    \bm{J}_{\mathrm{S}} = \sum_{\mathclap{n_{x - 1}, n_{x + 1} = 0}}^{1} \bm{e}_{n_{x - 1}} \bm{e}_{n_{x + 1}}^{\mathrm{T}} \otimes \bm{I} \otimes \bm{e}_{n_{x + 1}} \bm{e}_{n_{x - 1}}^{\mathrm{T}},
\end{equation}
where $\bm{e}_{0}$ and $\bm{e}_{1}$ denote the elementary basis vectors,
\begin{equation}
    \bm{e}_{0} =
    \begin{bmatrix}
        1 \\
        0 \\
    \end{bmatrix}.
    \qquad \bm{e}_{1} =
    \begin{bmatrix}
        0 \\
        1 \\
    \end{bmatrix}.
\end{equation}
The matrix $\bm{J}_{\mathrm{S}}$, therefore, reads,
\begin{equation}
    \bm{J}_{\mathrm{S}} = 
    \begin{bmatrix}
        1 & & & & & & & \\
        & & & & 1 & & & \\
        & & 1 & & & & & \\
        & & & & & & 1 & \\
        & 1 & & & & & & \\
        & & & & & 1 & & \\
        & & & 1 & & & & \\
        & & & & & & & 1 \\
    \end{bmatrix},
\end{equation}
from which the first equality in Eq.~\eqref{eq:symmetries-local-symmetries} follows directly. Next, we consider the generator $\bm{J}_{\mathrm{T}}$ whose action reverses the direction of time, $t + 1 \to t - 1$. It, therefore, follows that the operator $\bm{J}_{\mathrm{T}}$ must necessarily satisfy the relation $\smash{\bm{J}_{\mathrm{T}} \U \bm{J}_{\mathrm{T}}^{-1} = \U^{-1}}$. Recalling that the local time evolution operator $\U$ is an involutory matrix (i.e., $\U = \U^{-1}$) then returns the second condition in Eq.~\eqref{eq:symmetries-local-symmetries}. This allows us to freely choose
\begin{equation}\label{eq:symmetries-local-time-reversal-symmetry}
    \bm{J}_{\mathrm{T}} = \U,
\end{equation}
such that, for simplicity, we can define
\begin{equation}
    \bm{J}_{\mathrm{T}} = 
    \begin{bmatrix}
        1 & & & & & & & \\
        & & & 1 & & & & \\
        & & 1 & & & & & \\
        & 1 & & & & & & \\
        & & & & & & 1 & \\
        & & & & & 1 & & \\
        & & & & 1 & & & \\
        & & & & & & & 1 \\
    \end{bmatrix}.
\end{equation}
Finally, we consider the particle-hole symmetry operator, whose generator is defined by its action on the vector $\p^{t}$, namely, it exchanges the binary variables $0$ and $1$,
so
\begin{equation}\label{eq:symmetries-local-particle-hole-symmetry}
    \bm{J}_{\mathrm{P}} = \sum_{\mathclap{n_{x - 1}, n_{x}, n_{x + 1} = 0}}^{1} \bm{e}_{n_{x - 1}, n_{x}, n_{x + 1}} \bm{e}_{1 - n_{x - 1}, 1 - n_{x}, 1 - n_{x + 1}}^{\mathrm{T}},
\end{equation}
where $\bm{e}_{n_{x - 1}, n_{x}, n_{x + 1}} = \bm{e}_{n_{x - 1}} \otimes \bm{e}_{n_{x}} \otimes \bm{e}_{n_{x + 1}}$ is an element of the standard basis of the vector space $(\mathbb{R}^{2})^{\otimes 3}$. It then follows immediately that $\bm{J}_{\mathrm{P}}$ is the exchange matrix,
\begin{equation}
    \bm{J}_{\mathrm{P}} =
    \begin{bmatrix}
        & & & & & & & 1 \\
        & & & & & & 1 & \\
        & & & & & 1 & & \\
        & & & & 1 & & & \\
        & & & 1 & & & & \\
        & & 1 & & & & & \\
        & 1 & & & & & & \\
        1 & & & & & & & \\
    \end{bmatrix}.
\end{equation}

At the level of the master equation~\eqref{eq:model-floquet-master-equation}, the local symmetries manifest in the dynamics as a combined spatial-inversion and time-reversal symmetry and a particle-hole symmetry [cf. Eq.~\eqref{eq:model-global-symmetry-conditions}]. Particularly,
\begin{equation}\label{eq:symmetries-global-symmetries}
    [\M{}, \bm{\mathcal{J}}_{\mathrm{ST}}] = 0, \qquad [\M{}, \bm{\mathcal{J}}_{\mathrm{P}}] = 0,
\end{equation}
with $\bm{\mathcal{J}}_{\mathrm{ST}}$ and $\bm{\mathcal{J}}_{\mathrm{P}}$ 
the $2^{2N} \times 2^{2N}$ matrix representations of the respective symmetry generators. We note that we can write the operator $\smash{\bm{\mathcal{J}}_{\mathrm{ST}} \equiv \bm{\mathcal{J}}_{\mathrm{S}} \bm{\mathcal{J}}_{\mathrm{T}}}$, that is, as a product of operators that generalize the local matrices in Eqs.~\eqref{eq:symmetries-local-spatial-inversion-symmetry} and~\eqref{eq:symmetries-local-time-reversal-symmetry}, which allows us to write
\begin{equation}
    \bm{\mathcal{J}}_{\mathrm{S}} = \sum_{\mathclap{n_{1}, \ldots, n_{2N} = 0}}^{1} \bm{e}_{n_{2N}, \ldots, n_{1}} \bm{e}_{n_{1}, \ldots, n_{2N}}^{\mathrm{T}},
\end{equation}
with $\bm{e}_{n_{1}, \ldots, n_{2N}} = \bm{e}_{n_{1}} \otimes \cdots \otimes \bm{e}_{n_{2N}}$ denoting a basis state of the vector space $(\mathbb{R}^{2})^{\otimes 2N}$. As was required of the local symmetry operator $\bm{J}_{\mathrm{T}}$, we demand that $\bm{\mathcal{J}}_{\mathrm{T}}$ satisfies the identity $\smash{\bm{\mathcal{J}}_{\mathrm{T}} \M{} \bm{\mathcal{J}}_{\mathrm{T}}^{-1} = \M{}^{-1}}$ 
which, after recalling that the operator $\M{} = \M{O} \M{E}$, immediately allows us to set
\begin{equation}
    \bm{\mathcal{J}}_{\mathrm{T}} = \M{E}.
\end{equation}
Similarly, it follows straightforwardly that
\begin{equation}
    \bm{\mathcal{J}}_{\mathrm{P}} = \sum_{\mathclap{n_{1}, \ldots, n_{2N} = 0}}^{1} \bm{e}_{1 - n_{1}, \ldots, 1 - n_{2N}} \bm{e}_{n_{1}, \ldots, n_{2N}}^{\mathrm{T}}.
\end{equation}
Finally, as mentioned, the dynamics additionally exhibits a symmetry that conserves the parity of the total number of empty and occupied sites [cf. Eq.~\eqref{eq:model-global-number-parity-symmetry}],
\begin{equation}
    [\M{}, \bm{\mathcal{J}}_{\mathrm{N}}] = 0,
\end{equation}
where the symmetry generator $\bm{\mathcal{J}}_{\mathrm{N}}$ is trivially given by
\begin{equation}
    \bm{\mathcal{J}}_{\mathrm{N}} = \sum_{\mathclap{n_{1}, \ldots, n_{2N} = 0}}^{1} (-1)^{\sum_{x} n_{x}} \bm{e}_{n_{1}, \ldots, n_{2N}} \bm{e}_{n_{1}, \ldots, n_{2N}}^{\mathrm{T}}.
\end{equation}

\section{Quasiparticle number constraint}\label{app:quasiparticle-constraint}

\begin{figure}[t]
    \centering
    \includegraphics{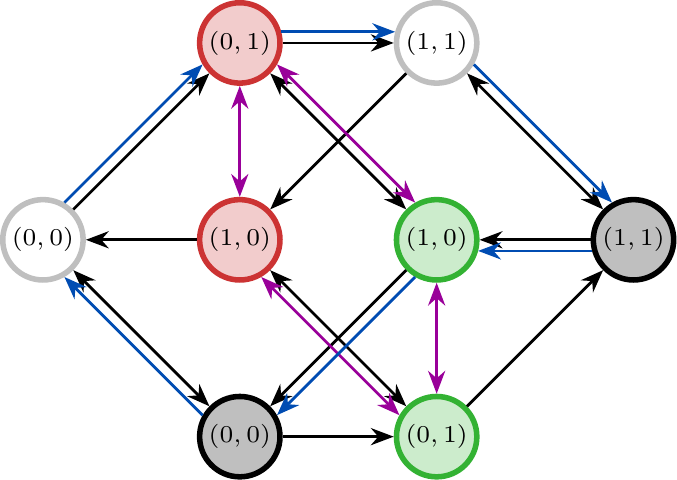}
    \caption{\textbf{Quasiparticle constraint.} Graph representation of the lattice that illustrates the physical constraint imposed on the numbers of positive and negative quasiparticles by the even size of the system and PBC~\eqref{eq:quasiparticle-constraint}. Vertices associated to configurations that start on sites with even or odd space-time parity [i.e., $x + t \pmod{2}$] are, respectively, colored black and white; configurations with positive or negative quasiparticles are highlighted green and red. Black arrows then indicate the directed edges which connect vertices obtained by shifting one site in space [i.e., $(n^{t}_{x}, n^{t}_{x + 1}) \leftrightarrow (n^{t}_{x + 1}, n^{t}_{x + 2})$], or equivalently, one step in time [i.e., $(n^{t}_{x}, n^{t}_{x + 1}) \leftrightarrow (n^{t + 1}_{x}, n^{t + 1}_{x + 1})$], while purple arrows correspond to the subset of contracted directed edges of the subgraph. To illustrate the basic concepts of this graph representation, the closed walk associated to the configuration $\n = (0, 0, 1, 1, 1, 0)$ is mapped by blue arrows.}
    \label{fig:quasiparticle-constraint}
\end{figure}

To prove the constraint on the numbers of positive and negative quasiparticles in a given configuration $\n$ of even size $2N$ with PBC~\eqref{eq:quasiparticle-constraint}, we introduce a convenient graph theoretic representation for the lattice, shown in Figure~\ref{fig:quasiparticle-constraint}. The graph, a \textit{directed bipartite graph}, is composed of two disjoint and independent sets of vertices, that are labelled by binary strings and represent the subconfigurations of adjacent sites with even and odd space and time indices [i.e., $x + t \pmod{2}$], respectively. Vertices of the even or odd set are then connected to vertices of the odd or even set by directed edges that correspond to either shifting one site in space (i.e., $x \leftrightarrow x + 1$) or evolving one step in time (i.e., $t \leftrightarrow t + 1$). We remark that we can simplify the graph significantly by contracting the paths between vertices whose labels represent quasiparticles~\eqref{eq:quasiparticle-detection}. Doing so yields a \textit{symmetric directed subgraph} with four vertices [i.e., a pair of vertices from each vertex set corresponding to the subconfigurations $(0, 1)$ and $(1, 0)$], that are each connected to exactly two other vertices. From this, it can be straightforwardly verified that each and every cycle of the subgraph is of even length and so satisfies Eq.~\eqref{eq:quasiparticle-constraint}.

\section{\label{app:maximum-entropy-state} Maximum entropy state}

For the case where $\xi = \omega = 1$, the stationary states, $\p$ and $\pp$, correspond to the \textit{maximum entropy state}. That is, the state for which the probabilities for each and every configuration are equal. In this limit, the MPS representation for the state simplifies such that the components of the probability state vectors can be written as
\begin{equation}\label{eq:maximum-entropy-state}
    \lim_{\xi, \omega \to 1} p_{n} = \frac{1}{Z} \Tr\big( \bm{X}_{n_{1}} \bm{X}_{n_{2}} \cdots \bm{X}_{n_{2N - 1}} \bm{X}_{n_{2N}}\big),
\end{equation}
where the auxiliary space matrices,
\begin{equation}
    \bm{X}_{0} =
    \begin{bmatrix}
        1 & 1 \\
        0 & 0 \\
    \end{bmatrix}, \qquad
    \bm{X}_{1} =
    \begin{bmatrix}
        0 & 0 \\
        1 & 1 \\
    \end{bmatrix}.
\end{equation}
To prove the stationarity of the maximum entropy state, we introduce a cubic algebraic relation, analog to that in Eq.~\eqref{eq:mps-algebraic-relation}, which reads
\begin{equation}
    \bm{X}_{n_{x - 1}} \bm{X}_{f_{x}} \bm{X}_{n_{x + 1}} = \bm{X}_{n_{x - 1}} \bm{X}_{n_{x}} \bm{X}_{n_{x + 1}},
\end{equation}
where we have utilised the simplifications in Eq.~\eqref{eq:matrix-product-simplification} and the fact that, in the limit $\xi, \omega \to 1$, the auxiliary matrices trivialise, explicitly, $\V_{n_{x}}, \W_{n_{x}} \to \bm{X}_{n_{x}}$. Noting that this identity is solved by the following relation,
\begin{equation}
    \bm{X}_{n_{x}} \bm{X}_{n_{x + 1}} = \bm{X}_{n_{x}},
\end{equation}
which holds for all $n_{x} = 0, 1$, then proves the invariance of the state. An identical proof holds for the state $\pp$ as, in the limit $\xi, \omega \to 1$, $\pp \equiv \p$.

\section{State counting function}\label{app:partition-function} 
 
To show that the state counting function $\Omega$ in Eq.~\eqref{eq:partition-function-entropy} really counts the number of states with $\q^{+}$ positive and $\q^{-}$ negative quasiparticles, we prove that the expressions for the grand canonical partition functions~\eqref{eq:mps-partition-function} and \eqref{eq:combinatoric-partition-function} are equivalent. To start, we write the product of transfer matrices as a recursion relation, specifically,
\begin{equation}
    \T^{N} = \T^{N - 1} \T,
\end{equation}
with elements $\smash{T_{ij}^{N} \equiv (\T^{N})_{i, j}}$ given by
\begin{equation}\label{eq:transfer-matrix-equations}
    T_{ij}^{N} = \sum_{k = 0}^{1} T_{ik}^{N - 1} T_{kj}^{\phantom{N}}.
\end{equation}
Substituting this parametrization for the transfer matrix components into $Z$~\eqref{eq:mps-partition-function}, admits the following expression for the grand canonical partition function,
\begin{equation}
    Z = \sum_{i = 0}^{1} T_{ii}^{N}.
\end{equation}
Introducing the following parametrization of the transfer matrix components,
\begin{equation}\label{eq:transfer-matrix-reductions}
    T_{11} = T_{00} = 1 + \xi \omega, \qquad T_{10} = T_{01} = \omega + \xi,
\end{equation}
then allows us to reduce the system of equations~\eqref{eq:transfer-matrix-equations} to just two recursive relations,
\begin{equation}\label{eq:transfer-matrix-solutions}
\begin{aligned}
    T^{N}_{00} & = (1 + \xi \omega) T^{N - 1}_{00} + (\omega + \xi) T^{N - 1}_{01}, \\
    T^{N}_{01} & = (1 + \xi \omega) T^{N - 1}_{01} + (\omega + \xi) T^{N - 1}_{00},
\end{aligned}
\end{equation}
yielding an expression for $Z$ in terms of just one recursive parameter,
\begin{equation}\label{eq:recursive-partition-function}
    Z = 2 T^{N}_{00},
\end{equation}
which can subsequently be expressed as a one-parameter second-order recurrence relation,
\begin{equation}
    T^{N}_{00} = 2 (1 + \xi \omega) T^{N - 1}_{00} - (1 - \xi^{2}) (1 - \omega^{2}) T^{N - 2}_{00}.
\end{equation}

In order to relate the MPS representation of the grand canonical partition function~\eqref{eq:recursive-partition-function} to the expression for $Z$, obtained by normalizing the thermodynamic ensemble, in Eq.~\eqref{eq:combinatoric-partition-function}, we look for a combinatoric formulation for $T^{N}_{00}$. We start by noting that the recursive relations~\eqref{eq:transfer-matrix-solutions} can be rewritten in terms of the following summations,
\begin{equation}
\begin{aligned}
    T^{N}_{00} & = \sum_{i = 0}^{\lfloor \frac{N}{2} \rfloor} \binom{N}{2i} (1 + \xi \omega)^{N - 2i} (\omega + \xi)^{2i}, \\
    T^{N}_{01} & = \sum_{i = 0}^{\lfloor \frac{N}{2} \rfloor} \binom{N}{2i} (\omega + \xi)^{N - 2i} (1 + \xi \omega)^{2i}. \\
\end{aligned}
\end{equation}
Expanding the binomials and rearranging for the spectral parameters then gives the following expression for $T^{N}_{00}$,
\begin{equation}
    T^{N}_{00} = \sum_{i = 0}^{\lfloor \frac{N}{2} \rfloor} \binom{N}{2i} \sum_{j = 0}^{\mathclap{N - 2i}} \binom{N - 2i}{j} \sum_{k = 0}^{2i} \binom{2i}{k} \xi^{2i + j - k} \omega^{j + k}.
\end{equation}
To make further progress, we split the expression for $\smash{T^{N}_{00}}$ into three separate summations (i.e., $k < i$, $k = i$, $k > i$) which independently count the sets of configurations with $\smash{\q^{+} > \q^{-}}$, $\smash{\q^{+} = \q^{-}}$, and $\smash{\q^{+} < \q^{-}}$ quasiparticles. By proving the equivalence of each of these to the associated part of Eq.~\eqref{eq:combinatoric-partition-function}, we necessarily prove the equivalence of the grand canonical partition functions~\eqref{eq:mps-partition-function} and~\eqref{eq:combinatoric-partition-function} and the correctness of the state counting function $\Omega$~\eqref{eq:partition-function-entropy}. In what follows, it will prove helpful to refer to the following binomial coefficient identities for increasing or decreasing the integers $n$ and $k$,
\begin{equation}\label{eq:partition-function-binomial-coefficient-identities}
\begin{aligned}
    \binom{n}{k} & = \frac{k + 1}{n - k} \binom{n}{k + 1}, \\
    \binom{n}{k} & = \frac{n + 1 - k}{n + 1} \binom{n + 1}{k},
\end{aligned}
\end{equation}
and Vandermonde's identity,
\begin{equation}\label{eq:partition-function-vandermonde-identity}
        \sum_{k = 0}^{j} \binom{m}{k} \binom{n}{j - k} = \binom{m + n}{j}.
\end{equation}

\begin{widetext}
    To start, we consider the summation for $k < i$,
    \begin{equation}
        T^{N}_{00} = \sum_{i = 1}^{\lfloor \frac{N}{2} \rfloor} \sum_{j = 0}^{N - 2i} \sum_{k = 0}^{i - 1} \binom{N}{2i} \binom{N - 2i}{j} \binom{2i}{k} \xi^{2i + j - k} \omega^{j + k}.
    \end{equation}
    In order to obtain the desired expression, we first shift the summation index $j \mapsto j - k$ and rearrange the order of the summations for $j$ and $k$ such that the expression reads
    \begin{equation}
        T^{N}_{00} = \sum_{i = 1}^{\lfloor \frac{N}{2} \rfloor} \sum_{k = 0}^{i - 1} \sum_{j = k}^{N - 2i - k} \binom{N}{2i} \binom{N - 2i}{j - k} \binom{2i}{k} \xi^{2i + j - 2k} \omega^{j}.
    \end{equation}
    Next, we rearrange the summations for $i$ and $k$ and subsequently shift the index of summation $i \mapsto i + k$ to give
    \begin{equation}
        T^{N}_{00} = \sum_{k = 0}^{\lfloor \frac{N - 2}{2} \rfloor} \sum_{i = 1}^{\lfloor \frac{N - 2k}{2} \rfloor} \sum_{j = k}^{N - 2i - k} \binom{N}{2i + 2k} \binom{N - 2i - 2k}{j - k} \binom{2i + 2k}{k} \xi^{2i + j} \omega^{j}. 
    \end{equation}
    Now, we use the binomial identities in Eq.~\eqref{eq:partition-function-binomial-coefficient-identities} to transform the coefficients such that we have
    \begin{equation}
        T^{N}_{00} = \sum_{k = 0}^{\lfloor \frac{N - 2}{2} \rfloor} \sum_{i = 1}^{\lfloor \frac{N - 2k}{2} \rfloor} \sum_{j = k}^{N - 2i - k} \binom{N}{2i + j} \binom{N - 2i - j}{k} \binom{2i + j}{j - k} \xi^{2i + j} \omega^{j}.
    \end{equation}
    which after rearranging the order of the summations $i$ and $k$ followed by $j$ and $k$ reads
    \begin{equation}
        T^{N}_{00} = \sum_{i = 1}^{\lfloor \frac{N}{2} \rfloor} \sum_{j = 0}^{N - 2i} \sum_{k = 0}^{j} \binom{N}{2i + j} \binom{N - 2i - j}{k} \binom{2i + j}{j - k} \xi^{2i + j} \omega^{j},
    \end{equation}
    where, in the summation over $k$, we use the identity $\binom{n < k}{k} = 0$. Finally, we apply Vandermonde's identity~\eqref{eq:partition-function-vandermonde-identity} to sum over $k$ to obtain,
    \begin{equation}
        T^{N}_{00} = \sum_{i = 1}^{\lfloor \frac{N}{2} \rfloor} \sum_{j = 0}^{N - 2i} \binom{N}{2i + j} \binom{N}{j}  \xi^{2i + j} \omega^{j}.
    \end{equation}
    Identifying the numbers of positive and negative quasiparticles as $\q^{+} \equiv 2i + j$ and $\q^{-} \equiv j$, respectively, it follows directly that this expression is exactly equivalent to Eq.~\eqref{eq:partition-function-entropy} for $\q^{+} > \q^{-}$, with the constraints in Eqs.~\eqref{eq:quasiparticle-constraint} and~\eqref{eq:quasiparticle-size-constraint} imposed by the bounds of the summations and the factor of $2$ from Eq.~\eqref{eq:recursive-partition-function}.
    
    Next, we consider the summation for $k = i$,
    \begin{equation}
        T^{N}_{00} = \sum_{i = 0}^{\lfloor \frac{N}{2} \rfloor} \sum_{j = 0}^{N - 2i} \binom{N}{2i} \binom{N - 2i}{j} \binom{2i}{i} \xi^{i + j} \omega^{i + j}.
    \end{equation}
    To start, we again shift the summation index $j \mapsto j - i$ to give
    \begin{equation}
        T^{N}_{00} = \sum_{i = 0}^{\lfloor \frac{N}{2} \rfloor} \sum_{j = i}^{N - i} \binom{N}{2i} \binom{N - 2i}{j - i} \binom{2i}{i} \xi^{j} \omega^{j}.
    \end{equation}
    Subsequently applying the identities~\eqref{eq:partition-function-binomial-coefficient-identities} yields
    \begin{equation}
        T^{N}_{00} = \sum_{i = 0}^{\lfloor \frac{N}{2} \rfloor} \sum_{j = i}^{N - i} \binom{N}{j} \binom{N - j}{i} \binom{j}{i} \xi^{j} \omega^{j},
    \end{equation}
    which after rearranging the order of the summations $i$ and $j$ reads
    \begin{equation}
        T^{N}_{00} = \sum_{j = 0}^{N} \sum_{i}^{j} \binom{N}{j} \binom{N - j}{i} \binom{j}{i} \xi^{j} \omega^{j},
    \end{equation}
    where again, in the summation over $i$, we have used the identity $\binom{n < k}{k} = 0$. Further noting the identity $\binom{n}{k} = \binom{n}{n - k}$, then applying Vandermonde's identity~\eqref{eq:partition-function-vandermonde-identity} to sum over $i$ then gives
    \begin{equation}
        T^{N}_{00} = \sum_{j = 0}^{N} \binom{N}{j} \binom{N}{j} \xi^{j} \omega^{j},
    \end{equation}
    which, substituted into Eq.~\eqref{eq:recursive-partition-function}, is exactly equal to Eq.~\eqref{eq:partition-function-entropy} for $\q^{+} = \q^{-} = j$.
    
    Finally, we consider the summation for $k > i$,
    \begin{equation}
        T^{N}_{00} = \sum_{i = 1}^{\lfloor \frac{N}{2} \rfloor} \sum_{j = 0}^{N - 2i} \sum_{k = i + 1}^{2i} \binom{N}{2i} \binom{N - 2i}{j} \binom{2i}{k} \xi^{2i + j - k} \omega^{j + k},
    \end{equation}
    which after shifting the summation index $j \mapsto j - 2i + k$ and rearranging the order of the summations $j$ and $k$ reads
    \begin{equation}
        T^{N}_{00} = \sum_{i = 1}^{\lfloor \frac{N}{2} \rfloor} \sum_{k = i + 1}^{2i} \sum_{j = 2i - k}^{N - k} \binom{N}{2i} \binom{N - 2i}{j - 2i + k} \binom{2i}{k} \xi^{j} \omega^{-2i + j + 2k}.
    \end{equation}
    We now shift the index $k \mapsto k + i$,
    \begin{equation}
        T^{N}_{00} = \sum_{i = 1}^{\lfloor \frac{N}{2} \rfloor} \sum_{k = 1}^{i} \sum_{j = i - k}^{N - i - k} \binom{N}{2i} \binom{N - 2i}{j - i + k} \binom{2i}{i + k} \xi^{j} \omega^{j + 2k},
    \end{equation}
    and subsequently apply Eqs.~\eqref{eq:partition-function-binomial-coefficient-identities} to obtain
    \begin{equation}
        T^{N}_{00} = \sum_{i = 1}^{\lfloor \frac{N}{2} \rfloor} \sum_{k = 1}^{i} \sum_{j = i - k}^{N - i - k} \binom{N}{j + 2k} \binom{N - j - 2k}{i - k} \binom{j + 2k}{i + k} \xi^{j} \omega^{j + 2k}.
    \end{equation}
    Lastly, we rearrange the order of the summations $i$ and $k$ and then $i$ and $j$, which returns,
    \begin{equation}
        T^{N}_{00} = \sum_{k = 1}^{\lfloor \frac{N}{2} \rfloor} \sum_{j = 0}^{N - 2k} \sum_{i = k}^{j} \binom{N}{j + 2k} \binom{N - j - 2k}{i - k} \binom{j + 2k}{i + k} \xi^{j} \omega^{j + 2k},
    \end{equation}
    before applying Vandermonde's identity~\eqref{eq:partition-function-vandermonde-identity} to sum over $i$ to give
    \begin{equation}
        T^{N}_{00} = \sum_{k = 1}^{\lfloor \frac{N}{2} \rfloor} \sum_{j = 0}^{N - 2k} \binom{N}{j + 2k} \binom{N}{j} \xi^{j} \omega^{j + 2k}.
    \end{equation}
    Substituted into Eq.~\eqref{eq:recursive-partition-function}, this is precisely equivalent to Eq.~\eqref{eq:partition-function-entropy} for $\q^{+} < \q^{-}$ with $\q^{+} = j$ and $\q^{-} = j + 2k$, thus proving the equivalence of Eqs.~\eqref{eq:mps-partition-function} and~\eqref{eq:combinatoric-partition-function} and the correctness of Eq.~\eqref{eq:partition-function-entropy}.
\end{widetext}

\begin{figure}[t]
    \centering
    \includegraphics{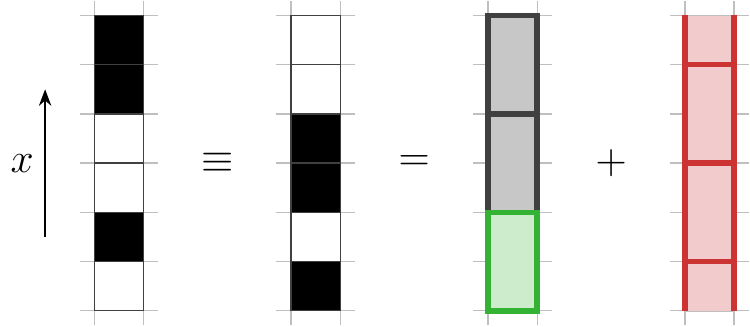}
    \caption{\textbf{State counting function.} The chain of $2N$ sites can be partitioned into two staggered overlapping lattices of $N$ blocks each. The blocks, which are composed of adjacent pairs of  sites, correspond to either quasiparticles, green for positive and red for negative, or vacua, indicated here by grey. As can be seen with the example configuration $\n = (0, 1, 0, 0, 1, 1)$, in this picture, configurations are equivalent under the exchange $\protect\cemp \leftrightarrow \protect\cocc$ (i.e., $0 \leftrightarrow 1$), hence the factor of $2$ in Eq.~\eqref{eq:partition-function-entropy}.}
    \label{fig:partition-function}
\end{figure}

Remarkably, we can also derive the expression for the counting function $\Omega$ directly from physical arguments by recalling the intrinsic properties of the quasiparticles. To start, we note that each and every quasiparticle occupies exactly two adjacent sites of the lattice and is statistically independent of each and every other quasiparticle. That is, the conditional probability of finding a quasiparticle at a pair of sites, given that the sites do not already contain a quasiparticle of that species does not depend on any of the other quasiparticles positions (see Sec.~\ref{sec:statistical-independence}). It then follows straightforwardly that the binomial coefficients in Eq.~\eqref{eq:partition-function-entropy} can be understood as independently counting the total number of possible ways to arrange $\q^{+}$ positive and $\q^{-}$ negative statistically independent quasiparticles of size $2$ in a system of size $2N$. An illustrative example highlighting the basic concepts of this argument, as well as an explanation for the multiplicative factor of $2$ which simply ensures that both subspecies of each quasiparticle are counted is presented in Figure~\ref{fig:partition-function}.

\section{Stochastic boundary driving constraint}\label{app:boundary-matrices-constraint}

In order to prove that the stochastic operators $\bm{R}$ and $\bm{L}$ must necessarily satisfy the particle-hole symmetry of the model, we show that the constraint~\eqref{eq:conditional-probability-symmetry} follows directly as a consequence of the boundary consistency condition. Solving separately the pair of equations~\eqref{eq:even-right-boundary-equation} and~\eqref{eq:odd-right-boundary-equation}, yields a unique solution for the spectral parameters $\xi$ and $\omega$ in terms of the conditional probabilities $R_{001}$ and $R_{011}$ and the normalization parameter $\Gamma \equiv \Gamma_{\text{R}} / \Gamma_{\text{L}}$. Namely,
\begin{align}
    \xi & = \frac{\Gamma - (1 - R_{001})}{R_{011}}, \label{eq:boundary-right-even-spectral-parameter-solutions} \\
    \omega & = \frac{\Gamma (1 - R_{011}) - (1 - R_{001} - R_{011})}{\Gamma R_{011}}. \label{eq:boundary-right-odd-spectral-parameter-solutions}
\end{align}
Similarly, solving separately the left boundary equations, Eqs.~\eqref{eq:even-left-boundary-equation} and~\eqref{eq:odd-left-boundary-equation}, returns the following unique solution for $\xi$ and $\omega$ in terms of the conditional probabilities $L_{100}$ and $L_{110}$ and normalization parameter $\Gamma$,
\begin{align}
    \xi & = \frac{(1 - L_{110}) - \Gamma(1 - L_{100} - L_{110})}{L_{110}}, \label{eq:boundary-left-even-spectral-parameter-solutions} \\
    \omega & = \frac{1 - \Gamma(1 - L_{100})}{\Gamma L_{110}}. \label{eq:boundary-left-odd-spectral-parameter-solutions}
\end{align}

We now demand that the expressions for $\xi$ in Eqs.~\eqref{eq:boundary-right-even-spectral-parameter-solutions} and~\eqref{eq:boundary-left-even-spectral-parameter-solutions} and for $\omega$ in Eqs.~\eqref{eq:boundary-right-odd-spectral-parameter-solutions} and~\eqref{eq:boundary-left-odd-spectral-parameter-solutions} are equivalent, respectively. Solving these coupled equations then yields a unique expression for the normalization parameter, 
\begin{equation}
    \Gamma = \frac{L_{110} (1 - R_{001}) + (1 - L_{110}) R_{011}}{R_{011} (1 - L_{100}) + (1 - R_{011}) L_{110}},
\end{equation}
which is equivalent to the expression derived in Eq.~\eqref{eq:ness-normalization-solution}, where the spectral parameters are given by
\begin{equation}
\begin{aligned}
    \xi & = \frac{R_{001} (1 - L_{110}) + (1 - R_{001}) L_{100}}{R_{011} (1 - L_{100}) + (1 - R_{011}) L_{110}}, \\
    \omega & = \frac{L_{100} (1 - R_{011}) + (1 - L_{100}) R_{001}}{L_{110} (1 - R_{001}) + (1 - L_{110}) R_{011}},
\end{aligned}
\end{equation}
which are precisely the solutions shown in Eq.~\eqref{eq:ness-spectral-parameters}, with $\xi, \omega \in \mathbb{R}^{+}$ for all $R_{001}, R_{011}, L_{100}, L_{110} \in [0, 1]$. Moreover,
solving the coupled equations for the spectral parameters also imposes a constraint on the conditional probabilities. Specifically, we have that,
\begin{widetext}
\begin{equation}
\begin{aligned}
    R_{110} & = \frac{\big(L_{100} (1 - R_{011}) + (1 - L_{100}) R_{001}\big) - (1 - R_{100})\big( R_{001} (1 - L_{110}) + (1 - R_{001}) L_{100} \big)}{R_{011} (1 - L_{100}) + (1 - R_{011}) L_{110}}, \\
    L_{011} & = \frac{\big(R_{001} (1 - L_{110}) + (1 - R_{001}) L_{100}\big) - (1 - L_{001})\big(L_{100} (1 - R_{011}) + (1 - L_{100}) R_{001}\big)}{L_{110} (1 - R_{001}) + (1 - L_{110}) R_{011}}.
\end{aligned}
\end{equation}
\end{widetext}
Requiring that each and every conditional probability is simultaneously bounded then returns a unique nontrivial solution for the $R_{n_{3}, n_{4}, n_{5}}$ and $L_{n_{0}, n_{1}, n_{2}}$, that is,
\begin{equation}
\begin{aligned}
    R_{100} & = R_{011}, \qquad & R_{110} & = R_{001}, \\
    L_{001} & = L_{110}, & L_{011} & = L_{100},
\end{aligned}
\end{equation}
which is exactly the constraint in Eq.~\eqref{eq:conditional-probability-symmetry}.

\section{Conditional probability factorization}\label{app:conditional-probabilities}

To prove the factorizations in Eqs.~\eqref{eq:right-conditional-probability-relations} and~\eqref{eq:left-conditional-probability-relations}, we must first clarify our notation. Specifically, let $\smash{p_{n_{1}, \ldots, n_{2N}}}$ and $\smash{p_{n_{1}, \ldots, n_{2N}}^{\prime}}$ denote the asymptotic probabilities for the configurations of even length $2N$ starting on either even sites at odd times or odd sites at even times, and either even sites at even times or odd sites at odd times. Then, let $\smash{p_{n_{1}, \ldots, n_{2N - 1}}}$ and $\smash{p_{n_{1}, \ldots, n_{2N - 1}}^{\prime}}$ denote the corresponding asymptotic probabilities for configurations of odd length $2N - 1$. Explicitly, these expressions read,
\begin{widetext}
\begin{align}
    p_{n_{1}, \ldots, n_{2N}} & = \lim_{M \to \infty} \frac{\Tr \big(\V_{n_{1}} \W_{n_{2}} \cdots \W_{n_{2N}} \T^{M - N}\big)}{\Tr \big(\T^{M}\big)} = \frac{\l \V_{n_{1}} \W_{n_{2}} \cdots \W_{n_{2N}} \r}{\chi^{N} \braket{l}{r}}, \label{eq:even-unprimed-configuration-probability} \\
    p_{n_{1}, \ldots, n_{2N}}^{\prime} & = \lim_{M \to \infty} \frac{\Tr \big(\W_{n_{1}} \V_{n_{2}} \cdots \V_{n_{2N}} \T^{M - N}\big)}{\Tr \big(\T^{M}\big)} = \frac{\l \W_{n_{1}} \V_{n_{2}} \cdots \V_{n_{2N}} \r}{\chi^{N} \braket{l}{r}}, \label{eq:even-primed-configuration-probability} \\
    p_{n_{1}, \ldots, n_{2N - 1}} & = \lim_{M \to \infty} \frac{\Tr \big(\V_{n_{1}} \W_{n_{2}} \cdots \V_{n_{2N - 1}} (\W_{0} + \W_{1}) \T^{M - N}\big)}{\Tr \big(\T^{M}\big)} = \frac{\l \V_{n_{1}} \W_{n_{2}} \cdots \V_{n_{2N - 1}} \r}{\chi^{N - 1} \l (\V_{0} + \V_{1}) \r}, \label{eq:odd-unprimed-configuration-probability} \\
    p_{n_{1}, \ldots, n_{2N - 1}}^{\prime} & = \lim_{M \to \infty} \frac{\Tr \big(\W_{n_{1}} \V_{n_{2}} \cdots \W_{n_{2N - 1}} (\V_{0} + \V_{1}) \T^{M - N}\big)}{\Tr \big(\T^{M}\big)} = \frac{\l \W_{n_{1}} \V_{n_{2}} \cdots \W_{n_{2N - 1}} \r}{\chi^{N - 1} \l (\W_{0} + \W_{1}) \r}, \label{eq:odd-primed-configuration-probability}
\end{align}
\end{widetext}
where we have used the facts that the following products of matrices and vectors hold,
\begin{equation}
\begin{aligned}
    (\W_{0} + \W_{1}) \r & = (1 + \omega) \r, \\
    (\V_{0} + \V_{1}) \r & = (1 + \xi) \r,
\end{aligned}
\end{equation}
and that both the transfer matrix $\T$ and vectors $\r$ and $\l$ are invariant under the exchange $\xi \leftrightarrow \omega$. That is,
\begin{equation}
    \T \equiv \T^{\prime}, \qquad \r \equiv \rp, \qquad \l \equiv \lp.
\end{equation}

In order to prove the relations in Eqs.~\eqref{eq:right-conditional-probability-relations} and~\eqref{eq:left-conditional-probability-relations}, we must show that the following products of vectors and matrices are \textit{linearly dependent}. Explicitly, that for each and every subconfiguration of two sites, there exist scalar coefficients $\smash{l_{n_{1}, n_{2}}}$, $\smash{l_{n_{1}, n_{2}}^{\prime}}$, $\smash{r_{n_{3}, n_{4}}}$, and $\smash{r_{n_{3}, n_{4}}^{\prime}}$, such that for the left boundary, the following identities hold,
\begin{align}
    \l \V_{n_{1}} \W_{n_{2}} & = l_{n_{1}, n_{2}} \l \W_{n_{2}}, \label{eq:left-unprimed-conditional-probability-relation} \\
    \l \W_{n_{1}} \V_{n_{2}} & = l_{n_{1}, n_{2}}^{\prime} \l \V_{n_{2}}, \label{eq:left-primed-conditional-probability-relation}
\end{align}
while for the right boundary, the identities read
\begin{align}
    \V_{n_{3}} \W_{n_{4}} \r & = r_{n_{3}, n_{4}} \V_{n_{3}} \r, \label{eq:right-unprimed-conditional-probability-relation} \\
    \W_{n_{3}} \V_{n_{4}} \r & = r_{n_{3}, n_{4}}^{\prime} \W_{n_{3}} \r. \label{eq:right-primed-conditional-probability-relation}
\end{align}
It can be straightforwardly demonstrated, by checking all four configurations for all four equations, that the scalar coefficients are given precisely by the tensors of the PSA in Eq.~\eqref{eq:psa-scalars} and MPS normalization constant in Eq.~\eqref{eq:ness-normalization-solution}. Explicitly, for the left boundary,
\begin{equation}
    l_{n_{1}, n_{2}} = \X_{n_{1}, n_{2}}, \qquad l_{n_{1}, n_{2}}^{\prime} = \Y_{n_{1}, n_{2}},
\end{equation}
and, similarly, for the right boundary,
\begin{equation}
    r_{n_{3}, n_{4}} = \frac{1}{\Gamma} \X_{n_{3}, n_{4}}, \qquad r_{n_{3}, n_{4}}^{\prime} = \Gamma \Y_{n_{3}, n_{4}}.
\end{equation}
From here, the factorization identities follow directly. To obtain the relations in Eq.~\eqref{eq:right-conditional-probability-relations}, we consecutively apply Eq.~\eqref{eq:left-unprimed-conditional-probability-relation} to Eqs.~\eqref{eq:even-unprimed-configuration-probability} and~\eqref{eq:odd-unprimed-configuration-probability} and, similarly, Eq.~\eqref{eq:left-primed-conditional-probability-relation} to Eqs.~\eqref{eq:even-primed-configuration-probability} and~\eqref{eq:odd-primed-configuration-probability}, while to acquire the equations in Eqs.~\eqref{eq:left-conditional-probability-relations}, we instead repeatedly apply~\eqref{eq:right-unprimed-conditional-probability-relation} and~\eqref{eq:right-primed-conditional-probability-relation}.

\section{\label{app:eigenvalue-degeneracy} Eigenvalue degeneracy conjecture}

To check the validity of the conjecture in Eq.~\eqref{eq:decay-eigenvalue-degeneracy} for the degeneracy $g$ of the eigenvalue $\Lambda$, we perform a simple calculation which counts the total number of eigenvalues. In particular, let $g(N) \equiv 2^{2N}$ denote the total number of eigenvalues of $\M{}$ for a system of even size $2N$. We argue that we can express this quantity as
\begin{equation}
    g(N) = 4 \sum_{p = 0}^{\mathclap{N - 1}} g(N, p), \quad g(N, p) = \sum_{q = 0}^{\mathclap{2N - 2}} g(N, p, q),
\end{equation}
which can intuitively be interpreted as counting the total number of degenerate eigenvalues by summing over every angle $q$, orbital $p$, and root $\lambda$ (cf. the multiplicative factor of $4$). The degeneracy $g \equiv g(N, p, q)$ then reads
\begin{equation}
    g(N, p, q) = \sum_{\mathclap{d | D}} \frac{d}{2N - 1} \sum_{\mathclap{d^{\prime} | D^{\prime}}} \mu(d^{\prime}) \binom{\frac{2N - 1}{d d^{\prime}}}{\frac{p}{d d^{\prime}}},
\end{equation}
where $\mu(\ \cdot\ )$ denotes the M\"{o}bius function and $j | k$ the set of positive integer divisors $j$ of the integer $k$, with
\begin{equation}
    D = \gcd(2N - 1, p, q), \qquad D^{\prime} = \frac{\gcd(2N - 1, p)}{q},
\end{equation}
where $\gcd(\ \cdot\ )$ denotes the greatest common divisor.

To start, we note that we can eliminate the summation over $q$ by expanding the summations over both $q$ and $d$, and then collecting terms in $d$ such that
\begin{equation}
    g(N, p) = \sum_{d | D^{\prime\prime}} \sum_{d^{\prime} | D^{\prime}} \mu(d^{\prime}) \binom{\frac{2N - 1}{d d^{\prime}}}{\frac{p}{d d^{\prime}}},
\end{equation}
with the integer
\begin{equation}
    D^{\prime\prime} = \gcd(2N - 1, p).
\end{equation}
Expanding the summations over $d$ and $d^{\prime}$, and collecting terms with similar binomial coefficients, we then obtain
\begin{equation}
    g(N, p) = \sum_{d | D^{\prime\prime}} \binom{\frac{2N - 1}{d}}{\frac{p}{d}} \sum_{d^{\prime} | d} \mu(d^{\prime}) = \binom{2N - 1}{p},
\end{equation}
where, to eliminate the summation over $d^{\prime}$, we have used the M\"{o}bius summation identity,
\begin{equation}
    \sum_{j | k} \mu(j) =
    \begin{cases}
        1, & k = 1, \\
        0, & k > 1. \\
    \end{cases}
\end{equation}
We now consider the summation over the orbital number $p$, which we can expand using Pascal's identity to read
\begin{equation}
\begin{aligned}
    g(N) & = 4 \sum_{p = 0}^{N - 1} \Bigg(\!\!\binom{2N - 2}{p - 1} + \binom{2N - 2}{p}\!\!\Bigg) \\
    & = 4 \Bigg(2 \sum_{p = 0}^{\mathclap{N - 2}} \binom{2N - 2}{p} + \binom{2N - 2}{N - 1}\!\!\Bigg),
\end{aligned}
\end{equation}
where, to obtain the latter equality, we used the binomial identity $\smash{\binom{j}{k < 0} = 0}$ to eliminate the term $\smash{\binom{2N - 1}{-1}}$. Finally, we split the first term into two separate summations over $p = 0, \ldots, N - 2$ and $p = N, \ldots, 2N - 2$ with the identity $\smash{\binom{j}{k} = \binom{j}{j - k}}$ to give
\begin{equation}
    g(N)= 4 \sum_{p = 0}^{\mathclap{2N - 2}} \binom{2N - 2}{p} = 4\big(2^{2N - 2}\big) = 2^{2N},
\end{equation}
as required where, to acquire the second equality, we used the binomial coefficient summation identity,
\begin{equation}
    \sum_{j = 0}^{k} \binom{k}{j} = 2^{k}.
\end{equation}

\section{Cumulants of long time observables}\label{app:cumulants-of-long-time-observables} 

To construct exact expressions for the cumulants $\kappa_{j}$ of the observable $K$ for all even system sizes $2N$ in the long time $T$ limit, we must first state Fa\`{a} di Bruno's formula, which generalizes the chain rule for higher derivatives. In particular, it states that if $z(y)$ and $y(x)$ are differentiable functions, then
\begin{equation}\label{eq:faa-di-brunos-forumla}
    \frac{\d^{p} z}{\d x^{p}} = \sum_{\mathcal{K}} \frac{p!}{k_{1}! \cdots k_{p}!} \frac{\d^{q} z}{\d y^{q}} \prod_{j = 1}^{p} \bigg( \frac{1}{j!} \frac{\d^{j} y}{\d x^{j}} \bigg)^{\ \mathclap{k_{j}}},
\end{equation}
where the summation is over every $p$-tuple of nonnegative integers $\mathcal{K} \equiv (k_{1}, \ldots, k_{p})$ satisfying the conditions,
\begin{equation}\label{eq:faa-di-brunos-constraints}
    \sum_{j = 1}^{p} j k_{j} = p, \qquad \sum_{j = 1}^{p} k_{j} = q.
\end{equation}
The exact expressions for the long time cumulants of the observable $K$ then follow directly from the application of Fa\`{a} di Bruno's formula~\eqref{eq:faa-di-brunos-forumla} to the SCGF~\eqref{eq:large-deviation-scaled-cumulant-generating-function}. To start, we consider the functions $\smash{z(y) = \theta_{N}(\tilde{\Lambda})}$ and $\smash{y(x) = \tilde{\Lambda}(s)}$, which after applying Eq.~\eqref{eq:faa-di-brunos-forumla} read
\begin{equation}
    \frac{\d^{p} \theta_{N}}{\d s^{p}} = \sum_{\mathcal{K}} \frac{p!}{k_{1}! \cdots k_{p}!} \frac{\d^{q} \theta_{N}}{\d \tilde{\Lambda}^{q}} \prod_{j = 1}^{p} \bigg( \frac{1}{j!} \frac{\d^{j} \tilde{\Lambda}}{\d s^{j}} \bigg)^{\ \mathclap{k_{j}}},
\end{equation}
with the intermediate derivative,
\begin{equation}
    \frac{\d^{q} \theta_{N}}{\d \tilde{\Lambda}^{q}} = \frac{(-1)^{q - 1} (q - 1)!}{\tilde{\Lambda}^{q}}.
\end{equation}
Now, we consider $\smash{z(y) = \tilde{\Lambda}(\tilde{\sigma})}$ and $\smash{y(x) = \tilde{\sigma}(s)}$, where we have introduced $\tilde{\sigma}(s) = \tilde{\eta}^{2}(s) - \tilde{\mu}(s)$, for which
\begin{equation}
    \frac{\d^{p} \tilde{\Lambda}}{\d s^{p}} = \frac{\d^{p} \tilde{\eta}}{\d s^{p}} + \sum_{\mathcal{K}} \frac{p!}{k_{1}! \cdots k_{p}!} \frac{\d^{q} \tilde{\Lambda}}{\d \tilde{\sigma}^{q}} \prod_{j = 1}^{p} \bigg( \frac{1}{j!} \frac{\d^{j} \tilde{\sigma}}{\d s^{j}} \bigg)^{\ \mathclap{k_{j}}},
\end{equation}
where the intermediate derivative,
\begin{equation}
     \frac{\d^{q} \tilde{\Lambda}}{\d \tilde{\sigma}^{q}} = \frac{(-1)^{q - 1} (2q)! \tilde{\sigma}^{-(2q - 1)/2}}{2^{2q} q!},
\end{equation}
and final derivative,
\begin{equation}
    \frac{\d^{p} \tilde{\eta}}{\d s^{p}} = \frac{1}{2} \sum_{k = 0}^{1} \sum_{j = 0}^{1} \psi_{j, k} (-N \zeta_{j, k})^{p} \exp(-N s \zeta_{j, k}).
\end{equation}
Finally, we consider $z = \tilde{\sigma}(\tilde{\eta})$ and $y = \tilde{\eta}(s)$, with
\begin{equation}
    \frac{\d^{p} \tilde{\sigma}}{\d s^{p}} = \sum_{\mathcal{K}} \frac{p!}{k_{1}! \cdots k_{p}!} \frac{\d^{q} \tilde{\sigma}}{\d \tilde{\eta}^{q}} \prod_{j = 1}^{p} \bigg( \frac{1}{j!} \frac{\d^{j} \tilde{\eta}}{\d s^{j}} \bigg)^{\ \mathclap{k_{j}}} - \frac{\d^{p} \tilde{\mu}}{\d s^{p}},
\end{equation}
where the intermediate derivative,
\begin{equation}
    \frac{\d \tilde{\sigma}}{\d \tilde{\eta}} = 2 \tilde{\eta}, \qquad \frac{\d^{2} \tilde{\sigma}}{\d \tilde{\eta}^{2}} = 2, \qquad \frac{\d^{q} \tilde{\sigma}}{\d \tilde{\eta}^{q}} = 0, \quad q \geq 3, \\
\end{equation}
and final derivative,
\begin{equation}
    \frac{\d^{p} \tilde{\mu}}{\d s^{p}} = \psi (-N \zeta)^{p} \exp(-N s \zeta).
\end{equation}

For completeness, let us now consider $\mathcal{K} = (k_{1}, \ldots, k_{p})$. It can be straightforwardly demonstrated that finding the $p$-tuples of nonnegative integers satisfying the constraint in Eq.~\eqref{eq:faa-di-brunos-constraints} is equivalent to finding every partition of the positive integer $p$ (i.e., every possible way of writing $p$ as a sum of positive integers, or equivalently, as a sum of $p$ nonnegative integers). The total number of partitions of a nonnegative integer $p$ is given by the partition function $P(p)$ from number theory~\cite{Andrews1976} and exhibits the following convenient recurrence relation,
\begin{equation}
    P(p) = \frac{1}{p} \sum_{\mathclap{j = 0}}^{\mathclap{p - 1}} \Sigma(p - j) P(j),
\end{equation}
where, by convention, $P(0) = 1$ and with $\Sigma(\ \cdot\ )$ the sum of divisors function,
\begin{equation}
    \Sigma(k) = \sum_{j | k} j,
\end{equation}
with $j | k$ denoting the set of positive integer divisors $j$ of the positive integer $k$. For reference, the integer sequence of the partition functions $P(p)$ can be found on Ref.~\cite{Sloane2021}.

\end{document}